\documentclass[twocolumn,numberedappendix,appendixfloats,twocolappendix]{openjournal}

\usepackage{lipsum}

\usepackage[usenames,dvipsnames]{xcolor}
\usepackage{textgreek}
\usepackage[utf8]{inputenc}
\usepackage[english]{babel}

\usepackage{hyperref}

\hypersetup{
    unicode, 
    colorlinks=true,
    linkcolor=linkcolor,
    citecolor=linkcolor,
    filecolor=linkcolor,
    urlcolor=linkcolor,
}
\usepackage{color,colortbl}
\definecolor{linkcolor}{RGB}{28, 96, 214}
\usepackage{tensind}
\tensordelimiter{?}
\DeclareGraphicsExtensions{.bmp,.png,.jpg,.pdf}
\usepackage{verbatim}
\usepackage[normalem]{ulem}
\usepackage{orcidlink}
\usepackage{soul}

\usepackage{newtxtext,newtxmath}
\usepackage{amsmath,amsfonts,amssymb,slashed,float,bm,algorithmic,bbold,wasysym}
\usepackage{array,multirow,multirow,dcolumn,booktabs}
\usepackage{physics}
\usepackage{savesym}
\savesymbol{tablenum}
\usepackage{siunitx}
\restoresymbol{SIX}{tablenum}
\usepackage{xspace}
\usepackage[capitalise]{cleveref}
\usepackage{aas_macros}

\makeatletter
\usepackage{etoolbox}
\patchcmd\H@refstepcounter{\protected@edef}{\protected@xdef}{}{}
\makeatother

\DeclareUnicodeCharacter{2212}{-}

\newcommand{\LCDM}{$\Lambda$CDM\xspace}
\newcommand{\Om}{\Omega_{\rm m}}
\newcommand{\planck}{\textit{Planck}\xspace}
\newcommand{\eg}{\textit{e.g.}\xspace}

\DeclareSIUnit\parsec{pc}
\DeclareSIUnit\sig{\ensuremath{\sigma}}
\DeclareSIUnit\h{\ensuremath{\mathit{h}}}
\DeclareSIUnit\iMpc{\per\mega\parsec}

\begin{document}
\title{Going beyond $S_8$: fast inference of the matter power spectrum from weak-lensing surveys}

\author{Cyrille Doux\orcidlink{0000-0003-4480-0096}}
\email{doux@lpsc.in2p3.fr}
\affiliation{Université Grenoble Alpes, CNRS, LPSC-IN2P3, 38000 Grenoble, France}

\author{Tanvi Karwal\orcidlink{0000-0002-1384-9949}}
\email{karwal@uchicago.edu}
\affiliation{Kavli Institute for Cosmological Physics, Enrico Fermi Institute, and Department of Astronomy \& Astrophysics, University of Chicago, Chicago IL 60637
}

\begin{abstract}
Weak lensing surveys are often summarized by constraints on the derived parameter ${S_8\equiv\sigma_8\sqrt{\Om/0.3}}$, obscuring the rich scale and redshift information encoded in the data, and limiting our ability to identify the origin of any tensions with $\Lambda$CDM predictions from the cosmic microwave background. 
In this work, we introduce a fast and flexible framework to extract the scale-dependent matter power spectrum $P(k, z)$ from cosmic shear and CMB lensing measurements, parameterizing deviations from the \planck $\Lambda$CDM prediction as a free function $\alpha(k)$. 
Using public data from DES Y3, KiDS-1000, HSC Y3, and ACT DR6, we constrain $\alpha(k)$ with fast Hamiltonian Monte Carlo inference, employing multipoles up to $\ell_{\rm max}\sim2000$ for the galaxy lensing surveys.
Our results show a consistent 15–30\% suppression in the matter power spectrum at intermediate scales ($k \sim \SIrange{0.1}{1}{\iMpc}$) in galaxy-lensing data relative to a \planck \LCDM prediction with a CDM-only (no baryonic feedback) power spectrum, with combined tensions reaching up to \SI{4}{\sig}\,. This is under a fixed cosmology and with analytic marginalization over shear and redshift calibration uncertainties.
In contrast, ACT CMB lensing is consistent with \LCDM at $k\lesssim\SI{0.1}{\iMpc}$.
We validate our method using mock data, quantify consistency between datasets, and demonstrate how the resulting $\alpha(k)$ likelihoods can be used to test specific models for the power spectrum. 
All code, data products, and derived likelihoods are publicly released.\footnote{The code, data, intermediate and final results, as well as a reproducible notebook for further model testing, are available in the GitHub repository: \url{https://github.com/xuod/cls2pk}.}
Our results highlight the importance of reporting lensing constraints on $P(k, z)$ and pave the way for model-agnostic test of growth of structure with upcoming surveys such as LSST, \textit{Euclid}, and \textit{Roman}.
\end{abstract}


\maketitle

\section{Introduction} \label{sec:Intro}

Understanding the distribution of matter in the Universe is central to cosmology, and a range of observational probes allows us to study it. 
The cosmic microwave background (CMB) offers precise constraints on cosmological parameters, including the amplitude of matter fluctuations, denoted by $\sigma_8$. 
Meanwhile, weak lensing---whether of background galaxies or of the CMB itself---offers independent, direct measurements of the large-scale structure of the Universe. 
However, these methods are not in perfect agreement.
Within the standard flat \LCDM cosmological model, lensing measurements from galaxy surveys generally yield lower values of matter clustering than those inferred from CMB observations, including CMB lensing~\citep{2105.13544,2105.13543,2203.07128,Dalal:2023olq,2007.15633,2305.17173,2304.05203,2304.05202,Planck:2018vyg,Dalal:2023olq,2503.19441}. 
This tension is often summarized by the derived parameter $S_8 \equiv \sigma_8 \sqrt{0.3/\Om}$~\citep{2209.12997}, which captures a combination of fluctuation amplitude and matter density in the parameter direction best constrained by galaxy weak-lensing experiments.
While $S_8$ provides a useful shorthand, it overlooks the fact that galaxy and CMB lensing probe different redshift ranges and physical scales~\citep{TZ}. 
Consequently, the observed discrepancies could indicate new physics beyond the standard \LCDM model. Therefore, a comprehensive analysis requires moving beyond the single parameter $S_8$ and fully exploiting the rich information contained within these datasets.

To do so, here we directly reconstruct the matter power spectrum $P(k, z)$ using galaxy and CMB lensing data. 
This spectrum encodes the statistical distribution of matter across different scales and redshifts and is a fundamental prediction of cosmological models computed by established Boltzmann solvers like \texttt{CLASS}~\citep{2011arXiv1104.2932L} and \texttt{CAMB}~\citep{2000ApJ...538..473L}. 
Using publicly-available galaxy-lensing data from the Dark Energy Survey~\citep[DES,][]{2005astro.ph.10346T}, the Kilo Degree Survey~\citep[KiDS,][]{2013ExA....35...25D}, and Hyper Suprime-Cam (HSC) Subaru Strategic Program~\citep{2018PASJ...70S...4A}, combined with CMB-lensing data from the Atacama Cosmology Telescope~\citep[ACT,][]{2011ApJS..194...41S}, we infer scale-dependent deviations from the matter power spectrum predicted by the best-fit \planck \LCDM model~\citep{Planck:2018vyg}.

Our approach introduces a linearized parameterization of these deviations, enabling a fast and flexible exploration of the matter power spectrum inferred from lensing data, bridging the gap between observations and theory. 
Importantly, we do not assume any specific physical origins for the deviations---whether they result from baryonic effects, intrinsic alignments, nonlinear structure formation, or genuinely new physics. Rather, we aim to characterize their presence and scale-dependence directly from the data.

Recent studies have explored modifications to the cold dark matter power spectrum as a potential solution to the apparent $S_8$ tension.  These approaches include 
empirically parameterizing a suppression of power on small scales~\citep{2206.11794,2305.09827,2409.13404} 
or at late times~\citep{2023PhRvD.107h3504A,2308.16183}, 
modeling baryonic feedback mechanisms~\citep{2025PhRvD.111f3509T}, 
introducing dark matter interactions~\citep{2505.02233,2024arXiv240906771Z,Khoury:2025txd,Poulin:2022sgp,Asghari:2019qld,BeltranJimenez:2021wbq}, 
using perturbative expansions~\citep{2411.07082}, 
and employing alternative reconstruction techniques~\citep{2404.18240,2502.04449,2502.06687}. 
Our method complements these efforts by offering a heuristic yet data-driven framework for testing consistency with \LCDM.

Our inference approach is presented in \cref{sec:discretization_TZ}.
Details of the datasets are provided in \cref{sec:data}. Our methodology---including binning choices, treatment of systematics, and validation with mock data---is described in \cref{sec:method}.
Our main results, quantifying individual and combined deviations of the lensing datasets from \planck \LCDM cosmology, are presented in \cref{sec:results}. 
\cref{sec:interpretation_application} outlines the scope and potential applications of these results, demonstrating how this approach can be used to constrain physics beyond \LCDM, with concrete examples based on two different prescriptions for the small-scale power spectrum. 
Finally, we conclude with a summary of our findings and our outlook in \cref{sec:conclusions}.

All code, intermediate data products, and posterior samples of $\alpha(k)$ used in this work are publicly available at \url{https://github.com/xuod/cls2pk}. We encourage lensing surveys to present similar constraints on the matter power spectrum $P(k, z)$, moving beyond the traditional $S_8$ summary statistic. In addition, we provide ready-to-use tools for theorists to test the viability of their models against the reconstructed $\alpha(k)$, capturing the scale-dependence of deviations. These are packaged as modular likelihoods compatible with \texttt{MontePython}~\citep{2013JCAP...02..001A}, enabling fast and flexible model comparison.

\section{Fast power spectrum inference}
\label{sec:discretization_TZ}

The galaxy and CMB lensing power spectra $C_\ell$ can be computed as line-of-sight integrals over the matter power spectrum $P(k,z)$ weighted by specific window functions. 
Within the Limber approximation~\citep{2008PhRvD..78l3506L,2018ARA&A..56..393M}, this relationship is given by
\begin{equation}
    C^{ab}_\ell = \int \dd{\chi} \frac{q_a(\chi) q_b(\chi)}{\chi^2} P\qty(k=\frac{\ell+1/2}{\chi}, z(\chi)),
    \label{eq:limber}
\end{equation}
where $\chi$ is the radial comoving distance from the observer, and the window functions $q_a$ encode information about the sources and cosmic distances. 
For CMB lensing, this window function is
\begin{equation}
    q(\chi) = \frac{3\Om H_0^2}{2c^2} 
                \frac{\chi}{a(\chi)}
                \frac{\chi-\chi_*}{\chi_*},
\end{equation}
where $\chi_*$ is the comoving distance to the surface of last scattering and $a$ is the scale factor. 
For galaxy lensing, it becomes
\begin{equation}
    q_a(\chi) = \frac{3 \Om H_0^2}{2c^2} \frac{\chi}{a(\chi)} \int_{\chi}^{\chi_*} \dd{\chi'} n_a(\chi') \frac{\chi-\chi'}{\chi'},
\end{equation}
where $n_a(\chi)$ is the redshift distribution of the $a$-th tomographic bin of source galaxies.

We introduce a scale-dependent, redshift-independent modulation $\alpha(k)$ of the matter power spectrum given by
\begin{equation}
    P(k,z) = [1 + \alpha(k)] P_{\rm fid}(k,z),
    \label{eq:alpha}
\end{equation}
where $P_{\rm fid}(k,z)$ is the nonlinear matter power spectrum computed using the \planck best-fit \LCDM cosmology~\citep{Planck:2018vyg}, employing the standard \texttt{halofit} prescription for nonlinearities~\citep{1208.2701}. 
By changing the integration variable in \cref{eq:limber} from $\chi$ to $k=(\ell+1/2)/\chi$ and then to $\log k$, this linear relation can be discretized into a sum over $n_k$ logarithmically-spaced $k$-bins with centers $\log k_i$ and width $\Delta_{\log k}$. 
Given a vector $\bm{C}^{ab}$ of length $n_\ell$ representing measurements of angular power spectra for tomographic bins $a$ and $b$ over a range of multipoles $\ell$, we obtain a linear model
\begin{equation}
    \bm{C}^{ab} = \bm{\mathsf{W}}^{ab} \qty(\bm{1} + \bm{\alpha}),
    \label{eq:model}
\end{equation}
where $\bm{\alpha}=\qty(\alpha(k_1),\dots,\alpha(k_{n_k}))$ represents deviations introduced in \cref{eq:alpha}, $\bm{1}=\qty(1,\cdots,1)$ is a vector of ones, and the window matrix $\bm{\mathsf{W}}^{ab}$ has $n_\ell \times n_k$ elements given by
\begin{equation}
    \bm{\mathsf{W}}^{ab}_{\ell,k_i} 
        =   \Delta_{\log k} 
            \frac{q_a(\chi_{\ell}^i) q_b(\chi_{\ell}^i)}{\chi_{\ell}^i} 
            P_{\rm fid}\qty(k_i, z(\chi_{\ell}^i)),
    \label{eq:window_matrix}
\end{equation}
with $\chi_{\ell}^i = (\ell+1/2)/k_i$. 
We then stack vectors and window matrices for every tomographic bin combination $ab$, and potentially for multiple data sets to combine them. 
To account for the fact that power spectra are measured in multipole bins (or bandpowers), we convolve the window matrices in \cref{eq:window_matrix} by the corresponding binning matrices~\citep{1809.09603,2010.09717}, such that the model remains linear, as in \cref{eq:model}.

The likelihoods for the galaxy- and CMB-lensing datasets used here are multivariate Gaussian distributions with fixed covariance matrices. 
Therefore, at fixed cosmology, $\bm{\mathsf{W}}$ is a constant matrix and using \cref{eq:model} to compute the means of these likelihoods results in the following statistical model 
\begin{equation}
    \mathcal{L}(\bm{C}|\bm{\alpha}) = \mathcal{N}\qty(\bm{C}; \bm{\mathsf{W}} \qty(\bm{1} + \bm{\alpha}), \bm{\mathsf{\Sigma}}),
\end{equation}
where $\bm{C}$ is the stacked data vector, $\bm{\mathsf{W}}$ the corresponding stacked window matrix, and $\bm{\mathsf{\Sigma}}$ the data covariance. This constitutes a differentiable model suitable for efficient Bayesian sampling using Hamiltonian Monte-Carlo techniques implemented in probabilistic programming libraries such as \texttt{PyMC}~\citep{PyMC}. 

We note that~\cite{TZ} proposed a direct, analytic inversion of \cref{eq:model} that was used \eg in~\cite{2203.07128}. 
However, it requires a strong regularization for $k$-modes that are not well-constrained by the data, adding a hyperparameter that depends on the choice of binning, and potentially leading to numerical instabilities---at least for the specific exercise at hand.
In contrast, with a Bayesian inversion, poorly constrained $k$-modes are simply prior-dominated, independent of binning.

\section{Data}
\label{sec:data}

\subsection{Weak-lensing survey data}

\subsubsection{Data sets}

Our analysis incorporates public data from three galaxy weak-lensing surveys and their recent cosmic shear analyses: DES Year 3\footnote{
DES Year 3 catalogs and real-space measurements are publicly available from \href{https://des.ncsa.illinois.edu/releases/y3a2}{NCSA}. Harmonic-space measurements used here are available upon request to Cyrille Doux.}~\citep[DES~Y3,][]{2203.07128}, KiDS-1000\footnote{KiDS-1000 data are publicly available from the \href{https://kids.strw.leidenuniv.nl/DR4/KiDS-1000_cosmicshear.php}{KiDS} website.}~\citep{2007.15633},
and HSC Year 3\footnote{HSC Year 3 data were provided by Roohi Dalal.}~\citep[HSC~Y3,][]{Dalal:2023olq}. 
From each survey, we use only the cosmic shear data, which includes the source redshift distributions $n_a(z)$ of the tomographic bins of source galaxies, the auto- and cross-correlated shear power spectra $C_\ell^{ab}$ between these bins, and their covariance matrices. 
DES~Y3 has 4 tomographic bins, hence 10 cross-correlation combinations for shear power spectra, as does HSC~Y3. KiDS-1000 has 5 tomographic bins for the redshift distributions of their sources and hence 15 combinations of these bins. 

For DES and KiDS, we use the entire range of scales available in measured power spectra, with $\ell_{\rm max} \simeq 2000$. 
HSC probes at far greater angular resolution, and we present results here making the scale cuts $300<\ell<1800$ recommended by their main analysis \citep{Dalal:2023olq} as well as a separate analysis including all scales up to $\ell_{\rm max} \simeq 15800$ (see \cref{app:syst}). 
{Note that these extremely small scales extend beyond the range where cosmological modeling is reliable due to uncertainties from baryonic effects and nonlinear evolution. 
Regardless, we include them in \cref{app:syst} to explore how the data itself constrains this regime of the projected power spectrum but caution against interpreting it for cosmological inference.}

\subsubsection{Marginalization over systematic effects}

For all three surveys, we include uncertainties $\Delta z_a$ in the redshift distributions $n_a(z)$ of source galaxies 
\begin{equation}
    n_a(z) \rightarrow n_a(z + \Delta z_a),
\end{equation}
where $z = z(\chi)$ as before. 
This allows for a translation in redshift of the entire redshift distribution of bin $a$. 
Here $\Delta z_a$ have Gaussian priors following the recommendations from each experiment, except for the last two redshift bins for HSC. \footnote{For DES~Y3, the priors are provided in Table~1 of~\cite{2203.07128}. For KiDS-1000, we use means and standard deviations from Table~1 of~\cite{2007.15633} and the correlation matrix shown in Fig.~6 of~\cite{2007.15635} to compute the correlated prior. For HSC~Y3, priors are given in Table~I of~\cite{Dalal:2023olq}. For bins 3 and 4, the prior is uniform in the range $[-1,+1]$, and we instead use the redshift parameter posteriors from Table~VI.}
For these bins, the original analysis posits that the source redshift calibrations are inaccurate, assumed to be shifted by a Gaussian with a mean and error that is marginalized over. 
Here, we adopt the posteriors of that marginalization as our priors for these two bins. 
This remains a conservative choice as the original analysis arrived at the posteriors while simultaneously marginalizing over cosmology, while we fix our cosmology to \planck \LCDM. 

For HSC and DES, we also include uncertainties $m_a$ in the shear calibration of each bin such that the overall shear amplitude of the bin can be rescaled as
\begin{equation}
    C_\ell^{ab} \rightarrow (1+m_a) (1+m_b) C_\ell^{ab}.
\end{equation}
These are assumed to be redshift independent within each bin following the analyses of the surveys, and we again adopt the Gaussian priors on these systematic parameters prescribed by the surveys. 
For KiDS, shear biases are included in the data covariances so we do not vary $m_a$ as additional parameters, following their analysis~\citep{2007.01844}.

To account for these uncertainties, we analytically marginalize over these biases, assuming Gaussian priors, following the method described in~\cite{0112114,1003.1136}. In essence, it adds a term to the data covariance matrix $\bm{\mathsf{\Sigma}}$ such that 
\begin{equation}
    \bm{\mathsf{\Sigma}} \rightarrow 
        \bm{\mathsf{\Sigma}} + 
        \sum_\theta \sigma^2_\theta \, \partial_\theta{\bm{C}^{ab}} 
        \cdot \partial_\theta{\bm{C}^{ab}}^{\mathsf{T}},
\end{equation}
where the sum runs over the parameters $\theta$ to be marginalized over, 
$\sigma_\theta$ is the standard deviation of the prior\footnote{
For KiDS, the redshift calibration errors are correlated, and $\sigma_\theta$ is straightforwardly promoted to a covariance matrix $\Sigma_\theta$.
} and 
$\partial_\theta{\bm{C}^{ab}}$ is the partial derivative of the shear power spectrum data vector with respect to $\theta$, 
which we evaluate analytically for shear biases and numerically for redshift distributions. 
Note that we estimate this extra covariance term as well as window matrices at the mean of the priors.

We do not include intrinsic alignment or baryonic effects in our treatment of systematics for any of the surveys in our analysis. 
Indeed, both will affect the projected power spectrum probed by weak-lensing surveys, which we attempt {to reconstruct rather than model explicitly} (see \cref{sec:results}).

\subsection{CMB lensing}

We use ACT DR6 data\footnote{
ACT DR6 CMB lensing data is publicly available from \href{https://lambda.gsfc.nasa.gov/product/act/actadv_dr6_lensing_lh_get.html}{LAMBDA}. }, the provided data covariance and follow their binning scheme for the CMB lensing power spectrum $C_\ell$~\citep{2304.05202,2304.05203}. 
We include multipoles in the range ${40<\ell<763}$, following the ACT DR6 analysis (see Section 6.1 of~\cite{2304.05202}).
The CMB lensing power spectrum has no systematics beyond astrophysical foregrounds, which are already accounted for in the provided covariance matrix.

\subsection{Sensitivity of data on the $(k,z)$ plane}

\begin{figure}
    \centering
    \includegraphics[scale=0.68]{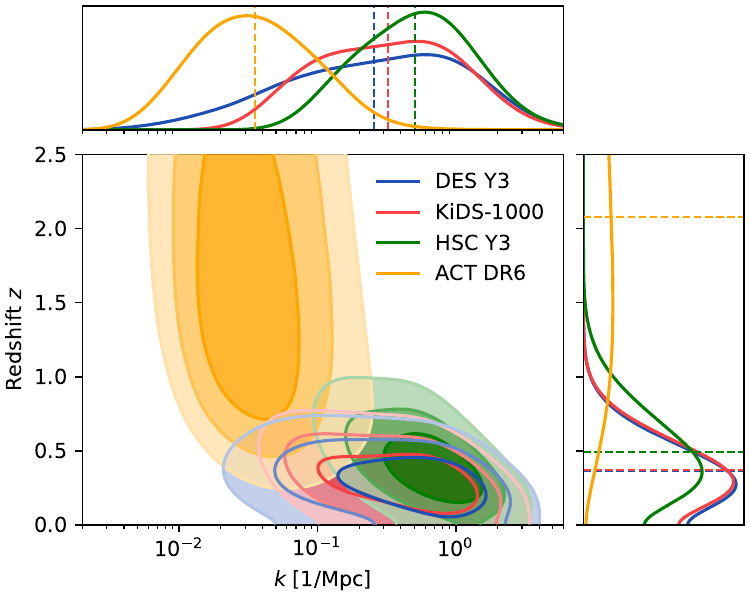}
    \caption{
    The contours in the main panel show the signal-to-noise ratio (SNR) defined in \cref{eq:snr} and hence the sensitivity of each experiment on the scale-redshift $(k,z)$ plane. 
    The upper and right panels show the SNR projected along each direction, 
    with dashed lines denoting the mean redshifts and (log) scales that the experiments are sensitive to (mean values are provided in \cref{tab:kzeff}).
    Where the contours overlap, they probe the same physics in redshift-independent $\alpha(k)$.
    However, as CMB- and galaxy-lensing surveys have little overlap, there are conceivable models that may impact the $P(k,z)$ probed by one observable but not the other. 
    Therefore, we urge caution when combining these data sets on $P(k,z)$ and $\alpha(k)$. 
    The HSC contours exclude scales following the original HSC analysis, such that all galaxy surveys have $\ell_{\rm max} \simeq 2000$.
    {When including all HSC scales, the sensitivity extends to smaller scales, as expected. We also account for the ACT fiducial scale cuts, with $\ell_{\rm max}=763$.}
    Note that the contours are simply linearly spaced from the maximum SNR to $0$, these do not represent $\sigma$ levels. 
    }
    \label{fig:kz_sensitivity}
\end{figure}

\begin{table}[t]
    \centering
    \caption{Effective scales and redshifts probed by the four lensing surveys considered in this work. They are computed as the means of the distributions of their signal-to-noise ratio in $(k,z)$ space, defined in \cref{eq:snr}.}
    \label{tab:kzeff}
    \begin{tabular}{l c c}
        \toprule
        Experiment & $k_{\rm eff}$ & $z_{\rm eff}$ \\
        \midrule
        DES Y3      & \SI{2.6e-1}{\iMpc} & 0.36 \\
        KiDS-1000   & \SI{3.2e-1}{\iMpc} & 0.37 \\
        HSC Y3      & \SI{5.1e-1}{\iMpc} & 0.49 \\
        ACT DR6     & \SI{3.5e-2}{\iMpc} & 2.08 \\
        \bottomrule
    \end{tabular}
\end{table}

Each galaxy-lensing survey has different sky footprints, survey areas, depths and tomographic redshift bins. 
CMB lensing has a different window function entirely. 
They each hence probe cosmology at different scales and redshifts. 
We delineate the sensitivity of each survey on the $(k,z)$ plane in \cref{fig:kz_sensitivity}. 
This is determined by calculating 
\begin{equation}
    \pdv{C_\ell^{ab}}{\log k}{z} = \frac{q_a(\chi) q_b(\chi)}{\chi}P(k,z(\chi)) \delta\qty(k = \frac{\ell + 1/2}{\chi})    
\end{equation}
for each tomographic bin pair and stacking them. We then take the resultant vector ${\partial^2_{\log k \, z}\bm{C}}$ and compute the signal-to-noise ratio (SNR) as 
\begin{equation}
    {\rm SNR}(\log k,z) = \sqrt{{\partial^2_{\log k \, z}\bm{C}} \cdot \bm{\mathsf{\Sigma}}^{-1} \cdot {\partial^2_{\log k \, z}\bm{C}}^{\mathsf{T}}},
    \label{eq:snr}
\end{equation}
where $\bm{\mathsf{\Sigma}}$ is the covariance of the $C_\ell^{ab}$ including systematics. 

{Because the Limber relation $k = (\ell+1/2)/\chi(z)$ links angular multipoles to physical wavenumbers, each multipole~$\ell$ samples a slice in the $(k,z)$ plane. 
When combined across multipoles and tomographic bins, the surveys accumulate sensitivity through this plane, as shown in \cref{fig:kz_sensitivity}. 
{\Cref{tab:kzeff} indicates the corresponding effective scales and redshifts of peak sensitivity for each experiment. }
As a consequence, while a scale-dependent modification such as $\alpha(k)$ is formally a function of $k$ alone, constraints on $\alpha(k)$ at different wavenumbers are informed by different redshift ranges. This redshift dependence arises from the survey geometry and window functions, rather than from the theoretical definition of $\alpha(k)$ itself.}

DES and KiDS almost completely overlap, and share a significant overlap with HSC, which is a deeper survey and hence probes slightly higher redshifts. 
The survey area for HSC is smaller, hence the compression of its contour towards smaller scales and larger $k$. 
However, galaxy-lensing sensitivity has little overlap with CMB lensing. 
The CMB-lensing kernel peaks at $z \simeq 2$, while all galaxy lensing kernels peak at $z \lesssim 1$, leading to minimal redshift overlap. 
These hence also probe physics at different scales, with CMB lensing probing entirely linear scales and galaxy lensing extending into quasi-linear and nonlinear scales. 
For this reason, we will refrain from claiming a reconstruction of the present-day matter power spectrum, $P(k, z=0)$, and focus on our specific, yet general, scale-dependent parametrization via $\alpha(k)$ of $P(k,z)$.

\subsection{Survey correlations}

Whenever possible, we aim to combine data sets to tighten constraints on $\alpha(k)$. Given the overlap in $(k,z)$ space and on the sky, we account for correlations between the experiments with an approximate cross-covariance. To do so, we start from the analytic expression for the unbinned spectra \citep{1906.11765,2012.08568} 
\begin{equation}
    {\rm Cov}\qty(C_\ell^{ab},C_{\ell^\prime}^{a^\prime b^\prime}) \approx \delta_{\ell\ell^\prime} \frac{C_\ell^{aa^\prime} C_\ell^{bb^\prime} + C_\ell^{ab^\prime} C_\ell^{ba^\prime}}{(2\ell+1)} \frac{f_{\rm sky}^{\rm overlap}}{f_{\rm sky}f_{\rm sky}^\prime},
\end{equation}
where $f_{\rm sky}$ and $f_{\rm sky}^\prime$ are the respective sky fractions of the surveys being correlated, and $f_{\rm sky}^{\rm overlap}$ the sky fraction of their overlap. We amended the formula from~\cite{1906.11765,2012.08568}, which has a term $1/f_{\rm sky}$ instead of the ratio we propose, to account for the fact that for fixed survey areas, the cross-covariance reduces if the overlap decreases. We then finally convolve this expression with the binning window matrices.

While this is an approximation, the overlap on sky is generally small---except when combining galaxy surveys with CMB lensing data, in which case the overlap in $(k,z)$ space becomes small---meaning that cross-covariance should be relatively small and only have a minor impact on $\alpha(k)$ constraints. In practice, we find that neglecting this cross-covariance shifts the $\alpha(k)$ tension by less than \SI{0.1}{\sig} (see definitions in the next section and column $n_\sigma$ of \cref{tab:individual_results}), validating our approach.

\section{Implementation and validation}
\label{sec:method}

\begin{table}
    \centering
    \caption{Fiducial cosmological parameters from \planck 2018 (TT,TE,EE+lowE) used in this work.}
    \label{tab:fiducial_cosmology}
    \begin{tabular}{lc}
        \toprule
        Parameter ([unit]) & Value \\
        \midrule
        $\Omega_{\rm b} h^2$ & 0.02236 \\
        $\Omega_{\rm c} h^2$ & 0.1202 \\
        $\ln(10^{10} A_{\rm s})$ & 3.045 \\
        $n_{\rm s}$ & 0.9649 \\
        $H_0$ [km/s/Mpc] & 67.27 \\
        $N_{\rm eff}$ & 3.046 \\
        $m_\nu$ [eV] & 0.06 (1 massive, 2 massless) \\
        \bottomrule
    \end{tabular}
\end{table}

Our analysis adopts the \planck 2018 cosmology~\citep{Planck:2018vyg} (TT,TE,EE+lowE) as the fiducial model. The parameters are listed in \cref{tab:fiducial_cosmology}. We assume one massive neutrino with mass $m_\nu = \SI{0.06}{\eV}$, two massless species, and $N_{\rm eff} = 3.046$.
All cosmological quantities are computed using the Core Cosmology Library~\citep[\texttt{CCL},][]{2019ApJS..242....2C}.

\subsection{Binning and smoothing}
\label{sec:smoothing}

\begin{figure}
    \centering
    \includegraphics[scale=0.68]{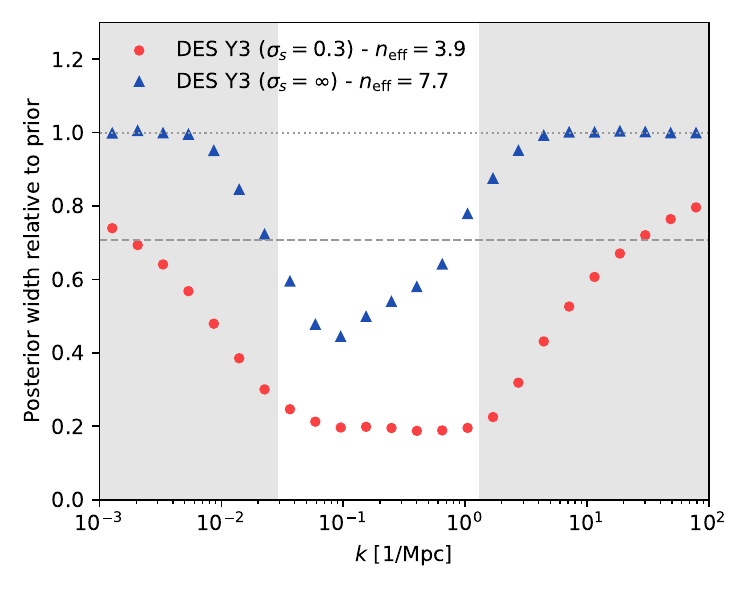}
    \caption{Ratios of the posterior and prior widths for $\alpha(k)$ for DES using $n_k=24$ $k$-bins, without (blue triangles) and with (red points) a smoothing prior. 
    From the case without smoothing, we can determine the $k$-range that is constrained $\sim \SI{3e-2}{\iMpc}<k<\SI{1}{\iMpc}$ shown by the unshaded band between the gray areas. 
    We use the prior and posterior covariances to compute the effective number of constrained parameters, of order $n_{\rm eff}\simeq6.5$ for DES, on par with the 9~bins in the constrained range. 
    Smoothing reduces the relative impact of data and the effective number of constrained bins to $n_{\rm eff}\simeq4.0$, but yields more physical curves.}
    \label{fig:DES_constraints}
\end{figure}

For our analysis to be reasonably accurate, the model in \cref{eq:model} must approximate the full Limber integral in \cref{eq:limber} to a fraction of the error bars. 
Therefore, the $k$-bins must cover the full range of $k$-modes that contribute to the signal for all measurements, and be sufficiently thin. 
In practice, we find that these conditions are reached for all surveys for $n_k \gtrsim 20$ logarithmically-spaced bins in the range \SIrange{e-3}{e2}{\iMpc}, which we fix at $n_k = 24$ throughout this work.

Another sensible requirement to ease the interpretation of our results is to choose $n_k$ such that the number of $k$-bins in the range that is constrained by the data (see \cref{fig:kz_sensitivity}) does not greatly exceed the effective number of modes that are actually constrained by the data. 
To do so, we performed tests varying this value and comparing the width of the posterior to that of the $[-1,+1]$ uniform prior, as shown on \cref{fig:DES_constraints}. We find that $n_k=24$ allows us to meet this requirement as well.

\begin{figure*}
    \centering
    \includegraphics[scale=0.68]{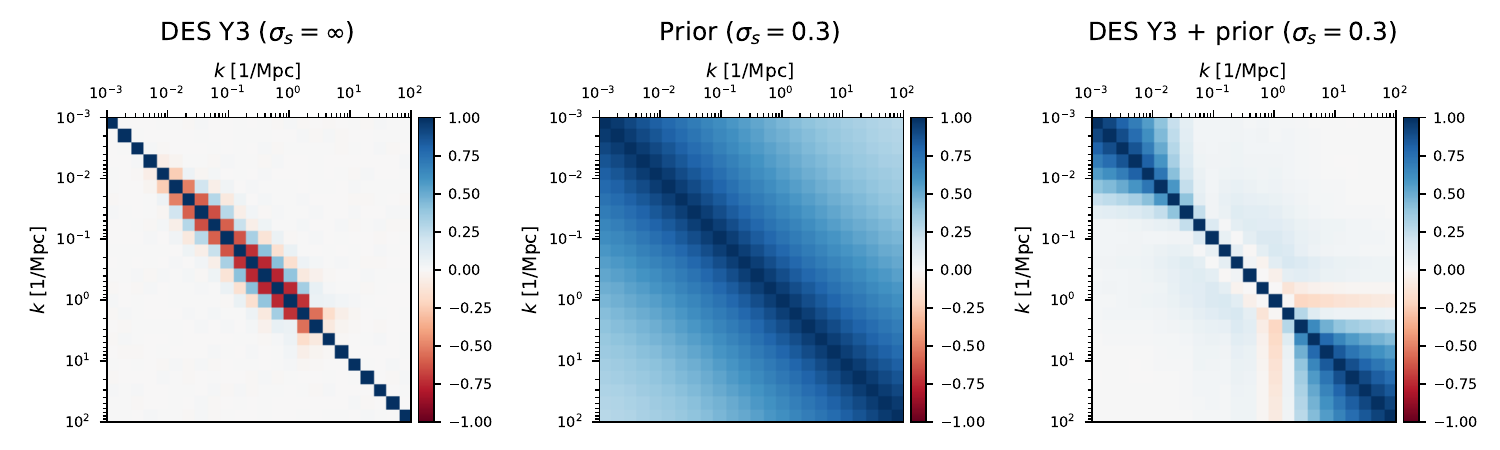}
    \caption{
    Correlation matrices of the posterior distribution of $\bm{\alpha}$'s for DES without smoothing (left), of the smoothing prior with smoothing strength $\sigma_{\rm s}=0.3$ (center), and of the posterior distribution for DES when including this smoothing prior (right).
    The middle and right panels demonstrate the strong correlations introduced across bins where data loses sensitivity at small and large $k$, in the regions beyond where $\alpha(k)$ are well-constrained.
    }
    \label{fig:DES_corr}
\end{figure*}

Undesirably, adjacent $k$-bins are strongly anticorrelated in this configuration. 
This is evident in the leftmost panel of \cref{fig:DES_corr} showing the correlation matrix of the posterior distribution, determined using a Gaussian likelihood with the model in \cref{eq:model}, the data covariance plus the systematics marginalization term, and a uniform prior in the range $[-1,+1]$. 
This is expected, as adjacent bins may compensate each other in \cref{eq:model}, allowing oscillating patterns. 
However, such patterns may be unphysical and they limit the interpretability of our results.

We therefore enforce a certain level of smoothness on the reconstructed $\alpha(k)$, penalizing rapid variations while avoiding biasing the reconstruction towards the fiducial \LCDM model, loosely inspired by~\cite{1902.01366}. 
To do so, we impose a Gaussian prior $\Pi_s(\bm{\alpha})$ on the differences of adjacent $\alpha(k)$'s 
\begin{equation}
    \Pi_s(\bm{\alpha}) = \prod_{i=1}^{n_k-1} \mathcal{N}\qty(\alpha(k_{i+1})-\alpha(k_i); 0, \Delta_{\log k} \sigma_{\rm s}).
    \label{eq:prior}
\end{equation}
The coefficient $\sigma_{\rm s}$ allows us to vary the constraint on the derivative of the reconstructed $\alpha(k)$, such that ${\abs{\Delta\alpha/\Delta_{\log k}} \lesssim\sigma_{\rm s}}$.
This setup then allows smooth deviations from the fiducial \LCDM but suppresses rapid oscillations between adjacent bins, decorrelating them. 

We adopt a heuristic approach to choose $\sigma_{\rm s}$, picking the value that approximately decorrelates adjacent bins. 
In other words, we balance the data and prior such that variations in $\alpha(k)$ are roughly independent in the posterior distribution, allowing one to read ``$\chi$-by-eye" on plots, at the price of reducing the effective number $n_{\rm eff}$ of constrained parameters. 
For the covariance matrices of the prior $\mathcal{C}_{\Pi_s}$ and the posterior $\mathcal{C}_p$, this can be estimated as
\begin{equation}
    n_{\rm eff}=\Tr(\mathbb{I} -\mathcal{C}_{\Pi_s}^{-1} \mathcal{C}_p) = n_k-\Tr(\mathcal{C}_{\Pi_s}^{-1} \mathcal{C}_p),
    \label{eq:neff}
\end{equation}
where $\Tr$ is the trace operator~\citep{2020PhRvD.101j3527R}. For DES, we find $n_{\rm eff}\simeq7.8$ without a smoothing prior (corresponding to $\sigma_{\rm s}=\infty$), and $n_{\rm eff}\simeq3.6$ with a smoothing prior of strength $\sigma_{\rm s}=0.3$ that yields an approximately diagonal posterior in the constrained $k$-range. 

With these choices for $n_k = 24$ bins and smoothing scales $\sigma_{\rm s}=\{0.3, \infty\}$, we use \cref{eq:model} to obtain constraints on $\alpha(k)$ through Hamiltonian Monte Carlo sampling using the \texttt{PyMC} package. 
While the smoothed $\alpha(k)$ form our main results, the unsmoothed $\alpha(k)$ help determine the scales for which $\alpha$ is constrained over its prior, as illustrated by the unshaded region in \cref{fig:DES_constraints}. 
This range is determined by selecting bins in which the variance of the posterior distribution is less than half that of the prior (which is $1/3$ for a uniform prior over $[-1,+1]$), and the number of constrained bins $\tilde{n}_k$ are reported in \cref{tab:individual_results} for all data comnbinations. In practice, we find that this range captures roughly 90\% of the total constraining power estimated with \cref{eq:neff}.

\subsection{Recovery of deviations}
\label{sec:validation}

In this section, we validate that our approach can recover injected deviations from \LCDM. To do so, we employ the DES setup, but use data vectors computed with the full Limber equation (see \cref{eq:limber}) for three specific models. 
We employ the scalar $A_{\rm mod}$ parametrization suggested in~\cite{2206.11794} to model deviations from the standard power spectrum as a function of the linear $P_{\rm lin}$ and nonlinear power spectra $P_{\rm nl}$ given by
\begin{equation}
    P(k,z) = P_{\rm lin}(k,z) + A_{\rm mod} (P_{\rm nl}(k,z) - P_{\rm lin}(k,z)).
\end{equation}
In our next test, we vary the prescription for the nonlinear power spectrum and use \texttt{HMCode}~\citep{2021MNRAS.502.1401M} to compute a power spectrum that includes a model for baryonic effects that suppress fluctuations around $k\sim\SI{1}{\iMpc}$ and enhances them on smaller scales. {In this model, a higher AGN temperature $T_{\rm AGN}$ causes less power suppression because the more energetic feedback ejects gas farther from halos, reducing its long-term impact on suppressing small-scale matter clustering.} Finally, we test a scale-dependent modified-gravity model with the $(\Sigma,\mu)$ parametrization, which differentiates the gravitational potentials experienced by matter and by light, with values $\Sigma_0=0$, $\mu_0=0.5$ and $c_1^\mu=2$ as implemented in \texttt{CCL}~\citep{2019ApJS..242....2C}.

\begin{figure}
    \centering
    \includegraphics[scale=0.68]{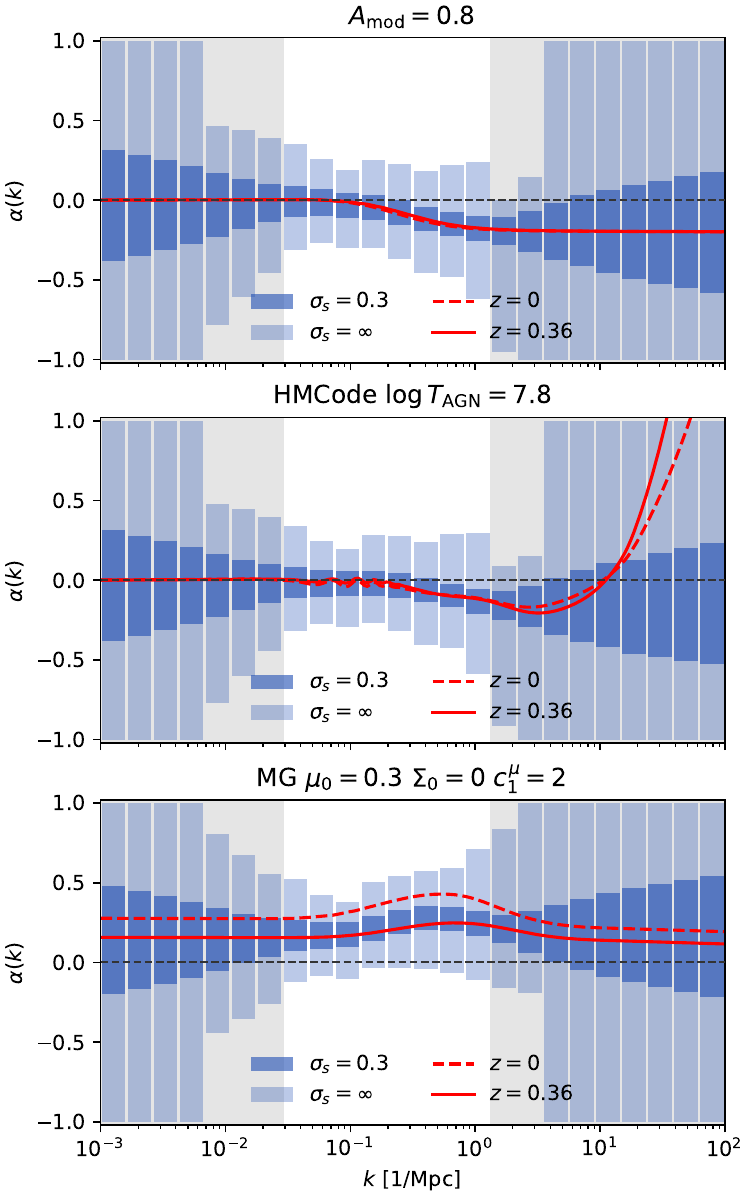}
    \caption{
    Validation tests performed using mock data generated at \planck \LCDM cosmological parameters, incorporating various deviations to the standard nonlinear power spectrum.
    For each panel, the posterior of each component of $\bm{\alpha}=\qty(\alpha(k_1),\dots,\alpha(k_{n_k}))$ is visualized as a box. 
    The vertical extent of the boxes represents the 68\% limits of the posterior, computed with \texttt{GetDist}~\citep{2019arXiv191013970L}.
    The horizontal extent corresponds to the width of the logarithmic $k$-bins. 
    Dark blue boxes represent the posterior with smoothing ($\sigma_{\rm s}=0.3$), while light blue boxes show the posterior without smoothing ($\sigma_{\rm s}=\infty$). 
    {The red lines indicate the relative differences between the injected and fiducial power spectra, evaluated at $z=0$ and at the effective DES redshift $z_{\rm eff}=0.36$.}
    Note that any divergence between the posterior and smoothed posterior boxes in regions of low data sensitivity does not necessarily invalidate the recovery of the input cosmology. 
    In the gray regions, the posterior distribution is dominated by the prior, and it is either uniform (no smoothing) or favors flat curves that satisfy the smoothing prior.
    \textit{Top panel}: a deviation following the $A_{\rm mod}$ parametrization of~\cite{2206.11794} is successfully recovered. 
    \textit{Middle panel}: baryonic feedback was modelled using \texttt{HMCode}~\citep{2021MNRAS.502.1401M} and the injected deviations were recovered within the constrained $k$-range, denoted by the unshaded band between gray regions. 
    \textit{Bottom panel}: a $(\Sigma,\mu)$ modified-gravity model was implemented. While the deviation at the effective DES redshift ($z_{\rm eff}=0.36$) was recovered, the redshift-dependent deviation from the \LCDM power spectrum was not captured by our only-scale-dependent model, as expected. 
    }
    \label{fig:DES_validation}
\end{figure}

The results shown in \cref{fig:DES_validation} demonstrate that our method recovers the input deviations in the $k$-range that is constrained by the data.
{The red lines show the relative differences between the tested power spectrum model and the fiducial one, evaluated at both redshift $z=0$ and at the effective DES redshift, $z_{\rm eff}=0.36$.
}
Posterior distributions are represented by boxes with vertical extent showing the 68\% limits on $\alpha(k)$ components computed with \texttt{GetDist}~\citep{2019arXiv191013970L}, and horizontal extent showing the width of the $k$-bins.
Importantly, our model is only scale-dependent, such that if a deviation is also redshift-dependent, we will only recover the scale-dependent part at an effective redshift probed by the data. 
This is illustrated in the lowest panel for the modified-gravity case, where we recover the deviations not at $z=0$, but at the approximate effective redshift $z\simeq0.36$ probed by DES data, as illustrated in \cref{fig:kz_sensitivity}.
{The smoothing prior is also shown in \cref{fig:DES_validation} through the comparison of smoothed and unsmoothed posteriors. 
It regularizes the reconstruction by damping oscillations between adjacent $k$-bins, as motivated in \cref{sec:smoothing}. 
This prior leaves the posterior mean unchanged but reduces bin-to-bin correlations, tightening the marginalized constraints.}

\section{Results}
\label{sec:results}

\subsection{Constraints from individual surveys}
\label{sec:res_ind}

\begin{figure*}
    \centering
    \includegraphics[width=0.99\textwidth]{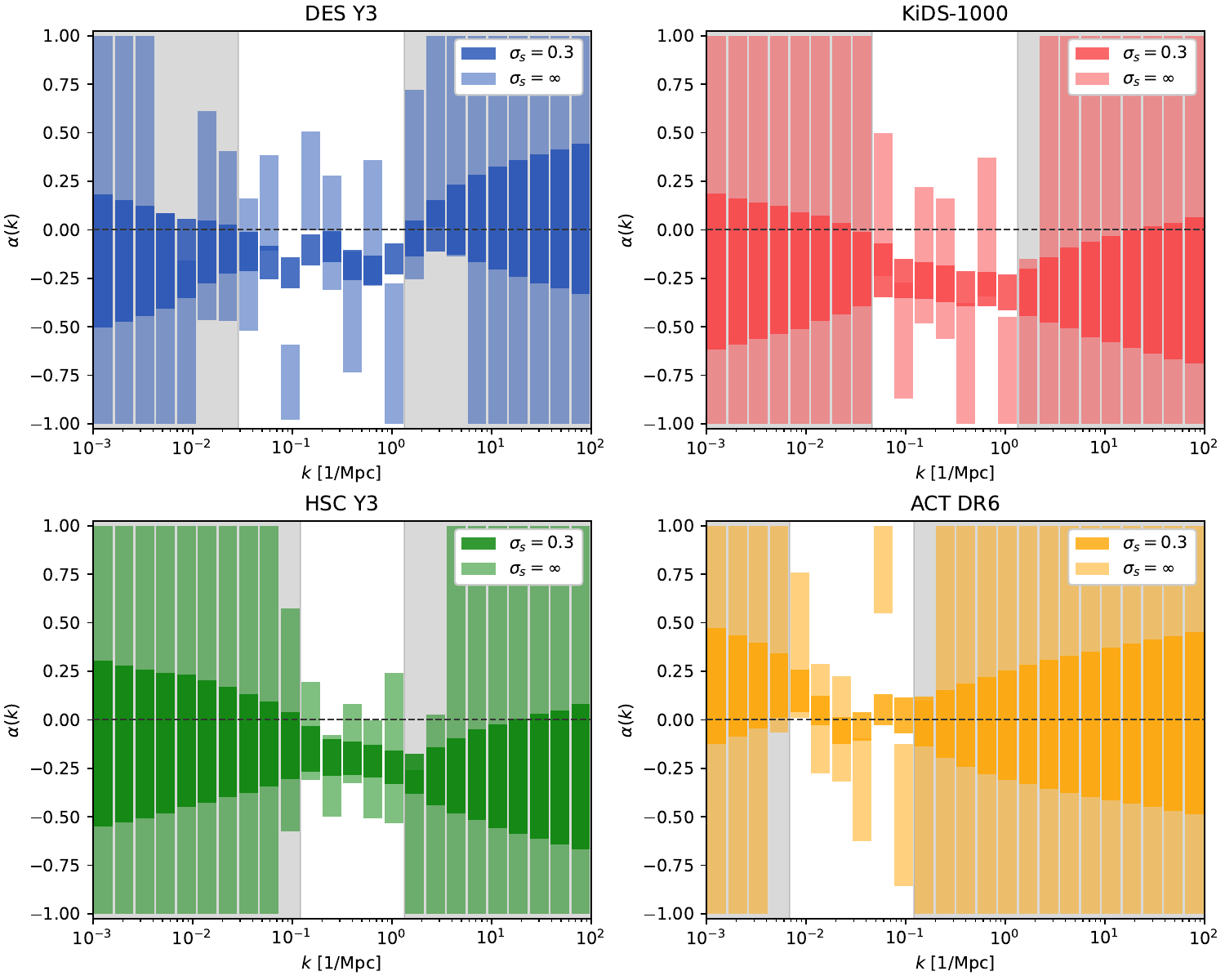}
    \caption{
    Constraints from individual lensing surveys on $\alpha(k)$ defined as $P(k,z) \equiv (1+\alpha(k)) P_{\rm fid}(k,z)$ where $P_{\rm fid}(k,z)$ is the nonlinear \LCDM matter power spectrum at \planck cosmology (see \cref{tab:fiducial_cosmology}) computed with \texttt{halofit}. 
    \Cref{tab:kzeff} shows the effective scales and redshifts probed by these experiments.
    {Boxes are used to represent the 68\% credible intervals derived from the posterior distributions of $\alpha(k)$ computed in 24~logarithmically-spaced $k$-bins.}
    Lighter shaded boxes represent the unsmoothed, more strongly anti-correlated $k$-bins, while darker boxes show the smoothed, decorrelated bins (smoothing scale $\sigma_{\rm s}=0.3$). 
    The gray shaded regions indicate where each dataset loses sensitivity, and where the points become strongly correlated (see \cref{fig:DES_constraints,fig:DES_corr}).
    We note that our model does not include intrinsic alignments and baryonic feedback, which could contribute to the lower power spectrum observed by cosmic shear surveys.
    {Results adopting a DES~Y3 cosmology~\citep{2203.07128} {as fiducial} are shown in \cref{app:des_cosmo}.}
    }
    \label{fig:all_Pk}
\end{figure*}

\Cref{fig:all_Pk} shows the constraints on $\alpha(k)$ obtained from each lensing survey independently. 
We report constraints for $n_k=24$ logarithmically spaced $k$-bins in the range \SIrange{e-3}{e2}{\iMpc}, with and without a smoothing prior of strength $\sigma_{\rm s}=0.3$, as per \cref{eq:prior}. 
These deviations should only be noted in the regions where surveys actually constrain the matter power spectrum (see \cref{fig:kz_sensitivity,fig:DES_constraints}). 
Outside these regions, in the gray bands, $\alpha(k)$ is unconstrained and a model need not have zero deviation from \LCDM to fit data. Also note that the regions are different for every survey, particularly between galaxy and CMB lensing surveys.

Notably, all three galaxy lensing surveys find similar suppression in $\alpha(k)$ in their individual constrained regions. 
We find that these data favor a 15-to-30\% suppression of the matter power spectrum in the range ${k\sim\SIrange{0.1}{1}{\iMpc}}$ than extrapolated from \planck CMB observations at high redshift, in agreement with constraints reported by galaxy surveys~\citep{2203.07128,Dalal:2023olq,2007.15633}.
\Cref{app:syst} illustrates the robustness of these constraints when systematics marginalization is ignored, or using less conservative scale cuts.

These results are quantified in \cref{tab:individual_results}. 
We report the level of tension between the $\alpha(k)$ constraints and their fiducial zero-value if agreement with \planck was perfect, 
computing a $\Delta\chi^2_\alpha$ using the mean and covariance of the sample, excluding the gray regions in $k$ where each individual experiment loses sensitivity as shown in \cref{fig:all_Pk}. 
Given the probability $p$ to exceed the observed value, its significance is given in units of Gaussian standard deviations by ${n_\sigma=\sqrt{2} \erf^{-1}(1-p)}$, 
and we consider that there is no significant tension for $p>0.01$, corresponding to $n_\sigma<\SI{2.6}{\sig}$.
Using this metric, we find that DES and HSC have a mild tension (below \SI{3}{\sig}) with the fiducial model, while ACT is in excellent agreement. 
KiDS seems to strongly reject \planck \LCDM, with the caveat that our model itself provides a poor fit to KiDS data (see below).

{The posterior uncertainties shown here are evaluated at fixed \planck \LCDM cosmological parameters. 
While one could in principle propagate the \planck parameter uncertainties into $\alpha(k)$, these are significantly smaller than the constraints from the lensing surveys themselves. 
Including this effect would therefore lead to only a subdominant broadening of the inferred posteriors and would not qualitatively change the tension levels quoted in \cref{tab:individual_results}. 
For completeness, we repeat the inference with the DES~Y3 harmonic-space cosmology~\citep{2203.07128} as fiducial; \cref{app:des_cosmo} shows that beyond an expected overall amplitude shift, the scale dependence within the constrained $k$-range is very similar.
{A full accounting of parametric uncertainties in the chosen fiducial cosmology is beyond the scope of this paper. 
}
}

\Cref{tab:individual_results} also provides the goodness-of-fit of the model described by \cref{eq:model}, and its significance computed with a number of degrees of freedom given by the data-vector size minus the effective number $n_{\rm eff}$ of constrained bins from \cref{eq:neff}. 
We find that our linear model generally provides a good fit to data, except for KiDS-1000. 
The best-fit of our model $n_\sigma\simeq\SI{2.7}{\sig}$ is only marginally worse than the one reported in~\cite{2007.15633} with a $p$-value of \num{0.013}, corresponding to $n_\sigma\simeq\SI{2.5}{\sig}$.\footnote{Removing bin 2 which was found to be inconsistent with the others bins, albeit with negligible impact on cosmology, does not help to improve the fit.}
This poor fit may be indicative of a tension requiring a redshift-dependent model, unlike the one used here, or systematic effects that are not accounted for. We also note that recent results from the complete KiDS-Legacy dataset, with improved redshift calibration, show better internal consistency and are in agreement with \planck~\citep{2503.19441}, supporting the need for a redshift-dependent model.

\begin{table*}
    \centering
    \caption{\normalfont{Comparison of $\alpha(k)$ posteriors using a fiducial \planck \LCDM power spectrum, excluding intrinsic alignments and baryonic feedback. The columns show: 
    (1) the datasets used; 
    (2) the number of $C_\ell$ data points; 
    (3) the number $\tilde{n}_k$ of constrained $k$-bins; 
    (4) the smoothing level $\sigma_{\rm s}$ (see \cref{sec:method} for definitions of $\tilde{n}_k$ and smoothing); 
    (5) the $\Delta\chi^2_\alpha$ computed using only the constrained $k$-bins, and
    (6) the significance $n_\sigma$ of the tension with \planck in the $\alpha(k)$ projection; 
    and the goodness-of-fit of our linear model, including 
    (7) the $\chi^2_{\rm min}$ at the best-fit, 
    (8) the effective number $n_{\rm eff}$ of constrained parameters (see \cref{eq:neff}), and 
    (9) the significance $n_\sigma$ of the deviation in Gaussian standard deviations.
    {We note that poor fits ($n_\sigma\gtrsim\SI{2.5}{\sig}$) in the last column may indicate that our redshift-independent model is insufficient to describe the data, that data favor different \LCDM parameters, or the presence of unknown systematic effects.}
    }}
    \begin{tabular}{lccldcddc}\toprule
    \multirow{2}{*}{Data combination} & \multirow{2}{*}{Data points} & \multirow{2}{*}{$\tilde{n}_k$} & \multirow{2}{*}{Smoothing} & \multicolumn{2}{c}{$\alpha(k)$ tension} & \multicolumn{3}{c}{Goodness-of-fit} \\ \cmidrule(lr){5-6} \cmidrule(lr){7-9}
     & & & & \multicolumn{1}{c}{$\Delta\chi^2_\alpha$} & $n_\sigma$ & \multicolumn{1}{c}{$\chi^2_{\rm min}$} & \multicolumn{1}{c}{$n_{\rm eff}$} & $n_\sigma$ \\ \midrule 
\multirow{2}{*}{DES~Y3}              & \multirow{2}{*}{320} & \multirow{2}{*}{ 8} & $\sigma_{\rm s}=\infty$ &  34.8 & \SI{4.2}{\sig} & 346.7 &  7.7 & \SI{1.7}{\sig} \\
                                  &                      &                     & $\sigma_{\rm s}=0.3$    &  23.9 & \SI{3.0}{\sig} & 351.5 &  3.9 & \SI{1.7}{\sig} \\
\midrule
\multirow{2}{*}{HSC~Y3}              & \multirow{2}{*}{ 60} & \multirow{2}{*}{ 5} & $\sigma_{\rm s}=\infty$ &  10.7 & \SI{1.9}{\sig} &  60.3 &  4.9 & \SI{1.0}{\sig} \\
                                  &                      &                     & $\sigma_{\rm s}=0.3$    &  14.4 & \SI{2.5}{\sig} &  60.5 &  2.5 & \SI{0.9}{\sig} \\
\midrule
\multirow{2}{*}{KiDS-1000}             & \multirow{2}{*}{120} & \multirow{2}{*}{ 7} & $\sigma_{\rm s}=\infty$ &  37.4 & \SI{4.6}{\sig} & 156.7 &  5.1 & \SI{2.8}{\sig} \\
                                  &                      &                     & $\sigma_{\rm s}=0.3$    &  42.0 & \SI{5.0}{\sig} & 159.4 &  2.4 & \SI{2.7}{\sig} \\
\midrule
\multirow{2}{*}{ACT~DR6}              & \multirow{2}{*}{ 10} & \multirow{2}{*}{ 6} & $\sigma_{\rm s}=\infty$ &  10.4 & \SI{1.6}{\sig} &   4.0 &  4.7 & \SI{0.5}{\sig} \\
                                  &                      &                     & $\sigma_{\rm s}=0.3$    &   2.9 & \SI{0.2}{\sig} &   6.6 &  2.2 & \SI{0.6}{\sig} \\
\midrule\midrule
\multirow{2}{*}{DES+HSC}          & \multirow{2}{*}{380} & \multirow{2}{*}{ 8} & $\sigma_{\rm s}=\infty$ &  39.6 & \SI{4.6}{\sig} & 416.7 &  8.4 & \SI{1.9}{\sig} \\
                                  &                      &                     & $\sigma_{\rm s}=0.3$    &  37.1 & \SI{4.4}{\sig} & 419.4 &  4.4 & \SI{1.9}{\sig} \\
\midrule
\multirow{2}{*}{DES+KiDS}         & \multirow{2}{*}{440} & \multirow{2}{*}{ 8} & $\sigma_{\rm s}=\infty$ &  67.3 & \SI{6.7}{\sig} & 513.9 &  8.4 & \SI{2.9}{\sig} \\
                                  &                      &                     & $\sigma_{\rm s}=0.3$    &  59.9 & \SI{6.2}{\sig} & 519.5 &  4.1 & \SI{2.9}{\sig} \\
\midrule
\multirow{2}{*}{DES+ACT}          & \multirow{2}{*}{330} & \multirow{2}{*}{11} & $\sigma_{\rm s}=\infty$ &  37.5 & \SI{3.9}{\sig} & 363.8 &  9.2 & \SI{2.0}{\sig} \\
                                  &                      &                     & $\sigma_{\rm s}=0.3$    &  17.8 & \SI{1.7}{\sig} & 371.6 &  5.0 & \SI{2.1}{\sig} \\
\midrule
\multirow{2}{*}{HSC+KiDS}         & \multirow{2}{*}{180} & \multirow{2}{*}{ 8} & $\sigma_{\rm s}=\infty$ &  53.3 & \SI{5.7}{\sig} & 222.4 &  6.2 & \SI{2.7}{\sig} \\
                                  &                      &                     & $\sigma_{\rm s}=0.3$    &  62.4 & \SI{6.4}{\sig} & 222.7 &  3.3 & \SI{2.5}{\sig} \\
\midrule
\multirow{2}{*}{HSC+ACT}          & \multirow{2}{*}{ 70} & \multirow{2}{*}{11} & $\sigma_{\rm s}=\infty$ &  21.0 & \SI{2.1}{\sig} &  66.0 &  9.3 & \SI{1.0}{\sig} \\
                                  &                      &                     & $\sigma_{\rm s}=0.3$    &  16.9 & \SI{1.6}{\sig} &  68.1 &  4.8 & \SI{0.9}{\sig} \\
\midrule
\multirow{2}{*}{KiDS+ACT}         & \multirow{2}{*}{130} & \multirow{2}{*}{11} & $\sigma_{\rm s}=\infty$ &  43.5 & \SI{4.4}{\sig} & 166.1 &  8.6 & \SI{2.8}{\sig} \\
                                  &                      &                     & $\sigma_{\rm s}=0.3$    &  38.3 & \SI{4.0}{\sig} & 171.5 &  4.3 & \SI{2.9}{\sig} \\
\midrule
\multirow{2}{*}{DES+HSC+KiDS}     & \multirow{2}{*}{500} & \multirow{2}{*}{ 8} & $\sigma_{\rm s}=\infty$ &  69.8 & \SI{6.9}{\sig} & 581.7 &  8.7 & \SI{3.0}{\sig} \\
                                  &                      &                     & $\sigma_{\rm s}=0.3$    &  71.8 & \SI{7.0}{\sig} & 586.7 &  4.5 & \SI{3.0}{\sig} \\
\midrule
\multirow{2}{*}{DES+HSC+ACT}      & \multirow{2}{*}{390} & \multirow{2}{*}{11} & $\sigma_{\rm s}=\infty$ &  38.5 & \SI{4.0}{\sig} & 435.5 &  9.8 & \SI{2.2}{\sig} \\
                                  &                      &                     & $\sigma_{\rm s}=0.3$    &  29.2 & \SI{3.1}{\sig} & 441.8 &  5.3 & \SI{2.3}{\sig} \\
\midrule
\multirow{2}{*}{DES+KiDS+ACT}     & \multirow{2}{*}{450} & \multirow{2}{*}{11} & $\sigma_{\rm s}=\infty$ &  64.8 & \SI{6.1}{\sig} & 534.1 &  9.8 & \SI{3.2}{\sig} \\
                                  &                      &                     & $\sigma_{\rm s}=0.3$    &  49.0 & \SI{4.9}{\sig} & 542.6 &  5.1 & \SI{3.3}{\sig} \\
\midrule
\multirow{2}{*}{HSC+KiDS+ACT}     & \multirow{2}{*}{190} & \multirow{2}{*}{12} & $\sigma_{\rm s}=\infty$ &  59.7 & \SI{5.6}{\sig} & 231.4 &  9.6 & \SI{2.7}{\sig} \\
                                  &                      &                     & $\sigma_{\rm s}=0.3$    &  59.3 & \SI{5.5}{\sig} & 234.5 &  5.1 & \SI{2.7}{\sig} \\
\midrule
\multirow{2}{*}{DES+HSC+KiDS+ACT} & \multirow{2}{*}{510} & \multirow{2}{*}{11} & $\sigma_{\rm s}=\infty$ &  65.4 & \SI{6.1}{\sig} & 603.2 & 10.1 & \SI{3.3}{\sig} \\
                                  &                      &                     & $\sigma_{\rm s}=0.3$    &  60.2 & \SI{5.8}{\sig} & 610.2 &  5.6 & \SI{3.3}{\sig} \\ \bottomrule
    \end{tabular}
    \label{tab:individual_results}
\end{table*}

We reiterate that these $\alpha(k)$ denote all departures not just from \LCDM, but specifically from the assumption that lensing surveys measure the projected \LCDM nonlinear matter power spectrum computed at \planck cosmology with \texttt{halofit}. 
An overall suppression in the matter power spectrum therefore need not arise from new physics exclusively, but can also be explained by shifts in \LCDM parameters relative to \planck, or by contributions from intrinsic alignments and baryonic feedback that impact weak-lensing surveys. 
In particular, KiDS-1000 reported a \SI{2.5}{\sig} detection (for shear bandpowers) of redshift-independent intrinsic alignments with an amplitude $A_{\rm IA}\sim1$ that suppress power on all scales. When including this contribution in the Limber approximation and window matrices using a redshift-independent nonlinear alignment model~\citep[NLA,][]{2004PhRvD..70f3526H,2007NJPh....9..444B} with an amplitude of $A_{\rm IA}=1$, the tension reduces from \SI{4.8}{\sig} to \SI{3.7}{\sig} for KiDS-1000, without any marginalization, close to the \SI{3.4}{\sig} tension reported for COSEBIs in~\cite{2007.15633}. 
Again, we refer the reader to more recent results from the KiDS-Legacy dataset~\citep{2503.19441}.

\subsection{Combining data sets}
\label{sec:res_comb}

\begin{table}[t]
    \centering
    \caption{\normalfont{This table summarizes the consistency between all data subsets used in this analysis, based on the unsmoothed $\alpha(k)$ posteriors. 
    Consistency is quantified using the non-Gaussian parameter-shift metric $\Delta_{\rm NG}$ computed with \texttt{Tensiometer} on the full-dimensional, unsmoothed $\alpha(k)$ posteriors.
    }}
    \begin{tabular}{r@{ \textit{vs.} }l c c c}
        \toprule
        \multicolumn{2}{c}{Data subsets ($\sigma_{\rm s}=\infty$)} & $\Delta_{\rm NG}$ \\
        \midrule
        DES      & HSC          & \SI{0.02}{\sig} \\
        DES      & KiDS         & \SI{0.02}{\sig} \\
        DES      & ACT          & \SI{0.14}{\sig} \\
        HSC      & KiDS         & $<\SI{0.01}{\sig}$ \\
        HSC      & ACT          & $<\SI{0.01}{\sig}$ \\
        KiDS     & ACT          & $<\SI{0.01}{\sig}$ \\
        \midrule
        DES      & HSC+KiDS     & \SI{0.26}{\sig} \\
        DES      & HSC+ACT      & \SI{1.06}{\sig} \\
        DES      & KiDS+ACT     & \SI{0.89}{\sig} \\
        HSC      & DES+KiDS     & \SI{0.01}{\sig} \\
        HSC      & DES+ACT      & \SI{0.14}{\sig} \\
        HSC      & KiDS+ACT     & $<\SI{0.01}{\sig}$ \\
        KiDS     & DES+HSC      & \SI{0.01}{\sig} \\
        KiDS     & DES+ACT      & \SI{0.31}{\sig} \\
        KiDS     & HSC+ACT      & $<\SI{0.01}{\sig}$ \\
        ACT      & DES+HSC      & \SI{0.42}{\sig} \\
        ACT      & DES+KiDS     & \SI{0.52}{\sig} \\
        ACT      & HSC+KiDS     & $<\SI{0.01}{\sig}$ \\
        \midrule
        DES      & HSC+KiDS+ACT & \SI{1.50}{\sig} \\
        HSC      & DES+KiDS+ACT & \SI{0.02}{\sig} \\
        KiDS     & DES+HSC+ACT  & \SI{0.13}{\sig} \\
        ACT      & DES+HSC+KiDS & \SI{0.69}{\sig} \\
        \midrule
        DES+HSC  & KiDS+ACT     & \SI{0.72}{\sig} \\
        DES+KiDS & HSC+ACT      & \SI{1.03}{\sig} \\
        DES+ACT  & HSC+KiDS     & \SI{0.63}{\sig} \\
        \bottomrule
    \end{tabular}
    \label{tab:comb}
\end{table}

\begin{figure}[t]
    \centering
    \includegraphics[scale=0.68]{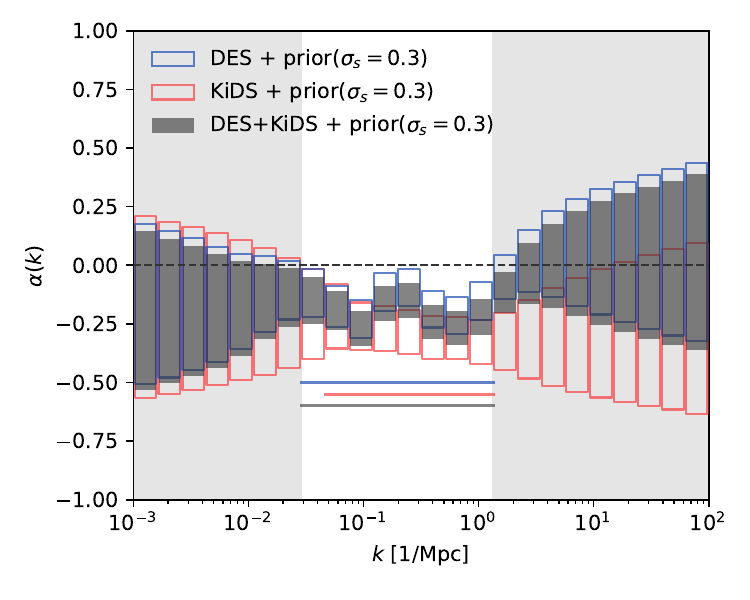}
    \caption{
    {Constraints on $\alpha(k)$ from DES~Y3 (red), KiDS-1000 (blue), and their combination (gray boxes), using the standard $\sigma_{\rm s}=0.3$ smoothing prior. Horizontal lines show the $k$-regions constrained by each of these data sets, while the unshaded region corresponds to constraints for DES+KiDS.}
    }
    \label{fig:DES_KiDS_Pk}
\end{figure}

There are two important caveats to combining lensing surveys and providing joint constraints on deviations from the \planck \LCDM matter power spectrum. 

Firstly, as with combining any two data sets, we check that the data sets are not in tension.
We employ \texttt{Tensiometer}~\citep{2020PhRvD.101j3527R,2021PhRvD.104d3504R} to test whether the unsmoothed $\alpha(k)$ posteriors are reasonably consistent across the $k$-range, such that any smoothing prior is not double counted.
In particular, we compute the non-Gaussian parameter-shift metric $\Delta_{\rm NG}$ introduced in~\cite{2021PhRvD.104d3504R}, which uses normalizing flows to model the parameter-difference distributions (see  \cref{app:full_posterior} for an illustration, comparing ACT \textit{vs}. DES+HSC). We report their significance $n_\sigma$ in units of Gaussian standard deviations, given this time by ${n_\sigma=\sqrt{2} \erf^{-1}(\Delta_{\rm NG})}$, in \cref{tab:comb}.\footnote{
We also computed the Gaussian parameter-shift metric $Q_{\rm UDM}$ in update form as defined in~\cite{2020PhRvD.101j3527R}. However, even though \texttt{Tensiometer} selects the bins where the posterior is at least 10\% tighter than the prior, where most of the signal is confined and where posteriors are more Gaussian, we found it remained unstable for such non-Gaussian posterior distributions.}
We find that all dataset combinations show reasonable consistency (tensions below $\sim\SI{2}{\sig}$) when considering the full-dimensional unsmoothed $\alpha(k)$ posteriors.
As expected from a visual inspection of \cref{fig:all_Pk}, the galaxy-lensing surveys can be readily combined. \Cref{fig:DES_KiDS_Pk} shows an example comparison of constraints from DES and KiDS, along with their combined constraint. 

A second important consideration, illustrated in \cref{fig:kz_sensitivity}, is the limited overlap in scale and redshift between CMB-lensing and galaxy-lensing surveys.  
Even the effective redshift for $\alpha(k)$ constraints differs slightly between HSC and KiDS/DES.  
Therefore, these surveys may probe different physics, and we advise caution when interpreting our results.  
Combining these datasets can yield meaningful constraints for new physics that causes either a truly redshift-independent shift in the matter power spectrum, or redshift-independent deviations within the surveys' redshift range (even if redshift-dependent outside of it). However, joint analyses should be avoided if the new physics introduces redshift-dependent modifications within the probed redshift range, as this would affect the different experiments unequally.

\begin{figure}
    \centering
    \includegraphics[scale=0.68]{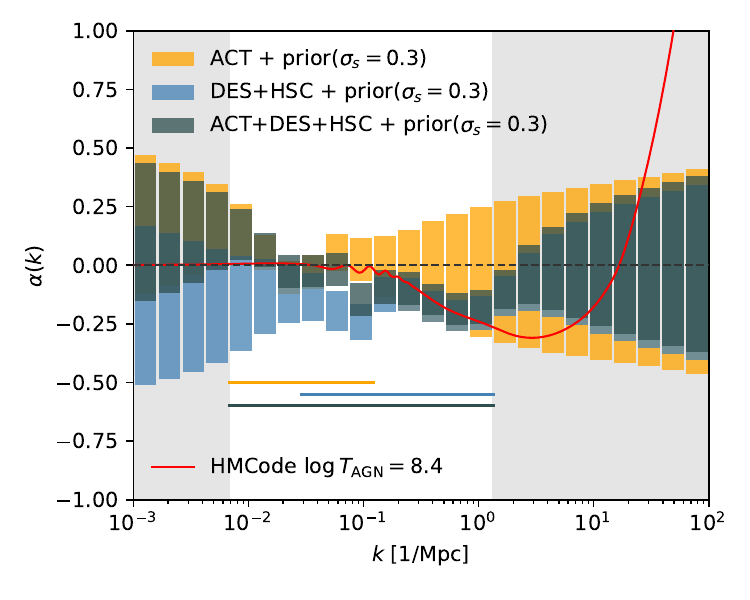}
    \caption{
    {Constraints on $\alpha(k)$ from ACT~DR6 (yellow), DES~Y3 combined with HSC~Y3 (blue), and their combination (dark gray boxes), using the standard $\sigma_{\rm s}=0.3$ smoothing prior. Horizontal lines show the $k$-regions constrained by each of these data sets, while the unshaded region corresponds to constraints for ACT+DES+KiDS. We overplot the theoretical prediction for baryonic feedback parameter $\log T_{\rm AGN}=8.4$ computed using \texttt{HMCode} in red, showing good agreement throughout the constrained range. See \cref{app:full_posterior} for a subset of the two-dimensional marginal distributions for these datasets.}
    }
    \label{fig:full_comb_Pk}
\end{figure}

\Cref{tab:individual_results} presents the $\alpha(k)$ tensions and goodness-of-fits for all dataset combinations.  
Our simple linear model provides a reasonable fit ($n_\sigma \lesssim \SI{2.5}{\sig}$) in all cases not involving KiDS, with or without smoothing, and the fit quality degrades slightly with the addition of more data.
\Cref{fig:DES_KiDS_Pk} compares constraints obtained individually from DES and KiDS and their combination, which were jointly analyzed in~\cite{2305.17173}.
\Cref{fig:full_comb_Pk} compares constraints from the combined DES and HSC galaxy surveys with those from ACT, and their joint constraints. 
The galaxy surveys prefer a 20\% lower projected power spectrum than the fiducial, dark-matter-only power spectrum. 
The ACT data, consistent with the fiducial model, pulls $\alpha(k)$ towards zero in the range $k\sim\SIrange{0.01}{0.1}{\iMpc}$, thus reducing the tension from \SI{4.4}{\sig} to \SI{3.1}{\sig}, at the price of a slightly worse fit. 
We also show the theoretical prediction obtained with \texttt{HMCode} to account for the effects of baryonic feedback, and find that the tension diminishes to \SI{0.8}{\sig} {for DES+HSC+ACT (\SI{2.4}{\sig} for DES+HSC)} with a baryonic feedback parameter of $\log T_{\rm AGN}=8.4$. 
Although we state once again that these constraints should be interpreted with caution, the combination of these data is able to constrain two orders of magnitudes of physical scales, from \SIrange{0.01}{1}{\iMpc}.

\section{Applications}
\label{sec:interpretation_application}

In the previous sections, we reconstructed scale-dependent deviations $\alpha(k)$ to the \planck \LCDM matter power spectrum, at the effective redshifts probed by galaxy- and CMB-lensing observations. 
We now describe how these results can be used to constrain extensions of the \LCDM model that modify matter density perturbations. This section outlines a general methodology for such an analysis, highlights important caveats, and demonstrates its application to specific physical models.

\subsection{Constraining power spectrum models}

The core idea is to use the posterior distributions of $\alpha(k)$ as a derived likelihood function to test alternative models of structure formation---analogous to how BAO analyses first extract the BAO scale along and across the line of sight before fitting cosmological models to those measurements~\citep{2025JCAP...02..021A}. 
We construct this likelihood directly from the $\alpha(k)$ posteriors and recommend restricting comparisons to the $k$-range where the data meaningfully constrain $\alpha(k)$. 
Our results show that, when a smoothing prior of width $\sigma_{\rm s}=0.3$ is applied, the $\alpha(k)$ posteriors are well approximated by a multivariate Gaussian within the constrained $k$-range. For unsmoothed posteriors, a truncated multivariate Gaussian may be more appropriate.

This likelihood can then be used to test the predictions of extended cosmological models. Compared to traditional approaches that reduce lensing data to a single summary statistic like $S_8$, this method preserves the full scale-dependent information available in the data. Since $\alpha(k)$ is defined at the effective redshift of the lensing data, comparisons must be made at the corresponding redshifts. For combined datasets with differing redshift sensitivities, a first approximation is to treat the $\alpha(k)$ constraints from each experiment as independent likelihoods, thereby constraining the model at multiple redshifts.

\subsubsection{Caveats}

\begin{figure}
    \centering
    \includegraphics[scale=0.68]{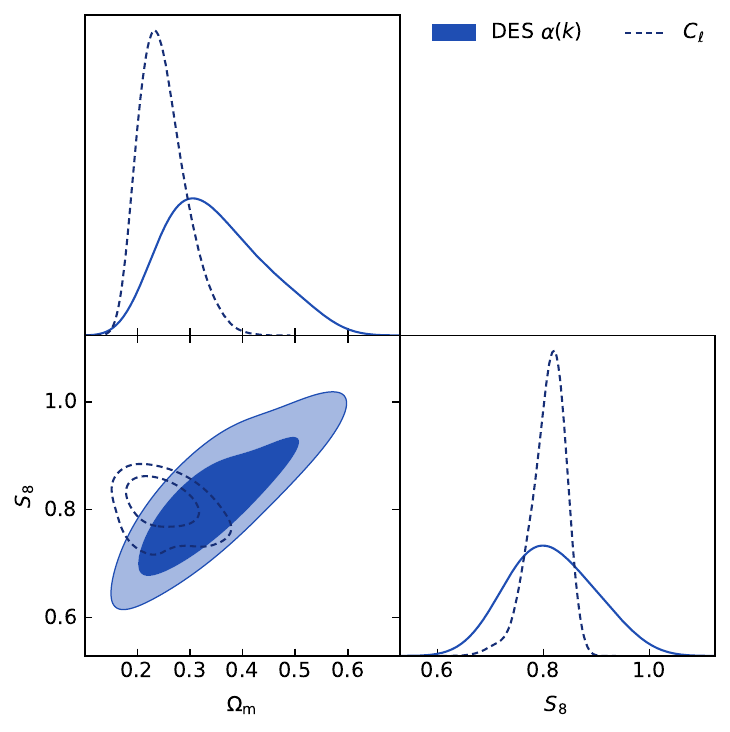}
    \caption{
    Constraints on the \LCDM parameters $S_8$ and $\Om$ derived from the $\alpha(k)$ likelihood function (solid, filled posteriors)
    are compared to those obtained using standard power spectrum analyses (dashed posteriors) for DES Y3 data (using a scale cut at $k_{\rm max}=\SI{1}{\h\per\mega\parsec}$, see section 3.5.2 of~\cite{2203.07128}). 
    While the mean parameter values are broadly consistent, we caution that varying background cosmological parameters is formally inconsistent with our inference framework, which assumes a fixed \LCDM cosmology---particularly in the construction of the window functions (see \cref{eq:window_matrix}).
    }
    \label{fig:LCDM_corner}
\end{figure}

While this methodology is powerful, it is subject to several caveats. The reconstruction of $\alpha(k)$ relies on a fiducial cosmology, and thus a specific projection from the angular power spectra $C_\ell$ into $P(k, z)$, as per \cref{eq:model}. 
This projection enables intuitive comparison between inferred deviations and theoretical models but assumes that the \textit{background} cosmology is fixed. If an extended model also alters the background expansion (e.g., changes in $\Om$, $H_0$), the window functions used in the $\alpha(k)$ inference in \cref{eq:window_matrix} will no longer be valid.
{The practical impact of switching the fiducial baseline is illustrated in \cref{app:des_cosmo}.}
In addition, the relationship between physical scales $k$ and angular scales $\ell$ would also be altered. 
We illustrate this in \cref{fig:LCDM_corner} by varying background \LCDM parameters\footnote{Note that here we follow the DES modeling of neutrinos as three massive species with degenerate mass.} and showing that the resulting constraints differ from those obtained directly from the lensing $C_\ell$ spectra. 
Although it is in principle possible to repeat the full $\alpha(k)$ inference at a different fiducial background, 
these caveats still remain, they are simply offset by a different fiducial cosmology. 

Further limitations arise, for instance, in modified gravity models parameterized by $\mu$ and $\Sigma$, where matter and light respond to different gravitational potentials. In such cases, the lensing-inferred $\alpha(k)$ may not directly correspond to modifications in the matter power spectrum, breaking the assumption that a single $P(k, z)$ governs both observables.

Finally, if the model does not induce an overall rescaling of the matter power spectrum, it may not fit both large and small scale $\alpha(k)$ equally well. In such cases, we recommend allowing the amplitude of scalar perturbations (e.g., $A_s$ or $\sigma_8$) to vary in the model comparison. This ensures that any scale-dependent deviation provides a globally consistent fit to the reconstructed $\alpha(k)$. In general, one must verify that the extended model yields a good fit to the data-constrained $\alpha(k)$, with the result depending on which model parameters are allowed to vary.

\subsubsection{Constraints}

\begin{figure}
    \centering
    \includegraphics[scale=0.68]{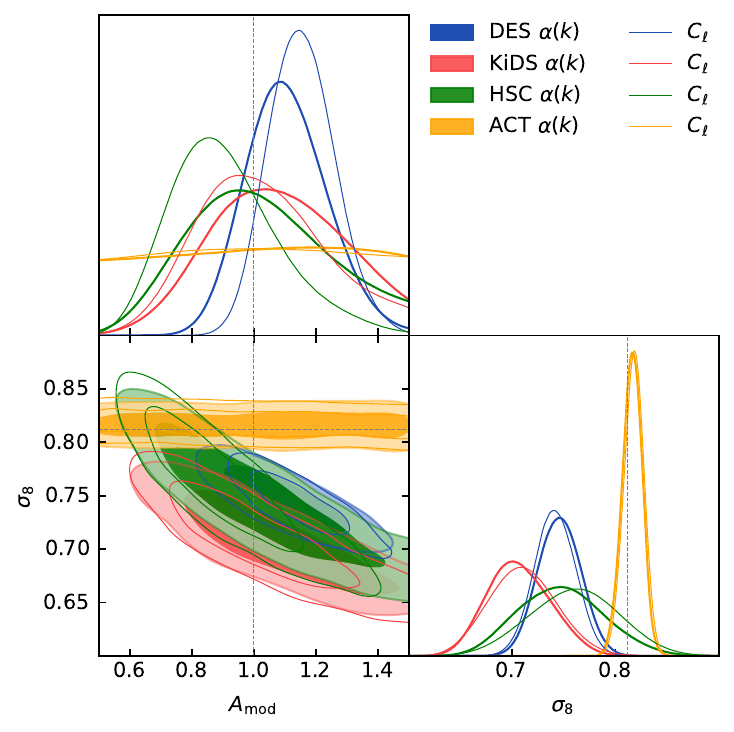}
    \caption{Constraints on the $A_{\rm mod}$ model derived from the $\alpha(k)$-based likelihood (filled contours), with $\sigma_8$ varied and all other parameters fixed to their \planck values. For comparison, constraints obtained directly from the angular power spectra are shown as unfilled contours (and thin lines in one-dimensional marginal panels). Dashed lines indicate the fiducial value $A_{\rm mod} = 1$ and the \planck best-fit value of $\sigma_8$.
    }
    \label{fig:Amod_corner}
\end{figure}

We demonstrate the utility of this approach by applying it to two physical scenarios. Model predictions are generated using the \texttt{CCL} cosmology library, with parameter inference performed via the \texttt{Nautilus} nested-sampling algorithm~\citep{2023MNRAS.525.3181L}.

First, we constrain a phenomenological amplitude modifier $A_{\rm mod}$ which suppresses power at small scales, while jointly varying $\sigma_8$ to accommodate the large-scale amplitude. 
Our constraints from individual surveys are consistent with those obtained from direct $C_\ell$ analyses at fixed background cosmology, as shown in \cref{fig:Amod_corner}. 
This agreement holds across smoothing priors and when restricting to only the well-constrained $k$-range. All three galaxy surveys prefer lower values of $\sigma_8$, with constraints that are anticorrelated with $A_{\rm mod}$. In contrast, ACT does not tightly constrain $A_{\rm mod}$, as it primarily probes the matter power spectrum at $z \sim 2$ and on scales where nonlinear effects are subdominant. Reconciling the lower $\sigma_8$ preferred by galaxy lensing within \LCDM with the higher value inferred from ACT thus requires a suppression of small-scale power, corresponding to $A_{\rm mod} < 1$.

\begin{figure}
    \centering
    \includegraphics[scale=0.68]{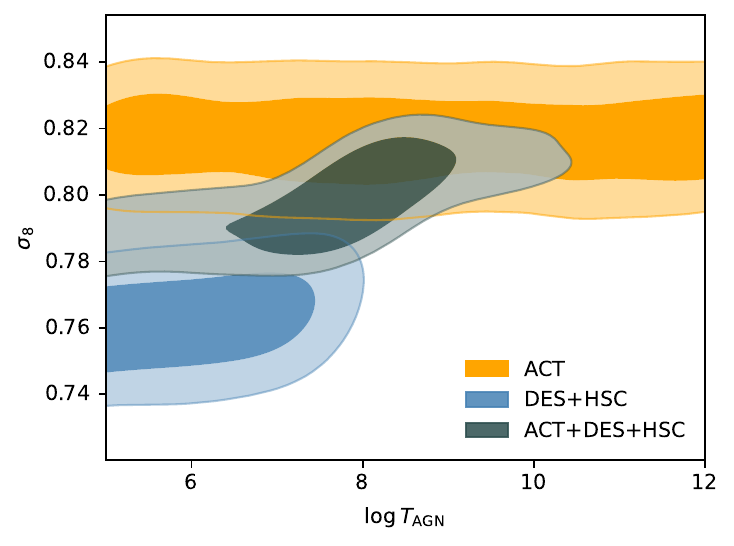}
    \caption{Constraints on baryonic feedback as modeled in \texttt{HMCode}, derived from the $\alpha(k)$-based likelihoods of ACT, DES, and HSC treated independently. 
    Rather than using the joint constraints on $\alpha(k)$ shown in \cref{fig:full_comb_Pk}, this result combines the single-experiment posteriors as independent likelihoods.
    }
    \label{fig:HMCode_corner}
\end{figure}

Second, we apply our method to constrain baryonic feedback as modeled in \texttt{HMCode}, using the AGN heating parameter $\log T_{\rm AGN}$. 
Individual experiments do not independently constrain both $\sigma_8$ and $\log T_{\rm AGN}$, but combining their $\alpha(k)$ likelihoods allows for joint inference. We consider the combination of DES, HSC and ACT data, for which our model provided an acceptable fit (see \cref{tab:individual_results}).
While DES and HSC favor a lower $\sigma_8$ value seemingly inconsistent with ACT as shown in \cref{fig:HMCode_corner}, their combined likelihoods constrain both parameters with a reasonable fit, and we obtain
\[
\log T_{\rm AGN} = 7.76^{+0.87}_{-0.69},
\]
demonstrating the power of combining datasets with complementary redshift and scale sensitivities.

\subsection{Plug-and-play implementations}

To facilitate reproducibility and encourage further exploration, we provide a comprehensive GitHub repository accompanying this work. The repository includes all the data and code used in our analysis, along with intermediate products such as the computed window functions and the covariances marginalized over observational systematics. It also contains posterior samples for $\alpha(k)$ and a Jupyter notebook that walks through the complete analysis workflow.

The demonstration notebook is designed to guide users through each stage of the pipeline, allowing them to reproduce the main results of our study with minimal setup. Specifically, it covers:
\begin{itemize}
\item Loading and pre-processing the weak-lensing data, with results saved for direct reuse.
\item Extracting the deviations $\alpha(k)$ from the matter power spectrum $P(k, z)$ using efficient Hamiltonian Monte Carlo (HMC) inference.
\item Comparing the inferred $\alpha(k)$ with predictions from the Planck \LCDM model.
\item Exploring extensions to the \LCDM model's predictions, as discussed in the previous section.
\end{itemize}

These files enable the user to reproduce our entire analysis at the fiducial cosmology of their choosing, potentially mitigating one of the principal caveats of this approach. 
By shifting the fiducial cosmology to the best-fit parameters of an extended model inferred from external data (\eg CMB data), or to theoretically motivated values, the resulting $\alpha(k)$ posteriors will reflect deviations relative to a new baseline. 
The user can then draw conclusions based on the level of tension between the updated $\alpha(k)$ posteriors and the new zero-point defined by their chosen model.
Note that this approach remains tied to the choice of a single fiducial cosmology.

In addition, we provide a set of pre-constructed $\alpha(k)$ likelihood modules compatible with the widely-used cosmological inference code \texttt{MontePython} \citep{2013JCAP...02..001A,Brinckmann:2018cvx}. 
Each module includes the necessary components to define a multivariate Gaussian likelihood over the constrained $k$ range: the $k$ vector, the mean $\alpha(k)$ vector, the inverse covariance matrix, the effective redshift of the measurement, and the fiducial power spectrum at that redshift for those $k$. 
{We also include a notebook that extracts these components from the $\alpha(k)$ posteriors computed above, and an example input file for using this likelihood with \texttt{MontePython}. 
Our provided framework is compatible with modifying the fiducial cosmology and producing alternate $\alpha(k)$ likelihoods. 
}
This setup allows for straightforward integration of our results into broader cosmological parameter-estimation efforts.

\section{Conclusions}
\label{sec:conclusions}

In this work, we introduced a framework for directly inferring scale-dependent deviations from the matter power spectrum $P(k, z)$ using galaxy- and CMB-lensing surveys.
By extracting this information beyond the traditional $S_8$ summary statistic, our method identifies where current observations may deviate from \planck \LCDM predictions as a function of scale at the effective redshift of each survey. 
This enables a more detailed and flexible characterization of structure growth, capturing deviations that could otherwise be overlooked. The approach is fast, adaptable, and fully reproducible, and we have released an accompanying GitHub repository to support further exploration and integration into cosmological inference pipelines.

Specifically, we inferred ${\alpha(k)\equiv P(k)/P_{\rm fid}(k)-1}$ over 24 logarithmic $k$-bins in the range \SIrange{e-3}{e2}{\iMpc} using HMC sampling with a data-driven smoothing prior, and applied our method independently and jointly to lensing data from DES~Y3, KiDS-1000, HSC~Y3, and ACT~DR6. {In practice, this allows us to recover the \emph{projected} power spectrum probed by lensing surveys, which can include the combined effects of baryonic feedback and intrinsic alignments, without requiring explicit modeling of these processes.}
For galaxy lensing surveys, we used angular multipoles up to $\ell_{\rm max}\sim2000$, and all results assume fixed cosmology with analytic marginalization over shear calibration and redshift distribution uncertainties. All deviations are reported with respect to the \planck \LCDM prediction for a CDM-only (i.e., {neglecting} baryonic feedback {and intrinsic alignments}) matter power spectrum.
{Interpreting $\alpha(k)$ as the scale-dependent deviation in the projected matter power spectrum probed by lensing, our key findings are: }
\begin{itemize}
    \item DES, KiDS, and HSC all show a 15--30\% suppression of power at $k \sim \SIrange{0.1}{1}{\iMpc}$ relative to \planck, with combined tensions reaching \SI{4}{\sig} {for DES+HSC} at fixed cosmology (\cref{fig:all_Pk} and \cref{tab:individual_results}).
    \item ACT data are consistent with \LCDM, indicating that the suppression is either redshift-dependent and specific to lower $z$, or restricted to small scales ${k\gtrsim\SI{0.1}{\iMpc}}$.
    \item Dataset tensions assessed using non-Gaussian metrics from \texttt{Tensiometer} are mild and generally within $2\sigma$, allowing for robust joint analyses (\cref{tab:comb}).
    \item The combined ACT, DES and HSC datasets constrain $\alpha(k)$ over $k \sim \SIrange{0.01}{1}{\iMpc}$, extending across two decades in scale, consistent with \planck at ${k\lesssim\SI{0.1}{\iMpc}}$ and 10--25\% lower at smaller scales (\cref{fig:full_comb_Pk,fig:DES_HSC_ACT_comb_Pk_corner,fig:DES_HSC_ACT_comb_Pk_corner_diff}).
    \item Our results, which are agnostic to the physical origin of deviations, are both qualitatively and quantitatively consistent with previous studies.
\end{itemize}

We also demonstrated the utility of the derived $\alpha(k)$ likelihoods by constraining physical models such as baryonic feedback and a phenomenological suppression of power (\cref{fig:Amod_corner,fig:HMCode_corner}), recovering results consistent with direct $C_\ell$ analyses and enabling joint constraints across datasets with complementary scale and redshift sensitivity. 
To simplify reproducibility and implementation of our entire analysis, we provide a GitHub repository with code, data, results, and $\alpha(k)$ likelihoods compatible with \texttt{MontePython} which enable fast comparison with theoretical models. 
As the $\alpha(k)$ are constrained at a specific fiducial cosmology, switching this to a different theoretical cosmology will shift $\alpha(k)$ constraints{,as illustrated in \cref{app:des_cosmo} using a DES~Y3 cosmology}. 
These new constraints can then help the user determine if the new cosmology can explain lensing power spectra or if deviations are still preferred, effectively capturing how well lensing data are fit by the new model. 

Looking ahead, two key developments could substantially expand the reach and robustness of this framework. First, while our current method constrains deviations as a function of scale alone, future extensions could allow for joint inference of $\alpha(k, z)$ over both scale and redshift. 
This would capture the temporal evolution of structure growth, potentially disentangling physical effects (e.g., modified gravity or time-varying dark energy) from observational systematics that evolve differently with redshift.
It would also allow for a more complete comparison across datasets with different redshift sensitivities, such as galaxy and CMB lensing.

Second, the Limber projection from angular power spectra $C_\ell$ to $P(k,z)$ currently relies on a fiducial cosmology to relate redshifts to comoving distances. While this is reasonable for internal consistency, it implicitly assumes the background expansion history of $\Lambda$CDM. To increase model-independence, future versions of this analysis could replace this assumption with empirically determined distance-redshift relations, e.g., from Type Ia supernov{\ae} or BAO measurements. This would allow the lensing data to test the growth of structure with minimal reliance on background model assumptions, enhancing the interpretability of any detected deviations.

Together, these extensions would enable sharper identification of possible tensions with \LCDM as a function of both scale and cosmic time. However, current data may not provide sufficient sensitivity to fully realize this potential. 
The next generation of surveys---including the Rubin Observatory Legacy Survey of Space and Time (LSST), ESA's \textit{Euclid} mission, and NASA's \textit{Roman} Space Telescope---will dramatically improve weak-lensing power-spectrum measurements by delivering deeper, wider, and higher-resolution data. 
Combined with our scalable and model-flexible framework, these advances will significantly tighten constraints on $\alpha(k, z)$ and enhance our ability to probe the fundamental physics driving the growth of cosmic structure.

\section*{Acknowledgments}

We thank Marco Gatti, Wayne Hu, Bhuvnesh Jain, Meng-Xiang Lin, Mat Madhavacheril, Shivam Pandey, and Marco Raveri for stimulating and fruitful discussions. 
We are grateful to Roohi Dalal for providing the HSC cosmic shear data and thank Rafael C. H. Gomes for his input during the early stages of this analysis.
TK was supported by the Kavli Institute for Cosmological Physics at the University of
Chicago through an endowment from the Kavli Foundation. 
Computing resources were provided by the National Energy Research Scientific Computing Center (NERSC), and the University of Chicago Research Computing Center through the Kavli Institute for Cosmological Physics. 

\bibliographystyle{apsrev4-2}

\bibliography{bib}

@string{june = {June}}

@article{0112114,
 adsurl = {https://ui.adsabs.harvard.edu/abs/2002MNRAS.335.1193B},
 archiveprefix = {arXiv},
 author = {{Bridle}, S.~L. and {Crittenden}, R. and {Melchiorri}, A. and {Hobson}, M.~P. and {Kneissl}, R. and {Lasenby}, A.~N.},
 doi = {10.1046/j.1365-8711.2002.05709.x},
 eprint = {astro-ph/0112114},
 journal = {\mnras},
 month = {October},
 number = {4},
 pages = {1193-1200},
 primaryclass = {astro-ph},
 title = {{Analytic marginalization over CMB calibration and beam uncertainty}},
 volume = {335},
 year = {2002}
}

@article{1003.1136,
 adsurl = {https://ui.adsabs.harvard.edu/abs/2010MNRAS.408..865T},
 archiveprefix = {arXiv},
 author = {{Taylor}, A.~N. and {Kitching}, T.~D.},
 doi = {10.1111/j.1365-2966.2010.17201.x},
 eprint = {1003.1136},
 journal = {\mnras},
 month = {October},
 number = {2},
 pages = {865-875},
 primaryclass = {astro-ph.CO},
 title = {{Analytic methods for cosmological likelihoods}},
 volume = {408},
 year = {2010}
}

@article{1208.2701,
 adsurl = {https://ui.adsabs.harvard.edu/abs/2012ApJ...761..152T},
 archiveprefix = {arXiv},
 author = {{Takahashi}, Ryuichi and {Sato}, Masanori and {Nishimichi}, Takahiro and {Taruya}, Atsushi and {Oguri}, Masamune},
 doi = {10.1088/0004-637X/761/2/152},
 eid = {152},
 eprint = {1208.2701},
 journal = {\apj},
 month = {December},
 number = {2},
 pages = {152},
 primaryclass = {astro-ph.CO},
 title = {{Revising the Halofit Model for the Nonlinear Matter Power Spectrum}},
 volume = {761},
 year = {2012}
}

@article{1809.09603,
 adsurl = {https://ui.adsabs.harvard.edu/abs/2019MNRAS.484.4127A},
 archiveprefix = {arXiv},
 author = {{Alonso}, David and {Sanchez}, Javier and {Slosar}, An{\v{z}}e and {LSST Dark Energy Science Collaboration}},
 doi = {10.1093/mnras/stz093},
 eprint = {1809.09603},
 journal = {\mnras},
 month = {April},
 number = {3},
 pages = {4127-4151},
 primaryclass = {astro-ph.CO},
 title = {{A unified pseudo-C$_{{\ensuremath{\ell}}}$ framework}},
 volume = {484},
 year = {2019}
}

@article{1902.01366,
 adsurl = {https://ui.adsabs.harvard.edu/abs/2020PhRvD.101h3524R},
 archiveprefix = {arXiv},
 author = {{Raveri}, Marco},
 doi = {10.1103/PhysRevD.101.083524},
 eid = {083524},
 eprint = {1902.01366},
 journal = {\prd},
 month = {April},
 number = {8},
 pages = {083524},
 primaryclass = {astro-ph.CO},
 title = {{Reconstructing gravity on cosmological scales}},
 volume = {101},
 year = {2020}
}

@article{1906.11765,
 adsurl = {https://ui.adsabs.harvard.edu/abs/2019JCAP...11..043G},
 archiveprefix = {arXiv},
 author = {{Garc{\'\i}a-Garc{\'\i}a}, Carlos and {Alonso}, David and {Bellini}, Emilio},
 doi = {10.1088/1475-7516/2019/11/043},
 eid = {043},
 eprint = {1906.11765},
 journal = {\jcap},
 month = {November},
 number = {11},
 pages = {043},
 primaryclass = {astro-ph.CO},
 title = {{Disconnected pseudo-C$_{l}$ covariances for projected large-scale structure data}},
 volume = {2019},
 year = {2019}
}

@article{2000ApJ...538..473L,
 adsurl = {https://ui.adsabs.harvard.edu/abs/2000ApJ...538..473L},
 archiveprefix = {arXiv},
 author = {{Lewis}, Antony and {Challinor}, Anthony and {Lasenby}, Anthony},
 doi = {10.1086/309179},
 eprint = {astro-ph/9911177},
 journal = {\apj},
 month = {August},
 number = {2},
 pages = {473-476},
 primaryclass = {astro-ph},
 title = {{Efficient Computation of Cosmic Microwave Background Anisotropies in Closed Friedmann-Robertson-Walker Models}},
 volume = {538},
 year = {2000}
}

@article{2004PhRvD..70f3526H,
 adsurl = {https://ui.adsabs.harvard.edu/abs/2004PhRvD..70f3526H},
 archiveprefix = {arXiv},
 author = {{Hirata}, Christopher M. and {Seljak}, Uro{\v{s}}},
 doi = {10.1103/PhysRevD.70.063526},
 eid = {063526},
 eprint = {astro-ph/0406275},
 journal = {\prd},
 month = {September},
 number = {6},
 pages = {063526},
 primaryclass = {astro-ph},
 title = {{Intrinsic alignment-lensing interference as a contaminant of cosmic shear}},
 volume = {70},
 year = {2004}
}

@article{2005astro.ph.10346T,
 adsurl = {https://ui.adsabs.harvard.edu/abs/2005astro.ph.10346T},
 archiveprefix = {arXiv},
 author = {{The Dark Energy Survey Collaboration}},
 doi = {10.48550/arXiv.astro-ph/0510346},
 eid = {astro-ph/0510346},
 eprint = {astro-ph/0510346},
 journal = {arXiv e-prints},
 month = {October},
 pages = {astro-ph/0510346},
 primaryclass = {astro-ph},
 title = {{The Dark Energy Survey}},
 year = {2005}
}

@article{2007.01844,
 adsurl = {https://ui.adsabs.harvard.edu/abs/2021A&A...646A.129J},
 archiveprefix = {arXiv},
 author = {{Joachimi}, B. and {Lin}, C. -A. and {Asgari}, M. and {Tr{\"o}ster}, T. and {Heymans}, C. and {Hildebrandt}, H. and {K{\"o}hlinger}, F. and {S{\'a}nchez}, A.~G. and {Wright}, A.~H. and {Bilicki}, M. and {Blake}, C. and {van den Busch}, J.~L. and {Crocce}, M. and {Dvornik}, A. and {Erben}, T. and {Getman}, F. and {Giblin}, B. and {Hoekstra}, H. and {Kannawadi}, A. and {Kuijken}, K. and {Napolitano}, N.~R. and {Schneider}, P. and {Scoccimarro}, R. and {Sellentin}, E. and {Shan}, H.~Y. and {von Wietersheim-Kramsta}, M. and {Zuntz}, J.},
 doi = {10.1051/0004-6361/202038831},
 eid = {A129},
 eprint = {2007.01844},
 journal = {\aap},
 month = {February},
 pages = {A129},
 primaryclass = {astro-ph.CO},
 title = {{KiDS-1000 methodology: Modelling and inference for joint weak gravitational lensing and spectroscopic galaxy clustering analysis}},
 volume = {646},
 year = {2021}
}

@article{2007.15633,
 adsurl = {https://ui.adsabs.harvard.edu/abs/2021A&A...645A.104A},
 archiveprefix = {arXiv},
 author = {{Asgari}, Marika and {Lin}, Chieh-An and {Joachimi}, Benjamin and {Giblin}, Benjamin and {Heymans}, Catherine and {Hildebrandt}, Hendrik and {Kannawadi}, Arun and {St{\"o}lzner}, Benjamin and {Tr{\"o}ster}, Tilman and {van den Busch}, Jan Luca and {Wright}, Angus H. and {Bilicki}, Maciej and {Blake}, Chris and {de Jong}, Jelte and {Dvornik}, Andrej and {Erben}, Thomas and {Getman}, Fedor and {Hoekstra}, Henk and {K{\"o}hlinger}, Fabian and {Kuijken}, Konrad and {Miller}, Lance and {Radovich}, Mario and {Schneider}, Peter and {Shan}, HuanYuan and {Valentijn}, Edwin},
 doi = {10.1051/0004-6361/202039070},
 eid = {A104},
 eprint = {2007.15633},
 journal = {\aap},
 month = {January},
 pages = {A104},
 primaryclass = {astro-ph.CO},
 title = {{KiDS-1000 cosmology: Cosmic shear constraints and comparison between two point statistics}},
 volume = {645},
 year = {2021}
}

@article{2007.15635,
 adsurl = {https://ui.adsabs.harvard.edu/abs/2021A&A...647A.124H},
 archiveprefix = {arXiv},
 author = {{Hildebrandt}, H. and {van den Busch}, J.~L. and {Wright}, A.~H. and {Blake}, C. and {Joachimi}, B. and {Kuijken}, K. and {Tr{\"o}ster}, T. and {Asgari}, M. and {Bilicki}, M. and {de Jong}, J.~T.~A. and {Dvornik}, A. and {Erben}, T. and {Getman}, F. and {Giblin}, B. and {Heymans}, C. and {Kannawadi}, A. and {Lin}, C. -A. and {Shan}, H. -Y.},
 doi = {10.1051/0004-6361/202039018},
 eid = {A124},
 eprint = {2007.15635},
 journal = {\aap},
 month = {March},
 pages = {A124},
 primaryclass = {astro-ph.CO},
 title = {{KiDS-1000 catalogue: Redshift distributions and their calibration}},
 volume = {647},
 year = {2021}
}

@article{2007NJPh....9..444B,
 adsurl = {https://ui.adsabs.harvard.edu/abs/2007NJPh....9..444B},
 archiveprefix = {arXiv},
 author = {{Bridle}, Sarah and {King}, Lindsay},
 doi = {10.1088/1367-2630/9/12/444},
 eprint = {0705.0166},
 journal = {New Journal of Physics},
 month = {December},
 number = {12},
 pages = {444},
 primaryclass = {astro-ph},
 title = {{Dark energy constraints from cosmic shear power spectra: impact of intrinsic alignments on photometric redshift requirements}},
 volume = {9},
 year = {2007}
}

@article{2008PhRvD..78l3506L,
 adsurl = {https://ui.adsabs.harvard.edu/abs/2008PhRvD..78l3506L},
 archiveprefix = {arXiv},
 author = {{LoVerde}, Marilena and {Afshordi}, Niayesh},
 doi = {10.1103/PhysRevD.78.123506},
 eid = {123506},
 eprint = {0809.5112},
 journal = {\prd},
 month = {December},
 number = {12},
 pages = {123506},
 primaryclass = {astro-ph},
 title = {{Extended Limber approximation}},
 volume = {78},
 year = {2008}
}

@article{2010.09717,
 adsurl = {https://ui.adsabs.harvard.edu/abs/2021JCAP...03..067N},
 archiveprefix = {arXiv},
 author = {{Nicola}, Andrina and {Garc{\'\i}a-Garc{\'\i}a}, Carlos and {Alonso}, David and {Dunkley}, Jo and {Ferreira}, Pedro G. and {Slosar}, An{\v{z}}e and {Spergel}, David N.},
 doi = {10.1088/1475-7516/2021/03/067},
 eid = {067},
 eprint = {2010.09717},
 journal = {\jcap},
 month = {March},
 number = {3},
 pages = {067},
 primaryclass = {astro-ph.CO},
 title = {{Cosmic shear power spectra in practice}},
 volume = {2021},
 year = {2021}
}

@article{2011ApJS..194...41S,
 adsurl = {https://ui.adsabs.harvard.edu/abs/2011ApJS..194...41S},
 archiveprefix = {arXiv},
 author = {{Swetz}, D.~S. and {Ade}, P.~A.~R. and {Amiri}, M. and {Appel}, J.~W. and {Battistelli}, E.~S. and {Burger}, B. and {Chervenak}, J. and {Devlin}, M.~J. and {Dicker}, S.~R. and {Doriese}, W.~B. and {D{\"u}nner}, R. and {Essinger-Hileman}, T. and {Fisher}, R.~P. and {Fowler}, J.~W. and {Halpern}, M. and {Hasselfield}, M. and {Hilton}, G.~C. and {Hincks}, A.~D. and {Irwin}, K.~D. and {Jarosik}, N. and {Kaul}, M. and {Klein}, J. and {Lau}, J.~M. and {Limon}, M. and {Marriage}, T.~A. and {Marsden}, D. and {Martocci}, K. and {Mauskopf}, P. and {Moseley}, H. and {Netterfield}, C.~B. and {Niemack}, M.~D. and {Nolta}, M.~R. and {Page}, L.~A. and {Parker}, L. and {Staggs}, S.~T. and {Stryzak}, O. and {Switzer}, E.~R. and {Thornton}, R. and {Tucker}, C. and {Wollack}, E. and {Zhao}, Y.},
 doi = {10.1088/0067-0049/194/2/41},
 eid = {41},
 eprint = {1007.0290},
 journal = {\apjs},
 month = {June},
 number = {2},
 pages = {41},
 primaryclass = {astro-ph.IM},
 title = {{Overview of the Atacama Cosmology Telescope: Receiver, Instrumentation, and Telescope Systems}},
 volume = {194},
 year = {2011}
}

@article{2011arXiv1104.2932L,
 adsurl = {https://ui.adsabs.harvard.edu/abs/2011arXiv1104.2932L},
 archiveprefix = {arXiv},
 author = {{Lesgourgues}, Julien},
 doi = {10.48550/arXiv.1104.2932},
 eid = {arXiv:1104.2932},
 eprint = {1104.2932},
 journal = {arXiv e-prints},
 month = {April},
 pages = {arXiv:1104.2932},
 primaryclass = {astro-ph.IM},
 title = {{The Cosmic Linear Anisotropy Solving System (CLASS) I: Overview}},
 year = {2011}
}

@article{2012.08568,
 adsurl = {https://ui.adsabs.harvard.edu/abs/2021MNRAS.508.3125F},
 archiveprefix = {arXiv},
 author = {{Friedrich}, O. and {Andrade-Oliveira}, F. and {Camacho}, H. and {Alves}, O. and {Rosenfeld}, R. and {Sanchez}, J. and {Fang}, X. and {Eifler}, T.~F. and {Krause}, E. and {Chang}, C. and {Omori}, Y. and {Amon}, A. and {Baxter}, E. and {Elvin-Poole}, J. and {Huterer}, D. and {Porredon}, A. and {Prat}, J. and {Terra}, V. and {Troja}, A. and {Alarcon}, A. and {Bechtol}, K. and {Bernstein}, G.~M. and {Buchs}, R. and {Campos}, A. and {Carnero Rosell}, A. and {Carrasco Kind}, M. and {Cawthon}, R. and {Choi}, A. and {Cordero}, J. and {Crocce}, M. and {Davis}, C. and {DeRose}, J. and {Diehl}, H.~T. and {Dodelson}, S. and {Doux}, C. and {Drlica-Wagner}, A. and {Elsner}, F. and {Everett}, S. and {Fosalba}, P. and {Gatti}, M. and {Giannini}, G. and {Gruen}, D. and {Gruendl}, R.~A. and {Harrison}, I. and {Hartley}, W.~G. and {Jain}, B. and {Jarvis}, M. and {MacCrann}, N. and {McCullough}, J. and {Muir}, J. and {Myles}, J. and {Pandey}, S. and {Raveri}, M. and {Roodman}, A. and {Rodriguez-Monroy}, M. and {Rykoff}, E.~S. and {Samuroff}, S. and {S{\'a}nchez}, C. and {Secco}, L.~F. and {Sevilla-Noarbe}, I. and {Sheldon}, E. and {Troxel}, M.~A. and {Weaverdyck}, N. and {Yanny}, B. and {Aguena}, M. and {Avila}, S. and {Bacon}, D. and {Bertin}, E. and {Bhargava}, S. and {Brooks}, D. and {Burke}, D.~L. and {Carretero}, J. and {Costanzi}, M. and {da Costa}, L.~N. and {Pereira}, M.~E.~S. and {De Vicente}, J. and {Desai}, S. and {Evrard}, A.~E. and {Ferrero}, I. and {Frieman}, J. and {Garc{\'\i}a-Bellido}, J. and {Gaztanaga}, E. and {Gerdes}, D.~W. and {Giannantonio}, T. and {Gschwend}, J. and {Gutierrez}, G. and {Hinton}, S.~R. and {Hollowood}, D.~L. and {Honscheid}, K. and {James}, D.~J. and {Kuehn}, K. and {Lahav}, O. and {Lima}, M. and {Maia}, M.~A.~G. and {Menanteau}, F. and {Miquel}, R. and {Morgan}, R. and {Palmese}, A. and {Paz-Chinch{\'o}n}, F. and {Plazas}, A.~A. and {Sanchez}, E. and {Scarpine}, V. and {Serrano}, S. and {Soares-Santos}, M. and {Smith}, M. and {Suchyta}, E. and {Tarle}, G. and {Thomas}, D. and {To}, C. and {Varga}, T.~N. and {Weller}, J. and {Wilkinson}, R.~D. and {Wilkinson}, R.~D. and {DES Collaboration}},
 doi = {10.1093/mnras/stab2384},
 eprint = {2012.08568},
 journal = {\mnras},
 month = {December},
 number = {3},
 pages = {3125-3165},
 primaryclass = {astro-ph.CO},
 title = {{Dark Energy Survey year 3 results: covariance modelling and its impact on parameter estimation and quality of fit}},
 volume = {508},
 year = {2021}
}

@article{2013ExA....35...25D,
 adsurl = {https://ui.adsabs.harvard.edu/abs/2013ExA....35...25D},
 archiveprefix = {arXiv},
 author = {{de Jong}, Jelte T.~A. and {Verdoes Kleijn}, Gijs A. and {Kuijken}, Konrad H. and {Valentijn}, Edwin A.},
 doi = {10.1007/s10686-012-9306-1},
 eprint = {1206.1254},
 journal = {Experimental Astronomy},
 month = {January},
 number = {1-2},
 pages = {25-44},
 primaryclass = {astro-ph.CO},
 title = {{The Kilo-Degree Survey}},
 volume = {35},
 year = {2013}
}

@article{2013JCAP...02..001A,
 adsurl = {https://ui.adsabs.harvard.edu/abs/2013JCAP...02..001A},
 archiveprefix = {arXiv},
 author = {{Audren}, Benjamin and {Lesgourgues}, Julien and {Benabed}, Karim and {Prunet}, Simon},
 doi = {10.1088/1475-7516/2013/02/001},
 eid = {001},
 eprint = {1210.7183},
 journal = {\jcap},
 month = {February},
 number = {2},
 pages = {001},
 primaryclass = {astro-ph.CO},
 title = {{Conservative constraints on early cosmology with MONTE PYTHON}},
 volume = {2013},
 year = {2013}
}

@article{2018ARA&A..56..393M,
 adsurl = {https://ui.adsabs.harvard.edu/abs/2018ARA&A..56..393M},
 archiveprefix = {arXiv},
 author = {{Mandelbaum}, Rachel},
 doi = {10.1146/annurev-astro-081817-051928},
 eprint = {1710.03235},
 journal = {\araa},
 month = {September},
 pages = {393-433},
 primaryclass = {astro-ph.CO},
 title = {{Weak Lensing for Precision Cosmology}},
 volume = {56},
 year = {2018}
}

@article{2018PASJ...70S...4A,
 adsurl = {https://ui.adsabs.harvard.edu/abs/2018PASJ...70S...4A},
 archiveprefix = {arXiv},
 author = {{Aihara}, Hiroaki and {Arimoto}, Nobuo and {Armstrong}, Robert and {Arnouts}, St{\'e}phane and {Bahcall}, Neta A. and {Bickerton}, Steven and {Bosch}, James and {Bundy}, Kevin and {Capak}, Peter L. and {Chan}, James H.~H. and {Chiba}, Masashi and {Coupon}, Jean and {Egami}, Eiichi and {Enoki}, Motohiro and {Finet}, Francois and {Fujimori}, Hiroki and {Fujimoto}, Seiji and {Furusawa}, Hisanori and {Furusawa}, Junko and {Goto}, Tomotsugu and {Goulding}, Andy and {Greco}, Johnny P. and {Greene}, Jenny E. and {Gunn}, James E. and {Hamana}, Takashi and {Harikane}, Yuichi and {Hashimoto}, Yasuhiro and {Hattori}, Takashi and {Hayashi}, Masao and {Hayashi}, Yusuke and {He{\l}miniak}, Krzysztof G. and {Higuchi}, Ryo and {Hikage}, Chiaki and {Ho}, Paul T.~P. and {Hsieh}, Bau-Ching and {Huang}, Kuiyun and {Huang}, Song and {Ikeda}, Hiroyuki and {Imanishi}, Masatoshi and {Inoue}, Akio K. and {Iwasawa}, Kazushi and {Iwata}, Ikuru and {Jaelani}, Anton T. and {Jian}, Hung-Yu and {Kamata}, Yukiko and {Karoji}, Hiroshi and {Kashikawa}, Nobunari and {Katayama}, Nobuhiko and {Kawanomoto}, Satoshi and {Kayo}, Issha and {Koda}, Jin and {Koike}, Michitaro and {Kojima}, Takashi and {Komiyama}, Yutaka and {Konno}, Akira and {Koshida}, Shintaro and {Koyama}, Yusei and {Kusakabe}, Haruka and {Leauthaud}, Alexie and {Lee}, Chien-Hsiu and {Lin}, Lihwai and {Lin}, Yen-Ting and {Lupton}, Robert H. and {Mandelbaum}, Rachel and {Matsuoka}, Yoshiki and {Medezinski}, Elinor and {Mineo}, Sogo and {Miyama}, Shoken and {Miyatake}, Hironao and {Miyazaki}, Satoshi and {Momose}, Rieko and {More}, Anupreeta and {More}, Surhud and {Moritani}, Yuki and {Moriya}, Takashi J. and {Morokuma}, Tomoki and {Mukae}, Shiro and {Murata}, Ryoma and {Murayama}, Hitoshi and {Nagao}, Tohru and {Nakata}, Fumiaki and {Niida}, Mana and {Niikura}, Hiroko and {Nishizawa}, Atsushi J. and {Obuchi}, Yoshiyuki and {Oguri}, Masamune and {Oishi}, Yukie and {Okabe}, Nobuhiro and {Okamoto}, Sakurako and {Okura}, Yuki and {Ono}, Yoshiaki and {Onodera}, Masato and {Onoue}, Masafusa and {Osato}, Ken and {Ouchi}, Masami and {Price}, Paul A. and {Pyo}, Tae-Soo and {Sako}, Masao and {Sawicki}, Marcin and {Shibuya}, Takatoshi and {Shimasaku}, Kazuhiro and {Shimono}, Atsushi and {Shirasaki}, Masato and {Silverman}, John D. and {Simet}, Melanie and {Speagle}, Joshua and {Spergel}, David N. and {Strauss}, Michael A. and {Sugahara}, Yuma and {Sugiyama}, Naoshi and {Suto}, Yasushi and {Suyu}, Sherry H. and {Suzuki}, Nao and {Tait}, Philip J. and {Takada}, Masahiro and {Takata}, Tadafumi and {Tamura}, Naoyuki and {Tanaka}, Manobu M. and {Tanaka}, Masaomi and {Tanaka}, Masayuki and {Tanaka}, Yoko and {Terai}, Tsuyoshi and {Terashima}, Yuichi and {Toba}, Yoshiki and {Tominaga}, Nozomu and {Toshikawa}, Jun and {Turner}, Edwin L. and {Uchida}, Tomohisa and {Uchiyama}, Hisakazu and {Umetsu}, Keiichi and {Uraguchi}, Fumihiro and {Urata}, Yuji and {Usuda}, Tomonori and {Utsumi}, Yousuke and {Wang}, Shiang-Yu and {Wang}, Wei-Hao and {Wong}, Kenneth C. and {Yabe}, Kiyoto and {Yamada}, Yoshihiko and {Yamanoi}, Hitomi and {Yasuda}, Naoki and {Yeh}, Sherry and {Yonehara}, Atsunori and {Yuma}, Suraphong},
 doi = {10.1093/pasj/psx066},
 eid = {S4},
 eprint = {1704.05858},
 journal = {\pasj},
 month = {January},
 pages = {S4},
 primaryclass = {astro-ph.IM},
 title = {{The Hyper Suprime-Cam SSP Survey: Overview and survey design}},
 volume = {70},
 year = {2018}
}

@article{2019ApJS..242....2C,
 adsurl = {https://ui.adsabs.harvard.edu/abs/2019ApJS..242....2C},
 archiveprefix = {arXiv},
 author = {{Chisari}, Nora Elisa and {Alonso}, David and {Krause}, Elisabeth and {Leonard}, C. Danielle and {Bull}, Philip and {Neveu}, J{\'e}r{\'e}my and {Villarreal}, Antonia Sierra and {Singh}, Sukhdeep and {McClintock}, Thomas and {Ellison}, John and {Du}, Zilong and {Zuntz}, Joe and {Mead}, Alexander and {Joudaki}, Shahab and {Lorenz}, Christiane S. and {Tr{\"o}ster}, Tilman and {Sanchez}, Javier and {Lanusse}, Francois and {Ishak}, Mustapha and {Hlozek}, Ren{\'e}e and {Blazek}, Jonathan and {Campagne}, Jean-Eric and {Almoubayyed}, Husni and {Eifler}, Tim and {Kirby}, Matthew and {Kirkby}, David and {Plaszczynski}, St{\'e}phane and {Slosar}, An{\v{z}}e and {Vrastil}, Michal and {Wagoner}, Erika L. and {LSST Dark Energy Science Collaboration}},
 doi = {10.3847/1538-4365/ab1658},
 eid = {2},
 eprint = {1812.05995},
 journal = {\apjs},
 month = {May},
 number = {1},
 pages = {2},
 primaryclass = {astro-ph.CO},
 title = {{Core Cosmology Library: Precision Cosmological Predictions for LSST}},
 volume = {242},
 year = {2019}
}

@article{2019arXiv191013970L,
 adsurl = {https://ui.adsabs.harvard.edu/abs/2019arXiv191013970L},
 archiveprefix = {arXiv},
 author = {{Lewis}, Antony},
 doi = {10.48550/arXiv.1910.13970},
 eid = {arXiv:1910.13970},
 eprint = {1910.13970},
 journal = {arXiv e-prints},
 month = {October},
 pages = {arXiv:1910.13970},
 primaryclass = {astro-ph.IM},
 title = {{GetDist: a Python package for analysing Monte Carlo samples}},
 year = {2019}
}

@article{2020PhRvD.101j3527R,
 adsurl = {https://ui.adsabs.harvard.edu/abs/2020PhRvD.101j3527R},
 archiveprefix = {arXiv},
 author = {{Raveri}, Marco and {Zacharegkas}, Georgios and {Hu}, Wayne},
 doi = {10.1103/PhysRevD.101.103527},
 eid = {103527},
 eprint = {1912.04880},
 journal = {\prd},
 month = {May},
 number = {10},
 pages = {103527},
 primaryclass = {astro-ph.CO},
 title = {{Quantifying concordance of correlated cosmological data sets}},
 volume = {101},
 year = {2020}
}

@article{2021MNRAS.502.1401M,
 adsurl = {https://ui.adsabs.harvard.edu/abs/2021MNRAS.502.1401M},
 archiveprefix = {arXiv},
 author = {{Mead}, A.~J. and {Brieden}, S. and {Tr{\"o}ster}, T. and {Heymans}, C.},
 doi = {10.1093/mnras/stab082},
 eprint = {2009.01858},
 journal = {\mnras},
 month = {March},
 number = {1},
 pages = {1401-1422},
 primaryclass = {astro-ph.CO},
 title = {{HMCODE-2020: improved modelling of non-linear cosmological power spectra with baryonic feedback}},
 volume = {502},
 year = {2021}
}

@article{2021PhRvD.104d3504R,
 adsurl = {https://ui.adsabs.harvard.edu/abs/2021PhRvD.104d3504R},
 archiveprefix = {arXiv},
 author = {{Raveri}, Marco and {Doux}, Cyrille},
 doi = {10.1103/PhysRevD.104.043504},
 eid = {043504},
 eprint = {2105.03324},
 journal = {\prd},
 month = {August},
 number = {4},
 pages = {043504},
 primaryclass = {astro-ph.CO},
 title = {{Non-Gaussian estimates of tensions in cosmological parameters}},
 volume = {104},
 year = {2021}
}

@article{2023MNRAS.525.3181L,
 adsurl = {https://ui.adsabs.harvard.edu/abs/2023MNRAS.525.3181L},
 archiveprefix = {arXiv},
 author = {{Lange}, Johannes U.},
 doi = {10.1093/mnras/stad2441},
 eprint = {2306.16923},
 journal = {\mnras},
 month = {October},
 number = {2},
 pages = {3181-3194},
 primaryclass = {astro-ph.IM},
 title = {{NAUTILUS: boosting Bayesian importance nested sampling with deep learning}},
 volume = {525},
 year = {2023}
}

@article{2024arXiv240906771Z,
 adsurl = {https://ui.adsabs.harvard.edu/abs/2024arXiv240906771Z},
 archiveprefix = {arXiv},
 author = {{Zhou}, Zilu and {Weiner}, Neal},
 doi = {10.48550/arXiv.2409.06771},
 eid = {arXiv:2409.06771},
 eprint = {2409.06771},
 journal = {arXiv e-prints},
 month = {September},
 pages = {arXiv:2409.06771},
 primaryclass = {hep-ph},
 title = {{Searching for Dark Matter Interactions with ACT, SPT and DES}},
 year = {2024}
}

@article{2025JCAP...02..021A,
 adsurl = {https://ui.adsabs.harvard.edu/abs/2025JCAP...02..021A},
 archiveprefix = {arXiv},
 author = {{Adame}, A.~G. and {Aguilar}, J. and {Ahlen}, S. and {Alam}, S. and {Alexander}, D.~M. and {Alvarez}, M. and {Alves}, O. and {Anand}, A. and {Andrade}, U. and {Armengaud}, E. and {Avila}, S. and {Aviles}, A. and {Awan}, H. and {Bahr-Kalus}, B. and {Bailey}, S. and {Baltay}, C. and {Bault}, A. and {Behera}, J. and {BenZvi}, S. and {Bera}, A. and {Beutler}, F. and {Bianchi}, D. and {Blake}, C. and {Blum}, R. and {Brieden}, S. and {Brodzeller}, A. and {Brooks}, D. and {Buckley-Geer}, E. and {Burtin}, E. and {Calderon}, R. and {Canning}, R. and {Carnero Rosell}, A. and {Cereskaite}, R. and {Cervantes-Cota}, J.~L. and {Chabanier}, S. and {Chaussidon}, E. and {Chaves-Montero}, J. and {Chen}, S. and {Chen}, X. and {Claybaugh}, T. and {Cole}, S. and {Cuceu}, A. and {Davis}, T.~M. and {Dawson}, K. and {de la Macorra}, A. and {de Mattia}, A. and {Deiosso}, N. and {Dey}, A. and {Dey}, B. and {Ding}, Z. and {Doel}, P. and {Edelstein}, J. and {Eftekharzadeh}, S. and {Eisenstein}, D.~J. and {Elliott}, A. and {Fagrelius}, P. and {Fanning}, K. and {Ferraro}, S. and {Ereza}, J. and {Findlay}, N. and {Flaugher}, B. and {Font-Ribera}, A. and {Forero-S{\'a}nchez}, D. and {Forero-Romero}, J.~E. and {Frenk}, C.~S. and {Garcia-Quintero}, C. and {Gazta{\~n}aga}, E. and {Gil-Mar{\'\i}n}, H. and {Gontcho a Gontcho}, S. and {Gonzalez-Morales}, A.~X. and {Gonzalez-Perez}, V. and {Gordon}, C. and {Green}, D. and {Gruen}, D. and {Gsponer}, R. and {Gutierrez}, G. and {Guy}, J. and {Hadzhiyska}, B. and {Hahn}, C. and {Hanif}, M.~M.~S. and {Herrera-Alcantar}, H.~K. and {Honscheid}, K. and {Howlett}, C. and {Huterer}, D. and {Ir{\v{s}}i{\v{c}}}, V. and {Ishak}, M. and {Juneau}, S. and {Kara{\c{c}}ayl{\i}}, N.~G. and {Kehoe}, R. and {Kent}, S. and {Kirkby}, D. and {Kremin}, A. and {Krolewski}, A. and {Lai}, Y. and {Lan}, T. -W. and {Landriau}, M. and {Lang}, D. and {Lasker}, J. and {Le Goff}, J.~M. and {Le Guillou}, L. and {Leauthaud}, A. and {Levi}, M.~E. and {Li}, T.~S. and {Linder}, E. and {Lodha}, K. and {Magneville}, C. and {Manera}, M. and {Margala}, D. and {Martini}, P. and {Maus}, M. and {McDonald}, P. and {Medina-Varela}, L. and {Meisner}, A. and {Mena-Fern{\'a}ndez}, J. and {Miquel}, R. and {Moon}, J. and {Moore}, S. and {Moustakas}, J. and {Mueller}, E. and {Mu{\~n}oz-Guti{\'e}rrez}, A. and {Myers}, A.~D. and {Nadathur}, S. and {Napolitano}, L. and {Neveux}, R. and {Newman}, J.~A. and {Nguyen}, N.~M. and {Nie}, J. and {Niz}, G. and {Noriega}, H.~E. and {Padmanabhan}, N. and {Paillas}, E. and {Palanque-Delabrouille}, N. and {Pan}, J. and {Penmetsa}, S. and {Percival}, W.~J. and {Pieri}, M.~M. and {Pinon}, M. and {Poppett}, C. and {Porredon}, A. and {Prada}, F. and {P{\'e}rez-Fern{\'a}ndez}, A. and {P{\'e}rez-R{\`a}fols}, I. and {Rabinowitz}, D. and {Raichoor}, A. and {Ram{\'\i}rez-P{\'e}rez}, C. and {Ramirez-Solano}, S. and {Rashkovetskyi}, M. and {Ravoux}, C. and {Rezaie}, M. and {Rich}, J. and {Rocher}, A. and {Rockosi}, C. and {Roe}, N.~A. and {Rosado-Marin}, A. and {Ross}, A.~J. and {Rossi}, G. and {Ruggeri}, R. and {Ruhlmann-Kleider}, V. and {Samushia}, L. and {Sanchez}, E. and {Saulder}, C. and {Schlafly}, E.~F. and {Schlegel}, D. and {Schubnell}, M. and {Seo}, H. and {Shafieloo}, A. and {Sharples}, R. and {Silber}, J. and {Slosar}, A. and {Smith}, A. and {Sprayberry}, D. and {Tan}, T. and {Tarl{\'e}}, G. and {Taylor}, P. and {Trusov}, S. and {Ure{\~n}a-L{\'o}pez}, L.~A. and {Vaisakh}, R. and {Valcin}, D. and {Valdes}, F. and {Vargas-Maga{\~n}a}, M. and {Verde}, L. and {Walther}, M. and {Wang}, B. and {Wang}, M.~S. and {Weaver}, B.~A. and {Weaverdyck}, N. and {Wechsler}, R.~H. and {Weinberg}, D.~H. and {White}, M. and {Yu}, J. and {Yu}, Y. and {Yuan}, S. and {Y{\`e}che}, C. and {Zaborowski}, E.~A. and {Zarrouk}, P. and {Zhang}, H. and {Zhao}, C. and {Zhao}, R. and {Zhou}, R. and {Zhuang}, T.},
 doi = {10.1088/1475-7516/2025/02/021},
 eid = {021},
 eprint = {2404.03002},
 journal = {\jcap},
 month = {February},
 number = {2},
 pages = {021},
 primaryclass = {astro-ph.CO},
 title = {{DESI 2024 VI: cosmological constraints from the measurements of baryon acoustic oscillations}},
 volume = {2025},
 year = {2025}
}

@article{2025PhRvD.111f3509T,
 adsurl = {https://ui.adsabs.harvard.edu/abs/2025PhRvD.111f3509T},
 archiveprefix = {arXiv},
 author = {{Terasawa}, Ryo and {Li}, Xiangchong and {Takada}, Masahiro and {Nishimichi}, Takahiro and {Tanaka}, Satoshi and {Sugiyama}, Sunao and {Kurita}, Toshiki and {Zhang}, Tianqing and {Shirasaki}, Masato and {Takahashi}, Ryuichi and {Miyatake}, Hironao and {More}, Surhud and {Nishizawa}, Atsushi J.},
 doi = {10.1103/PhysRevD.111.063509},
 eid = {063509},
 eprint = {2403.20323},
 journal = {\prd},
 month = {March},
 number = {6},
 pages = {063509},
 primaryclass = {astro-ph.CO},
 title = {{Exploring the baryonic effect signature in the Hyper Suprime-Cam Year 3 cosmic shear two-point correlations on small scales: The $S_8$ tension remains present}},
 volume = {111},
 year = {2025}
}

@article{2105.13543,
 adsurl = {https://ui.adsabs.harvard.edu/abs/2022PhRvD.105b3514A},
 archiveprefix = {arXiv},
 author = {{Amon}, A. and {Gruen}, D. and {Troxel}, M.~A. and {MacCrann}, N. and {Dodelson}, S. and {Choi}, A. and {Doux}, C. and {Secco}, L.~F. and {Samuroff}, S. and {Krause}, E. and {Cordero}, J. and {Myles}, J. and {DeRose}, J. and {Wechsler}, R.~H. and {Gatti}, M. and {Navarro-Alsina}, A. and {Bernstein}, G.~M. and {Jain}, B. and {Blazek}, J. and {Alarcon}, A. and {Fert{\'e}}, A. and {Lemos}, P. and {Raveri}, M. and {Campos}, A. and {Prat}, J. and {S{\'a}nchez}, C. and {Jarvis}, M. and {Alves}, O. and {Andrade-Oliveira}, F. and {Baxter}, E. and {Bechtol}, K. and {Becker}, M.~R. and {Bridle}, S.~L. and {Camacho}, H. and {Carnero Rosell}, A. and {Carrasco Kind}, M. and {Cawthon}, R. and {Chang}, C. and {Chen}, R. and {Chintalapati}, P. and {Crocce}, M. and {Davis}, C. and {Diehl}, H.~T. and {Drlica-Wagner}, A. and {Eckert}, K. and {Eifler}, T.~F. and {Elvin-Poole}, J. and {Everett}, S. and {Fang}, X. and {Fosalba}, P. and {Friedrich}, O. and {Gaztanaga}, E. and {Giannini}, G. and {Gruendl}, R.~A. and {Harrison}, I. and {Hartley}, W.~G. and {Herner}, K. and {Huang}, H. and {Huff}, E.~M. and {Huterer}, D. and {Kuropatkin}, N. and {Leget}, P. and {Liddle}, A.~R. and {McCullough}, J. and {Muir}, J. and {Pandey}, S. and {Park}, Y. and {Porredon}, A. and {Refregier}, A. and {Rollins}, R.~P. and {Roodman}, A. and {Rosenfeld}, R. and {Ross}, A.~J. and {Rykoff}, E.~S. and {Sanchez}, J. and {Sevilla-Noarbe}, I. and {Sheldon}, E. and {Shin}, T. and {Troja}, A. and {Tutusaus}, I. and {Tutusaus}, I. and {Varga}, T.~N. and {Weaverdyck}, N. and {Yanny}, B. and {Yin}, B. and {Zhang}, Y. and {Zuntz}, J. and {Aguena}, M. and {Allam}, S. and {Annis}, J. and {Bacon}, D. and {Bertin}, E. and {Bhargava}, S. and {Brooks}, D. and {Buckley-Geer}, E. and {Burke}, D.~L. and {Carretero}, J. and {Costanzi}, M. and {da Costa}, L.~N. and {Pereira}, M.~E.~S. and {De Vicente}, J. and {Desai}, S. and {Dietrich}, J.~P. and {Doel}, P. and {Ferrero}, I. and {Flaugher}, B. and {Frieman}, J. and {Garc{\'\i}a-Bellido}, J. and {Gaztanaga}, E. and {Gerdes}, D.~W. and {Giannantonio}, T. and {Gschwend}, J. and {Gutierrez}, G. and {Hinton}, S.~R. and {Hollowood}, D.~L. and {Honscheid}, K. and {Hoyle}, B. and {James}, D.~J. and {Kron}, R. and {Kuehn}, K. and {Lahav}, O. and {Lima}, M. and {Lin}, H. and {Maia}, M.~A.~G. and {Marshall}, J.~L. and {Martini}, P. and {Melchior}, P. and {Menanteau}, F. and {Miquel}, R. and {Mohr}, J.~J. and {Morgan}, R. and {Ogando}, R.~L.~C. and {Palmese}, A. and {Paz-Chinch{\'o}n}, F. and {Petravick}, D. and {Pieres}, A. and {Romer}, A.~K. and {Sanchez}, E. and {Scarpine}, V. and {Schubnell}, M. and {Serrano}, S. and {Smith}, M. and {Soares-Santos}, M. and {Tarle}, G. and {Thomas}, D. and {To}, C. and {Weller}, J. and {DES Collaboration}},
 doi = {10.1103/PhysRevD.105.023514},
 eid = {023514},
 eprint = {2105.13543},
 journal = {\prd},
 month = {January},
 number = {2},
 pages = {023514},
 primaryclass = {astro-ph.CO},
 title = {{Dark Energy Survey Year 3 results: Cosmology from cosmic shear and robustness to data calibration}},
 volume = {105},
 year = {2022}
}

@article{2105.13544,
 adsurl = {https://ui.adsabs.harvard.edu/abs/2022PhRvD.105b3515S},
 archiveprefix = {arXiv},
 author = {{Secco}, L.~F. and {Samuroff}, S. and {Krause}, E. and {Jain}, B. and {Blazek}, J. and {Raveri}, M. and {Campos}, A. and {Amon}, A. and {Chen}, A. and {Doux}, C. and {Choi}, A. and {Gruen}, D. and {Bernstein}, G.~M. and {Chang}, C. and {DeRose}, J. and {Myles}, J. and {Fert{\'e}}, A. and {Lemos}, P. and {Huterer}, D. and {Prat}, J. and {Troxel}, M.~A. and {MacCrann}, N. and {Liddle}, A.~R. and {Kacprzak}, T. and {Fang}, X. and {S{\'a}nchez}, C. and {Pandey}, S. and {Dodelson}, S. and {Chintalapati}, P. and {Hoffmann}, K. and {Alarcon}, A. and {Alves}, O. and {Andrade-Oliveira}, F. and {Baxter}, E.~J. and {Bechtol}, K. and {Becker}, M.~R. and {Brandao-Souza}, A. and {Camacho}, H. and {Carnero Rosell}, A. and {Carrasco Kind}, M. and {Cawthon}, R. and {Cordero}, J.~P. and {Crocce}, M. and {Davis}, C. and {Di Valentino}, E. and {Drlica-Wagner}, A. and {Eckert}, K. and {Eifler}, T.~F. and {Elidaiana}, M. and {Elsner}, F. and {Elvin-Poole}, J. and {Everett}, S. and {Fosalba}, P. and {Friedrich}, O. and {Gatti}, M. and {Giannini}, G. and {Gruendl}, R.~A. and {Harrison}, I. and {Hartley}, W.~G. and {Herner}, K. and {Huang}, H. and {Huff}, E.~M. and {Jarvis}, M. and {Jeffrey}, N. and {Kuropatkin}, N. and {Leget}, P. -F. and {Muir}, J. and {Mccullough}, J. and {Navarro Alsina}, A. and {Omori}, Y. and {Park}, Y. and {Porredon}, A. and {Rollins}, R. and {Roodman}, A. and {Rosenfeld}, R. and {Ross}, A.~J. and {Rykoff}, E.~S. and {Sanchez}, J. and {Sevilla-Noarbe}, I. and {Sheldon}, E.~S. and {Shin}, T. and {Troja}, A. and {Tutusaus}, I. and {Varga}, T.~N. and {Weaverdyck}, N. and {Wechsler}, R.~H. and {Yanny}, B. and {Yin}, B. and {Zhang}, Y. and {Zuntz}, J. and {Abbott}, T.~M.~C. and {Aguena}, M. and {Allam}, S. and {Annis}, J. and {Bacon}, D. and {Bertin}, E. and {Bhargava}, S. and {Bridle}, S.~L. and {Brooks}, D. and {Buckley-Geer}, E. and {Burke}, D.~L. and {Carretero}, J. and {Costanzi}, M. and {da Costa}, L.~N. and {De Vicente}, J. and {Diehl}, H.~T. and {Dietrich}, J.~P. and {Doel}, P. and {Ferrero}, I. and {Flaugher}, B. and {Frieman}, J. and {Garc{\'\i}a-Bellido}, J. and {Gaztanaga}, E. and {Gerdes}, D.~W. and {Giannantonio}, T. and {Gschwend}, J. and {Gutierrez}, G. and {Hinton}, S.~R. and {Hollowood}, D.~L. and {Honscheid}, K. and {Hoyle}, B. and {James}, D.~J. and {Jeltema}, T. and {Kuehn}, K. and {Lahav}, O. and {Lima}, M. and {Lin}, H. and {Maia}, M.~A.~G. and {Marshall}, J.~L. and {Martini}, P. and {Melchior}, P. and {Menanteau}, F. and {Miquel}, R. and {Mohr}, J.~J. and {Morgan}, R. and {Ogando}, R.~L.~C. and {Palmese}, A. and {Paz-Chinch{\'o}n}, F. and {Petravick}, D. and {Pieres}, A. and {Plazas Malag{\'o}n}, A.~A. and {Rodriguez-Monroy}, M. and {Romer}, A.~K. and {Sanchez}, E. and {Scarpine}, V. and {Schubnell}, M. and {Scolnic}, D. and {Serrano}, S. and {Smith}, M. and {Soares-Santos}, M. and {Suchyta}, E. and {Swanson}, M.~E.~C. and {Tarle}, G. and {Thomas}, D. and {To}, C. and {DES Collaboration}},
 doi = {10.1103/PhysRevD.105.023515},
 eid = {023515},
 eprint = {2105.13544},
 journal = {\prd},
 month = {January},
 number = {2},
 pages = {023515},
 primaryclass = {astro-ph.CO},
 title = {{Dark Energy Survey Year 3 results: Cosmology from cosmic shear and robustness to modeling uncertainty}},
 volume = {105},
 year = {2022}
}

@article{2203.07128,
 adsurl = {https://ui.adsabs.harvard.edu/abs/2022MNRAS.515.1942D},
 archiveprefix = {arXiv},
 author = {{Doux}, C. and {Jain}, B. and {Zeurcher}, D. and {Lee}, J. and {Fang}, X. and {Rosenfeld}, R. and {Amon}, A. and {Camacho}, H. and {Choi}, A. and {Secco}, L.~F. and {Blazek}, J. and {Chang}, C. and {Gatti}, M. and {Gaztanaga}, E. and {Jeffrey}, N. and {Raveri}, M. and {Samuroff}, S. and {Alarcon}, A. and {Alves}, O. and {Andrade-Oliveira}, F. and {Baxter}, E. and {Bechtol}, K. and {Becker}, M.~R. and {Bernstein}, G.~M. and {Campos}, A. and {Carnero Rosell}, A. and {Carrasco Kind}, M. and {Cawthon}, R. and {Chen}, R. and {Cordero}, J. and {Crocce}, M. and {Davis}, C. and {DeRose}, J. and {Dodelson}, S. and {Drlica-Wagner}, A. and {Eckert}, K. and {Eifler}, T.~F. and {Elsner}, F. and {Elvin-Poole}, J. and {Everett}, S. and {Fert{\'e}}, A. and {Fosalba}, P. and {Friedrich}, O. and {Giannini}, G. and {Gruen}, D. and {Gruendl}, R.~A. and {Harrison}, I. and {Hartley}, W.~G. and {Herner}, K. and {Huang}, H. and {Huff}, E.~M. and {Huterer}, D. and {Jarvis}, M. and {Krause}, E. and {Kuropatkin}, N. and {Leget}, P. -F. and {Lemos}, P. and {Liddle}, A.~R. and {MacCrann}, N. and {McCullough}, J. and {Muir}, J. and {Myles}, J. and {Navarro-Alsina}, A. and {Pandey}, S. and {Park}, Y. and {Porredon}, A. and {Prat}, J. and {Rodriguez-Monroy}, M. and {Rollins}, R.~P. and {Roodman}, A. and {Ross}, A.~J. and {Rykoff}, E.~S. and {S{\'a}nchez}, C. and {Sanchez}, J. and {Sevilla-Noarbe}, I. and {Sheldon}, E. and {Shin}, T. and {Troja}, A. and {Troxel}, M.~A. and {Tutusaus}, I. and {Varga}, T.~N. and {Weaverdyck}, N. and {Wechsler}, R.~H. and {Yanny}, B. and {Yin}, B. and {Zhang}, Y. and {Zuntz}, J. and {Abbott}, T.~M.~C. and {Aguena}, M. and {Allam}, S. and {Annis}, J. and {Bacon}, D. and {Bertin}, E. and {Bocquet}, S. and {Brooks}, D. and {Burke}, D.~L. and {Carretero}, J. and {Costanzi}, M. and {da Costa}, L.~N. and {Pereira}, M.~E.~S. and {De Vicente}, J. and {Desai}, S. and {Diehl}, H.~T. and {Doel}, P. and {Ferrero}, I. and {Flaugher}, B. and {Frieman}, J. and {Garc{\'\i}a-Bellido}, J. and {Gerdes}, D.~W. and {Giannantonio}, T. and {Gschwend}, J. and {Gutierrez}, G. and {Hinton}, S.~R. and {Hollowood}, D.~L. and {Honscheid}, K. and {James}, D.~J. and {Kim}, A.~G. and {Kuehn}, K. and {Lahav}, O. and {Marshall}, J.~L. and {Menanteau}, F. and {Miquel}, R. and {Morgan}, R. and {Ogando}, R.~L.~C. and {Palmese}, A. and {Paz-Chinch{\'o}n}, F. and {Pieres}, A. and {Plazas Malag{\'o}n}, A.~A. and {Reil}, K. and {Sanchez}, E. and {Scarpine}, V. and {Serrano}, S. and {Smith}, M. and {Suchyta}, E. and {Swanson}, M.~E.~C. and {Tarle}, G. and {Thomas}, D. and {To}, C. and {Weller}, J. and {DES Collaboration}},
 doi = {10.1093/mnras/stac1826},
 eprint = {2203.07128},
 journal = {\mnras},
 month = {September},
 number = {2},
 pages = {1942-1972},
 primaryclass = {astro-ph.CO},
 title = {{Dark energy survey year 3 results: cosmological constraints from the analysis of cosmic shear in harmonic space}},
 volume = {515},
 year = {2022}
}

@article{2206.11794,
 adsurl = {https://ui.adsabs.harvard.edu/abs/2022MNRAS.516.5355A},
 archiveprefix = {arXiv},
 author = {{Amon}, Alexandra and {Efstathiou}, George},
 doi = {10.1093/mnras/stac2429},
 eprint = {2206.11794},
 journal = {\mnras},
 month = {November},
 number = {4},
 pages = {5355-5366},
 primaryclass = {astro-ph.CO},
 title = {{A non-linear solution to the S$_{8}$ tension?}},
 volume = {516},
 year = {2022}
}

@article{2209.12997,
 adsurl = {https://ui.adsabs.harvard.edu/abs/2023PhRvD.107h3532S},
 archiveprefix = {arXiv},
 author = {{Secco}, Lucas F. and {Karwal}, Tanvi and {Hu}, Wayne and {Krause}, Elisabeth},
 doi = {10.1103/PhysRevD.107.083532},
 eid = {083532},
 eprint = {2209.12997},
 journal = {\prd},
 month = {April},
 number = {8},
 pages = {083532},
 primaryclass = {astro-ph.CO},
 title = {{Role of the Hubble scale in the weak lensing versus CMB tension}},
 volume = {107},
 year = {2023}
}

@article{2304.05202,
 adsurl = {https://ui.adsabs.harvard.edu/abs/2024ApJ...962..112Q},
 archiveprefix = {arXiv},
 author = {{Qu}, Frank J. and {Sherwin}, Blake D. and {Madhavacheril}, Mathew S. and {Han}, Dongwon and {Crowley}, Kevin T. and {Abril-Cabezas}, Irene and {Ade}, Peter A.~R. and {Aiola}, Simone and {Alford}, Tommy and {Amiri}, Mandana and {Amodeo}, Stefania and {An}, Rui and {Atkins}, Zachary and {Austermann}, Jason E. and {Battaglia}, Nicholas and {Battistelli}, Elia Stefano and {Beall}, James A. and {Bean}, Rachel and {Beringue}, Benjamin and {Bhandarkar}, Tanay and {Biermann}, Emily and {Bolliet}, Boris and {Bond}, J. Richard and {Cai}, Hongbo and {Calabrese}, Erminia and {Calafut}, Victoria and {Capalbo}, Valentina and {Carrero}, Felipe and {Carron}, Julien and {Challinor}, Anthony and {Chesmore}, Grace E. and {Cho}, Hsiao-mei and {Choi}, Steve K. and {Clark}, Susan E. and {C{\'o}rdova Rosado}, Rodrigo and {Cothard}, Nicholas F. and {Coughlin}, Kevin and {Coulton}, William and {Dalal}, Roohi and {Darwish}, Omar and {Devlin}, Mark J. and {Dicker}, Simon and {Doze}, Peter and {Duell}, Cody J. and {Duff}, Shannon M. and {Duivenvoorden}, Adriaan J. and {Dunkley}, Jo and {D{\"u}nner}, Rolando and {Fanfani}, Valentina and {Fankhanel}, Max and {Farren}, Gerrit and {Ferraro}, Simone and {Freundt}, Rodrigo and {Fuzia}, Brittany and {Gallardo}, Patricio A. and {Garrido}, Xavier and {Gluscevic}, Vera and {Golec}, Joseph E. and {Guan}, Yilun and {Halpern}, Mark and {Harrison}, Ian and {Hasselfield}, Matthew and {Healy}, Erin and {Henderson}, Shawn and {Hensley}, Brandon and {Herv{\'\i}as-Caimapo}, Carlos and {Hill}, J. Colin and {Hilton}, Gene C. and {Hilton}, Matt and {Hincks}, Adam D. and {Hlo{\v{z}}ek}, Ren{\'e}e and {Ho}, Shuay-Pwu Patty and {Huber}, Zachary B. and {Hubmayr}, Johannes and {Huffenberger}, Kevin M. and {Hughes}, John P. and {Irwin}, Kent and {Isopi}, Giovanni and {Jense}, Hidde T. and {Keller}, Ben and {Kim}, Joshua and {Knowles}, Kenda and {Koopman}, Brian J. and {Kosowsky}, Arthur and {Kramer}, Darby and {Kusiak}, Aleksandra and {La Posta}, Adrien and {Lague}, Alex and {Lakey}, Victoria and {Lee}, Eunseong and {Li}, Zack and {Li}, Yaqiong and {Limon}, Michele and {Lokken}, Martine and {Louis}, Thibaut and {Lungu}, Marius and {MacCrann}, Niall and {MacInnis}, Amanda and {Maldonado}, Diego and {Maldonado}, Felipe and {Mallaby-Kay}, Maya and {Marques}, Gabriela A. and {McMahon}, Jeff and {Mehta}, Yogesh and {Menanteau}, Felipe and {Moodley}, Kavilan and {Morris}, Thomas W. and {Mroczkowski}, Tony and {Naess}, Sigurd and {Namikawa}, Toshiya and {Nati}, Federico and {Newburgh}, Laura and {Nicola}, Andrina and {Niemack}, Michael D. and {Nolta}, Michael R. and {Orlowski-Scherer}, John and {Page}, Lyman A. and {Pandey}, Shivam and {Partridge}, Bruce and {Prince}, Heather and {Puddu}, Roberto and {Radiconi}, Federico and {Robertson}, Naomi and {Rojas}, Felipe and {Sakuma}, Tai and {Salatino}, Maria and {Schaan}, Emmanuel and {Schmitt}, Benjamin L. and {Sehgal}, Neelima and {Shaikh}, Shabbir and {Sierra}, Carlos and {Sievers}, Jon and {Sif{\'o}n}, Crist{\'o}bal and {Simon}, Sara and {Sonka}, Rita and {Spergel}, David N. and {Staggs}, Suzanne T. and {Storer}, Emilie and {Switzer}, Eric R. and {Tampier}, Niklas and {Thornton}, Robert and {Trac}, Hy and {Treu}, Jesse and {Tucker}, Carole and {Ullom}, Joel and {Vale}, Leila R. and {Van Engelen}, Alexander and {Van Lanen}, Jeff and {van Marrewijk}, Joshiwa and {Vargas}, Cristian and {Vavagiakis}, Eve M. and {Wagoner}, Kasey and {Wang}, Yuhan and {Wenzl}, Lukas and {Wollack}, Edward J. and {Xu}, Zhilei and {Zago}, Fernando and {Zheng}, Kaiwen},
 doi = {10.3847/1538-4357/acfe06},
 eid = {112},
 eprint = {2304.05202},
 journal = {\apj},
 month = {February},
 number = {2},
 pages = {112},
 primaryclass = {astro-ph.CO},
 title = {{The Atacama Cosmology Telescope: A Measurement of the DR6 CMB Lensing Power Spectrum and Its Implications for Structure Growth}},
 volume = {962},
 year = {2024}
}

@article{2304.05203,
 adsurl = {https://ui.adsabs.harvard.edu/abs/2024ApJ...962..113M},
 archiveprefix = {arXiv},
 author = {{Madhavacheril}, Mathew S. and {Qu}, Frank J. and {Sherwin}, Blake D. and {MacCrann}, Niall and {Li}, Yaqiong and {Abril-Cabezas}, Irene and {Ade}, Peter A.~R. and {Aiola}, Simone and {Alford}, Tommy and {Amiri}, Mandana and {Amodeo}, Stefania and {An}, Rui and {Atkins}, Zachary and {Austermann}, Jason E. and {Battaglia}, Nicholas and {Battistelli}, Elia Stefano and {Beall}, James A. and {Bean}, Rachel and {Beringue}, Benjamin and {Bhandarkar}, Tanay and {Biermann}, Emily and {Bolliet}, Boris and {Bond}, J. Richard and {Cai}, Hongbo and {Calabrese}, Erminia and {Calafut}, Victoria and {Capalbo}, Valentina and {Carrero}, Felipe and {Challinor}, Anthony and {Chesmore}, Grace E. and {Cho}, Hsiao-mei and {Choi}, Steve K. and {Clark}, Susan E. and {C{\'o}rdova Rosado}, Rodrigo and {Cothard}, Nicholas F. and {Coughlin}, Kevin and {Coulton}, William and {Crowley}, Kevin T. and {Dalal}, Roohi and {Darwish}, Omar and {Devlin}, Mark J. and {Dicker}, Simon and {Doze}, Peter and {Duell}, Cody J. and {Duff}, Shannon M. and {Duivenvoorden}, Adriaan J. and {Dunkley}, Jo and {D{\"u}nner}, Rolando and {Fanfani}, Valentina and {Fankhanel}, Max and {Farren}, Gerrit and {Ferraro}, Simone and {Freundt}, Rodrigo and {Fuzia}, Brittany and {Gallardo}, Patricio A. and {Garrido}, Xavier and {Givans}, Jahmour and {Gluscevic}, Vera and {Golec}, Joseph E. and {Guan}, Yilun and {Hall}, Kirsten R. and {Halpern}, Mark and {Han}, Dongwon and {Harrison}, Ian and {Hasselfield}, Matthew and {Healy}, Erin and {Henderson}, Shawn and {Hensley}, Brandon and {Herv{\'\i}as-Caimapo}, Carlos and {Hill}, J. Colin and {Hilton}, Gene C. and {Hilton}, Matt and {Hincks}, Adam D. and {Hlo{\v{z}}ek}, Ren{\'e}e and {Ho}, Shuay-Pwu Patty and {Huber}, Zachary B. and {Hubmayr}, Johannes and {Huffenberger}, Kevin M. and {Hughes}, John P. and {Irwin}, Kent and {Isopi}, Giovanni and {Jense}, Hidde T. and {Keller}, Ben and {Kim}, Joshua and {Knowles}, Kenda and {Koopman}, Brian J. and {Kosowsky}, Arthur and {Kramer}, Darby and {Kusiak}, Aleksandra and {La Posta}, Adrien and {Lague}, Alex and {Lakey}, Victoria and {Lee}, Eunseong and {Li}, Zack and {Limon}, Michele and {Lokken}, Martine and {Louis}, Thibaut and {Lungu}, Marius and {MacInnis}, Amanda and {Maldonado}, Diego and {Maldonado}, Felipe and {Mallaby-Kay}, Maya and {Marques}, Gabriela A. and {McMahon}, Jeff and {Mehta}, Yogesh and {Menanteau}, Felipe and {Moodley}, Kavilan and {Morris}, Thomas W. and {Mroczkowski}, Tony and {Naess}, Sigurd and {Namikawa}, Toshiya and {Nati}, Federico and {Newburgh}, Laura and {Nicola}, Andrina and {Niemack}, Michael D. and {Nolta}, Michael R. and {Orlowski-Scherer}, John and {Page}, Lyman A. and {Pandey}, Shivam and {Partridge}, Bruce and {Prince}, Heather and {Puddu}, Roberto and {Radiconi}, Federico and {Robertson}, Naomi and {Rojas}, Felipe and {Sakuma}, Tai and {Salatino}, Maria and {Schaan}, Emmanuel and {Schmitt}, Benjamin L. and {Sehgal}, Neelima and {Shaikh}, Shabbir and {Sierra}, Carlos and {Sievers}, Jon and {Sif{\'o}n}, Crist{\'o}bal and {Simon}, Sara and {Sonka}, Rita and {Spergel}, David N. and {Staggs}, Suzanne T. and {Storer}, Emilie and {Switzer}, Eric R. and {Tampier}, Niklas and {Thornton}, Robert and {Trac}, Hy and {Treu}, Jesse and {Tucker}, Carole and {Ullom}, Joel and {Vale}, Leila R. and {Van Engelen}, Alexander and {Van Lanen}, Jeff and {van Marrewijk}, Joshiwa and {Vargas}, Cristian and {Vavagiakis}, Eve M. and {Wagoner}, Kasey and {Wang}, Yuhan and {Wenzl}, Lukas and {Wollack}, Edward J. and {Xu}, Zhilei and {Zago}, Fernando and {Zheng}, Kaiwen},
 doi = {10.3847/1538-4357/acff5f},
 eid = {113},
 eprint = {2304.05203},
 journal = {\apj},
 month = {February},
 number = {2},
 pages = {113},
 primaryclass = {astro-ph.CO},
 title = {{The Atacama Cosmology Telescope: DR6 Gravitational Lensing Map and Cosmological Parameters}},
 volume = {962},
 year = {2024}
}

@article{2305.09827,
 adsurl = {https://ui.adsabs.harvard.edu/abs/2023MNRAS.525.5554P},
 archiveprefix = {arXiv},
 author = {{Preston}, Calvin and {Amon}, Alexandra and {Efstathiou}, George},
 doi = {10.1093/mnras/stad2573},
 eprint = {2305.09827},
 journal = {\mnras},
 month = {November},
 number = {4},
 pages = {5554-5564},
 primaryclass = {astro-ph.CO},
 title = {{A non-linear solution to the S$_{8}$ tension - II. Analysis of DES Year 3 cosmic shear}},
 volume = {525},
 year = {2023}
}

@article{2305.17173,
 adsurl = {https://ui.adsabs.harvard.edu/abs/2023OJAp....6E..36D},
 archiveprefix = {arXiv},
 author = {{Dark Energy Survey and Kilo-Degree Survey Collaboration} and {Abbott}, T.~M.~C. and {Aguena}, M. and {Alarcon}, A. and {Alves}, O. and {Amon}, A. and {Andrade-Oliveira}, F. and {Asgari}, M. and {Avila}, S. and {Bacon}, D. and {Bechtol}, K. and {Becker}, M.~R. and {Bernstein}, G.~M. and {Bertin}, E. and {Bilicki}, M. and {Blazek}, J. and {Bocquet}, S. and {Brooks}, D. and {Burger}, P. and {Burke}, D.~L. and {Camacho}, H. and {Campos}, A. and {Carnero Rosell}, A. and {Carrasco Kind}, M. and {Carretero}, J. and {Castander}, F.~J. and {Cawthon}, R. and {Chang}, C. and {Chen}, R. and {Choi}, A. and {Conselice}, C. and {Cordero}, J. and {Crocce}, M. and {da Costa}, L.~N. and {da Silva Pereira}, M.~E. and {Dalal}, R. and {Davis}, C. and {de Jong}, J.~T.~A. and {DeRose}, J. and {Desai}, S. and {Diehl}, H.~T. and {Dodelson}, S. and {Doel}, P. and {Doux}, C. and {Drlica-Wagner}, A. and {Dvornik}, A. and {Eckert}, K. and {Eifler}, T.~F. and {Elvin-Poole}, J. and {Everett}, S. and {Fang}, X. and {Ferrero}, I. and {Fert{\'e}}, A. and {Flaugher}, B. and {Friedrich}, O. and {Frieman}, J. and {Garc{\'\i}a-Bellido}, J. and {Gatti}, M. and {Giannini}, G. and {Giblin}, B. and {Gruen}, D. and {Gruendl}, R.~A. and {Gutierrez}, G. and {Harrison}, I. and {Hartley}, W.~G. and {Herner}, K. and {Heymans}, C. and {Hildebrandt}, H. and {Hinton}, S.~R. and {Hoekstra}, H. and {Hollowood}, D.~L. and {Honscheid}, K. and {Huang}, H. and {Huff}, E.~M. and {Huterer}, D. and {James}, D.~J. and {Jarvis}, M. and {Jeffrey}, N. and {Jeltema}, T. and {Joachimi}, B. and {Joudaki}, S. and {Kannawadi}, A. and {Krause}, E. and {Kuehn}, K. and {Kuijken}, K. and {Kuropatkin}, N. and {Lahav}, O. and {Leget}, P. -F. and {Lemos}, P. and {Li}, S. -S. and {Li}, X. and {Liddle}, A.~R. and {Lima}, M. and {Lin}, C. -A. and {Lin}, H. and {MacCrann}, N. and {Mahony}, C. and {Marshall}, J.~L. and {McCullough}, J. and {Mena-Fern{\'a}ndez}, J. and {Menanteau}, F. and {Miquel}, R. and {Mohr}, J.~J. and {Muir}, J. and {Myles}, J. and {Napolitano}, N. and {Navarro-Alsina}, A. and {Ogando}, R.~L.~C. and {Palmese}, A. and {Pandey}, S. and {Park}, Y. and {Paterno}, M. and {Peacock}, J.~A. and {Petravick}, D. and {Pieres}, A. and {Plazas Malag{\'o}n}, A.~A. and {Porredon}, A. and {Prat}, J. and {Radovich}, M. and {Raveri}, M. and {Reischke}, R. and {Robertson}, N.~C. and {Rollins}, R.~P. and {Romer}, A.~K. and {Roodman}, A. and {Rykoff}, E.~S. and {Samuroff}, S. and {S{\'a}nchez}, C. and {Sanchez}, E. and {Sanchez}, J. and {Schneider}, P. and {Secco}, L.~F. and {Sevilla-Noarbe}, I. and {Shan}, H. -Y. and {Sheldon}, E. and {Shin}, T. and {Sif{\'o}n}, C. and {Smith}, M. and {Soares-Santos}, M. and {St{\"o}lzner}, B. and {Suchyta}, E. and {Swanson}, M.~E.~C. and {Tarle}, G. and {Thomas}, D. and {To}, C. and {Troxel}, M.~A. and {Tr{\"o}ster}, T. and {Tutusaus}, I. and {van den Busch}, J.~L. and {Varga}, T.~N. and {Walker}, A.~R. and {Weaverdyck}, N. and {Wechsler}, R.~H. and {Weller}, J. and {Wiseman}, P. and {Wright}, A.~H. and {Yanny}, B. and {Yin}, B. and {Yoon}, M. and {Zhang}, Y. and {Zuntz}, J.},
 doi = {10.21105/astro.2305.17173},
 eid = {36},
 eprint = {2305.17173},
 journal = {The Open Journal of Astrophysics},
 month = {October},
 pages = {36},
 primaryclass = {astro-ph.CO},
 title = {{DES Y3 + KiDS-1000: Consistent cosmology combining cosmic shear surveys}},
 volume = {6},
 year = {2023}
}

@article{2308.16183,
 adsurl = {https://ui.adsabs.harvard.edu/abs/2024PhRvD.109f3523L},
 archiveprefix = {arXiv},
 author = {{Lin}, Meng-Xiang and {Jain}, Bhuvnesh and {Raveri}, Marco and {Baxter}, Eric J. and {Chang}, Chihway and {Gatti}, Marco and {Lee}, Sujeong and {Muir}, Jessica},
 doi = {10.1103/PhysRevD.109.063523},
 eid = {063523},
 eprint = {2308.16183},
 journal = {\prd},
 month = {March},
 number = {6},
 pages = {063523},
 primaryclass = {astro-ph.CO},
 title = {{Late time modification of structure growth and the S$_{8}$ tension}},
 volume = {109},
 year = {2024}
}

@article{2404.18240,
 adsurl = {https://ui.adsabs.harvard.edu/abs/2024MNRAS.533..621P},
 archiveprefix = {arXiv},
 author = {{Preston}, Calvin and {Amon}, Alexandra and {Efstathiou}, George},
 doi = {10.1093/mnras/stae1848},
 eprint = {2404.18240},
 journal = {\mnras},
 month = {September},
 number = {1},
 pages = {621-631},
 primaryclass = {astro-ph.CO},
 title = {{Reconstructing the matter power spectrum with future cosmic shear surveys}},
 volume = {533},
 year = {2024}
}

@article{2409.13404,
 adsurl = {https://ui.adsabs.harvard.edu/abs/2024A&A...692A.201B},
 archiveprefix = {arXiv},
 author = {{Broxterman}, Jeger C. and {Kuijken}, Konrad},
 doi = {10.1051/0004-6361/202452319},
 eid = {A201},
 eprint = {2409.13404},
 journal = {\aap},
 month = {December},
 pages = {A201},
 primaryclass = {astro-ph.CO},
 title = {{First step toward matter power spectrum reconstruction with Stage III weak gravitational lensing surveys}},
 volume = {692},
 year = {2024}
}

@article{2411.07082,
 adsurl = {https://ui.adsabs.harvard.edu/abs/2024arXiv241107082Y},
 archiveprefix = {arXiv},
 author = {{Ye}, Gen and {Jiang}, Jun-Qian and {Silvestri}, Alessandra},
 doi = {10.48550/arXiv.2411.07082},
 eid = {arXiv:2411.07082},
 eprint = {2411.07082},
 journal = {arXiv e-prints},
 month = {November},
 pages = {arXiv:2411.07082},
 primaryclass = {astro-ph.CO},
 title = {{A model-independent reconstruction of the matter power spectrum}},
 year = {2024}
}

@article{2502.04449,
 adsurl = {https://ui.adsabs.harvard.edu/abs/2025arXiv250204449S},
 archiveprefix = {arXiv},
 author = {{Simon}, Patrick and {Porth}, Lucas and {Burger}, Pierre and {Kuijken}, Konrad},
 doi = {10.48550/arXiv.2502.04449},
 eid = {arXiv:2502.04449},
 eprint = {2502.04449},
 journal = {arXiv e-prints},
 month = {February},
 pages = {arXiv:2502.04449},
 primaryclass = {astro-ph.CO},
 title = {{KiDS-1000: Detection of deviations from a purely cold dark matter power spectrum with tomographic weak gravitational lensing}},
 year = {2025}
}

@article{2502.06687,
 adsurl = {https://ui.adsabs.harvard.edu/abs/2025arXiv250206687P},
 archiveprefix = {arXiv},
 author = {{Perez Sarmiento}, Karen and {Lagu{\"e}}, Alex and {Madhavacheril}, Mathew and {Jain}, Bhuvnesh and {Sherwin}, Blake},
 doi = {10.48550/arXiv.2502.06687},
 eid = {arXiv:2502.06687},
 eprint = {2502.06687},
 journal = {arXiv e-prints},
 month = {February},
 pages = {arXiv:2502.06687},
 primaryclass = {astro-ph.CO},
 title = {{Reconstructing the shape of the non-linear matter power spectrum using CMB lensing and cosmic shear}},
 year = {2025}
}

@article{2503.19441,
 adsurl = {https://ui.adsabs.harvard.edu/abs/2025arXiv250319441W},
 archiveprefix = {arXiv},
 author = {{Wright}, Angus H. and {St{\"o}lzner}, Benjamin and {Asgari}, Marika and {Bilicki}, Maciej and {Giblin}, Benjamin and {Heymans}, Catherine and {Hildebrandt}, Hendrik and {Hoekstra}, Henk and {Joachimi}, Benjamin and {Kuijken}, Konrad and {Li}, Shun-Sheng and {Reischke}, Robert and {von Wietersheim-Kramsta}, Maximilian and {Yoon}, Mijin and {Burger}, Pierre and {Chisari}, Nora Elisa and {de Jong}, Jelte and {Dvornik}, Andrej and {Georgiou}, Christos and {Harnois-D{\'e}raps}, Joachim and {Jalan}, Priyanka and {William}, Anjitha John and {Joudaki}, Shahab and {Lesci}, Giorgio Francesco and {Linke}, Laila and {Loureiro}, Arthur and {Mahony}, Constance and {Maturi}, Matteo and {Miller}, Lance and {Moscardini}, Lauro and {Napolitano}, Nicola R. and {Porth}, Lucas and {Radovich}, Mario and {Schneider}, Peter and {Tr{\"o}ster}, Tilman and {Wittje}, Anna and {Yan}, Ziang and {Zhang}, Yun-Hao},
 doi = {10.48550/arXiv.2503.19441},
 eid = {arXiv:2503.19441},
 eprint = {2503.19441},
 journal = {arXiv e-prints},
 month = {March},
 pages = {arXiv:2503.19441},
 primaryclass = {astro-ph.CO},
 title = {{KiDS-Legacy: Cosmological constraints from cosmic shear with the complete Kilo-Degree Survey}},
 year = {2025}
}

@article{2505.02233,
 adsurl = {https://ui.adsabs.harvard.edu/abs/2025arXiv250502233P},
 archiveprefix = {arXiv},
 author = {{Preston}, Calvin and {Rogers}, Keir K. and {Amon}, Alexandra and {Efstathiou}, George},
 doi = {10.48550/arXiv.2505.02233},
 eid = {arXiv:2505.02233},
 eprint = {2505.02233},
 journal = {arXiv e-prints},
 month = {May},
 pages = {arXiv:2505.02233},
 primaryclass = {astro-ph.CO},
 title = {{Prospects for disentangling dark matter with weak lensing}},
 year = {2025}
}

@article{Brinckmann:2018cvx,
 adsurl = {https://ui.adsabs.harvard.edu/abs/2019PDU....24..260B},
 archiveprefix = {arXiv},
 author = {{Brinckmann}, Thejs and {Lesgourgues}, Julien},
 doi = {10.1016/j.dark.2018.100260},
 eid = {100260},
 eprint = {1804.07261},
 journal = {Physics of the Dark Universe},
 month = {March},
 pages = {100260},
 primaryclass = {astro-ph.CO},
 title = {{MontePython 3: Boosted MCMC sampler and other features}},
 volume = {24},
 year = {2019}
}

@article{Dalal:2023olq,
 adsurl = {https://ui.adsabs.harvard.edu/abs/2023PhRvD.108l3519D},
 archiveprefix = {arXiv},
 author = {{Dalal}, Roohi and {Li}, Xiangchong and {Nicola}, Andrina and {Zuntz}, Joe and {Strauss}, Michael A. and {Sugiyama}, Sunao and {Zhang}, Tianqing and {Rau}, Markus M. and {Mandelbaum}, Rachel and {Takada}, Masahiro and {More}, Surhud and {Miyatake}, Hironao and {Kannawadi}, Arun and {Shirasaki}, Masato and {Taniguchi}, Takanori and {Takahashi}, Ryuichi and {Osato}, Ken and {Hamana}, Takashi and {Oguri}, Masamune and {Nishizawa}, Atsushi J. and {Malag{\'o}n}, Andr{\'e}s A. Plazas and {Sunayama}, Tomomi and {Alonso}, David and {Slosar}, An{\v{z}}e and {Luo}, Wentao and {Armstrong}, Robert and {Bosch}, James and {Hsieh}, Bau-Ching and {Komiyama}, Yutaka and {Lupton}, Robert H. and {Lust}, Nate B. and {MacArthur}, Lauren A. and {Miyazaki}, Satoshi and {Murayama}, Hitoshi and {Nishimichi}, Takahiro and {Okura}, Yuki and {Price}, Paul A. and {Tait}, Philip J. and {Tanaka}, Masayuki and {Wang}, Shiang-Yu},
 doi = {10.1103/PhysRevD.108.123519},
 eid = {123519},
 eprint = {2304.00701},
 journal = {\prd},
 month = {December},
 number = {12},
 pages = {123519},
 primaryclass = {astro-ph.CO},
 title = {{Hyper Suprime-Cam Year 3 results: Cosmology from cosmic shear power spectra}},
 volume = {108},
 year = {2023}
}

@article{Khoury:2025txd,
 adsurl = {https://ui.adsabs.harvard.edu/abs/2025arXiv250316415K},
 archiveprefix = {arXiv},
 author = {{Khoury}, Justin and {Lin}, Meng-Xiang and {Trodden}, Mark},
 doi = {10.48550/arXiv.2503.16415},
 eid = {arXiv:2503.16415},
 eprint = {2503.16415},
 journal = {arXiv e-prints},
 month = {March},
 pages = {arXiv:2503.16415},
 primaryclass = {astro-ph.CO},
 title = {{Apparent $w<-1$ and a Lower $S_8$ from Dark Axion and Dark Baryons Interactions}},
 year = {2025}
}

@article{Planck:2018vyg,
 adsurl = {https://ui.adsabs.harvard.edu/abs/2020A&A...641A...6P},
 archiveprefix = {arXiv},
 author = {{Planck Collaboration} and {Aghanim}, N. and {Akrami}, Y. and {Ashdown}, M. and {Aumont}, J. and {Baccigalupi}, C. and {Ballardini}, M. and {Banday}, A.~J. and {Barreiro}, R.~B. and {Bartolo}, N. and {Basak}, S. and {Battye}, R. and {Benabed}, K. and {Bernard}, J. -P. and {Bersanelli}, M. and {Bielewicz}, P. and {Bock}, J.~J. and {Bond}, J.~R. and {Borrill}, J. and {Bouchet}, F.~R. and {Boulanger}, F. and {Bucher}, M. and {Burigana}, C. and {Butler}, R.~C. and {Calabrese}, E. and {Cardoso}, J. -F. and {Carron}, J. and {Challinor}, A. and {Chiang}, H.~C. and {Chluba}, J. and {Colombo}, L.~P.~L. and {Combet}, C. and {Contreras}, D. and {Crill}, B.~P. and {Cuttaia}, F. and {de Bernardis}, P. and {de Zotti}, G. and {Delabrouille}, J. and {Delouis}, J. -M. and {Di Valentino}, E. and {Diego}, J.~M. and {Dor{\'e}}, O. and {Douspis}, M. and {Ducout}, A. and {Dupac}, X. and {Dusini}, S. and {Efstathiou}, G. and {Elsner}, F. and {En{\ss}lin}, T.~A. and {Eriksen}, H.~K. and {Fantaye}, Y. and {Farhang}, M. and {Fergusson}, J. and {Fernandez-Cobos}, R. and {Finelli}, F. and {Forastieri}, F. and {Frailis}, M. and {Fraisse}, A.~A. and {Franceschi}, E. and {Frolov}, A. and {Galeotta}, S. and {Galli}, S. and {Ganga}, K. and {G{\'e}nova-Santos}, R.~T. and {Gerbino}, M. and {Ghosh}, T. and {Gonz{\'a}lez-Nuevo}, J. and {G{\'o}rski}, K.~M. and {Gratton}, S. and {Gruppuso}, A. and {Gudmundsson}, J.~E. and {Hamann}, J. and {Handley}, W. and {Hansen}, F.~K. and {Herranz}, D. and {Hildebrandt}, S.~R. and {Hivon}, E. and {Huang}, Z. and {Jaffe}, A.~H. and {Jones}, W.~C. and {Karakci}, A. and {Keih{\"a}nen}, E. and {Keskitalo}, R. and {Kiiveri}, K. and {Kim}, J. and {Kisner}, T.~S. and {Knox}, L. and {Krachmalnicoff}, N. and {Kunz}, M. and {Kurki-Suonio}, H. and {Lagache}, G. and {Lamarre}, J. -M. and {Lasenby}, A. and {Lattanzi}, M. and {Lawrence}, C.~R. and {Le Jeune}, M. and {Lemos}, P. and {Lesgourgues}, J. and {Levrier}, F. and {Lewis}, A. and {Liguori}, M. and {Lilje}, P.~B. and {Lilley}, M. and {Lindholm}, V. and {L{\'o}pez-Caniego}, M. and {Lubin}, P.~M. and {Ma}, Y. -Z. and {Mac{\'\i}as-P{\'e}rez}, J.~F. and {Maggio}, G. and {Maino}, D. and {Mandolesi}, N. and {Mangilli}, A. and {Marcos-Caballero}, A. and {Maris}, M. and {Martin}, P.~G. and {Martinelli}, M. and {Mart{\'\i}nez-Gonz{\'a}lez}, E. and {Matarrese}, S. and {Mauri}, N. and {McEwen}, J.~D. and {Meinhold}, P.~R. and {Melchiorri}, A. and {Mennella}, A. and {Migliaccio}, M. and {Millea}, M. and {Mitra}, S. and {Miville-Desch{\^e}nes}, M. -A. and {Molinari}, D. and {Montier}, L. and {Morgante}, G. and {Moss}, A. and {Natoli}, P. and {N{\o}rgaard-Nielsen}, H.~U. and {Pagano}, L. and {Paoletti}, D. and {Partridge}, B. and {Patanchon}, G. and {Peiris}, H.~V. and {Perrotta}, F. and {Pettorino}, V. and {Piacentini}, F. and {Polastri}, L. and {Polenta}, G. and {Puget}, J. -L. and {Rachen}, J.~P. and {Reinecke}, M. and {Remazeilles}, M. and {Renzi}, A. and {Rocha}, G. and {Rosset}, C. and {Roudier}, G. and {Rubi{\~n}o-Mart{\'\i}n}, J.~A. and {Ruiz-Granados}, B. and {Salvati}, L. and {Sandri}, M. and {Savelainen}, M. and {Scott}, D. and {Shellard}, E.~P.~S. and {Sirignano}, C. and {Sirri}, G. and {Spencer}, L.~D. and {Sunyaev}, R. and {Suur-Uski}, A. -S. and {Tauber}, J.~A. and {Tavagnacco}, D. and {Tenti}, M. and {Toffolatti}, L. and {Tomasi}, M. and {Trombetti}, T. and {Valenziano}, L. and {Valiviita}, J. and {Van Tent}, B. and {Vibert}, L. and {Vielva}, P. and {Villa}, F. and {Vittorio}, N. and {Wandelt}, B.~D. and {Wehus}, I.~K. and {White}, M. and {White}, S.~D.~M. and {Zacchei}, A. and {Zonca}, A.},
 doi = {10.1051/0004-6361/201833910},
 eid = {A6},
 eprint = {1807.06209},
 journal = {\aap},
 month = {September},
 pages = {A6},
 primaryclass = {astro-ph.CO},
 title = {{Planck 2018 results. VI. Cosmological parameters}},
 volume = {641},
 year = {2020}
}

@article{Poulin:2022sgp,
 adsurl = {https://ui.adsabs.harvard.edu/abs/2023PhRvD.107l3538P},
 archiveprefix = {arXiv},
 author = {{Poulin}, Vivian and {Bernal}, Jos{\'e} Luis and {Kovetz}, Ely D. and {Kamionkowski}, Marc},
 doi = {10.1103/PhysRevD.107.123538},
 eid = {123538},
 eprint = {2209.06217},
 journal = {\prd},
 month = {June},
 number = {12},
 pages = {123538},
 primaryclass = {astro-ph.CO},
 title = {{Sigma-8 tension is a drag}},
 volume = {107},
 year = {2023}
}

@article{PyMC,
 author = {Abril-Pla, Oriol and Andreani, Virgile and Carroll, Colin and Dong, Larry and Fonnesbeck, Christopher J and Kochurov, Maxim and Kumar, Ravin and Lao, Junpeng and Luhmann, Christian C and Martin, Osvaldo A and others},
 journal = {PeerJ Computer Science},
 pages = {e1516},
 publisher = {PeerJ Inc.},
 title = {PyMC: a modern, and comprehensive probabilistic programming framework in Python},
 volume = {9},
 year = {2023}
}

@article{TZ,
 adsurl = {https://ui.adsabs.harvard.edu/abs/2002PhRvD..66j3508T},
 archiveprefix = {arXiv},
 author = {{Tegmark}, Max and {Zaldarriaga}, Matias},
 doi = {10.1103/PhysRevD.66.103508},
 eid = {103508},
 eprint = {astro-ph/0207047},
 journal = {\prd},
 month = {November},
 number = {10},
 pages = {103508},
 primaryclass = {astro-ph},
 title = {{Separating the early universe from the late universe: Cosmological parameter estimation beyond the black box}},
 volume = {66},
 year = {2002}
}

@ARTICLE{2023PhRvD.107h3504A,
       author = {{Abbott}, T.~M.~C. and {Aguena}, M. and {Alarcon}, A. and {Alves}, O. and {Amon}, A. and {Andrade-Oliveira}, F. and {Annis}, J. and {Avila}, S. and {Bacon}, D. and {Baxter}, E. and {Bechtol}, K. and {Becker}, M.~R. and {Bernstein}, G.~M. and {Birrer}, S. and {Blazek}, J. and {Bocquet}, S. and {Brandao-Souza}, A. and {Bridle}, S.~L. and {Brooks}, D. and {Burke}, D.~L. and {Camacho}, H. and {Campos}, A. and {Carnero Rosell}, A. and {Carrasco Kind}, M. and {Carretero}, J. and {Castander}, F.~J. and {Cawthon}, R. and {Chang}, C. and {Chen}, A. and {Chen}, R. and {Choi}, A. and {Conselice}, C. and {Cordero}, J. and {Costanzi}, M. and {Crocce}, M. and {da Costa}, L.~N. and {Pereira}, M.~E.~S. and {Davis}, C. and {Davis}, T.~M. and {DeRose}, J. and {Desai}, S. and {Di Valentino}, E. and {Diehl}, H.~T. and {Dodelson}, S. and {Doel}, P. and {Doux}, C. and {Drlica-Wagner}, A. and {Eckert}, K. and {Eifler}, T.~F. and {Elsner}, F. and {Elvin-Poole}, J. and {Everett}, S. and {Fang}, X. and {Farahi}, A. and {Ferrero}, I. and {Fert{\'e}}, A. and {Flaugher}, B. and {Fosalba}, P. and {Friedel}, D. and {Friedrich}, O. and {Frieman}, J. and {Garc{\'\i}a-Bellido}, J. and {Gatti}, M. and {Giani}, L. and {Giannantonio}, T. and {Giannini}, G. and {Gruen}, D. and {Gruendl}, R.~A. and {Gschwend}, J. and {Gutierrez}, G. and {Hamaus}, N. and {Harrison}, I. and {Hartley}, W.~G. and {Herner}, K. and {Hinton}, S.~R. and {Hollowood}, D.~L. and {Honscheid}, K. and {Huang}, H. and {Huff}, E.~M. and {Huterer}, D. and {Jain}, B. and {James}, D.~J. and {Jarvis}, M. and {Jeffrey}, N. and {Jeltema}, T. and {Kovacs}, A. and {Krause}, E. and {Kuehn}, K. and {Kuropatkin}, N. and {Lahav}, O. and {Lee}, S. and {Leget}, P. -F. and {Lemos}, P. and {Leonard}, C.~D. and {Liddle}, A.~R. and {Lima}, M. and {Lin}, H. and {MacCrann}, N. and {Marshall}, J.~L. and {McCullough}, J. and {Mena-Fern{\'a}ndez}, J. and {Menanteau}, F. and {Miquel}, R. and {Miranda}, V. and {Mohr}, J.~J. and {Muir}, J. and {Myles}, J. and {Nadathur}, S. and {Navarro-Alsina}, A. and {Nichol}, R.~C. and {Ogando}, R.~L.~C. and {Omori}, Y. and {Palmese}, A. and {Pandey}, S. and {Park}, Y. and {Paterno}, M. and {Paz-Chinch{\'o}n}, F. and {Percival}, W.~J. and {Pieres}, A. and {Plazas Malag{\'o}n}, A.~A. and {Porredon}, A. and {Prat}, J. and {Raveri}, M. and {Rodriguez-Monroy}, M. and {Rogozenski}, P. and {Rollins}, R.~P. and {Romer}, A.~K. and {Roodman}, A. and {Rosenfeld}, R. and {Ross}, A.~J. and {Rykoff}, E.~S. and {Samuroff}, S. and {S{\'a}nchez}, C. and {Sanchez}, E. and {Sanchez}, J. and {Sanchez Cid}, D. and {Scarpine}, V. and {Scolnic}, D. and {Secco}, L.~F. and {Sevilla-Noarbe}, I. and {Sheldon}, E. and {Shin}, T. and {Smith}, M. and {Soares-Santos}, M. and {Suchyta}, E. and {Tabbutt}, M. and {Tarle}, G. and {Thomas}, D. and {To}, C. and {Troja}, A. and {Troxel}, M.~A. and {Tutusaus}, I. and {Varga}, T.~N. and {Vincenzi}, M. and {Walker}, A.~R. and {Weaverdyck}, N. and {Wechsler}, R.~H. and {Weller}, J. and {Yanny}, B. and {Yin}, B. and {Zhang}, Y. and {Zuntz}, J. and {DES Collaboration}},
        title = "{Dark Energy Survey Year 3 results: Constraints on extensions to {\ensuremath{\Lambda}} CDM with weak lensing and galaxy clustering}",
      journal = {\prd},
         year = 2023,
        month = apr,
       volume = {107},
       number = {8},
          eid = {083504},
        pages = {083504},
          doi = {10.1103/PhysRevD.107.083504},
archivePrefix = {arXiv},
       eprint = {2207.05766},
 primaryClass = {astro-ph.CO},
       adsurl = {https://ui.adsabs.harvard.edu/abs/2023PhRvD.107h3504A}
}

@article{Asghari:2019qld,
    author = "Asghari, Mahnaz and Beltr{\'a}n Jim{\'e}nez, Jose and Khosravi, Shahram and Mota, David F.",
    title = "{On structure formation from a small-scales-interacting dark sector}",
    eprint = "1902.05532",
    archivePrefix = "arXiv",
    primaryClass = "astro-ph.CO",
    doi = "10.1088/1475-7516/2019/04/042",
    journal = "JCAP",
    volume = "04",
    pages = "042",
    year = "2019"
}

@article{BeltranJimenez:2021wbq,
    author = "Beltr{\'a}n Jim{\'e}nez, Jose and Bettoni, Dario and Figueruelo, David and Teppa Pannia, Florencia Anabella and Tsujikawa, Shinji",
    title = "{Probing elastic interactions in the dark sector and the role of S8}",
    eprint = "2106.11222",
    archivePrefix = "arXiv",
    primaryClass = "astro-ph.CO",
    doi = "10.1103/PhysRevD.104.103503",
    journal = "Phys. Rev. D",
    volume = "104",
    number = "10",
    pages = "103503",
    year = "2021"
}

\begin{appendix}

\section{Impact of systematics and scale cuts}
\label{app:syst}

\begin{figure*}
    \centering
    \includegraphics[scale=0.68]{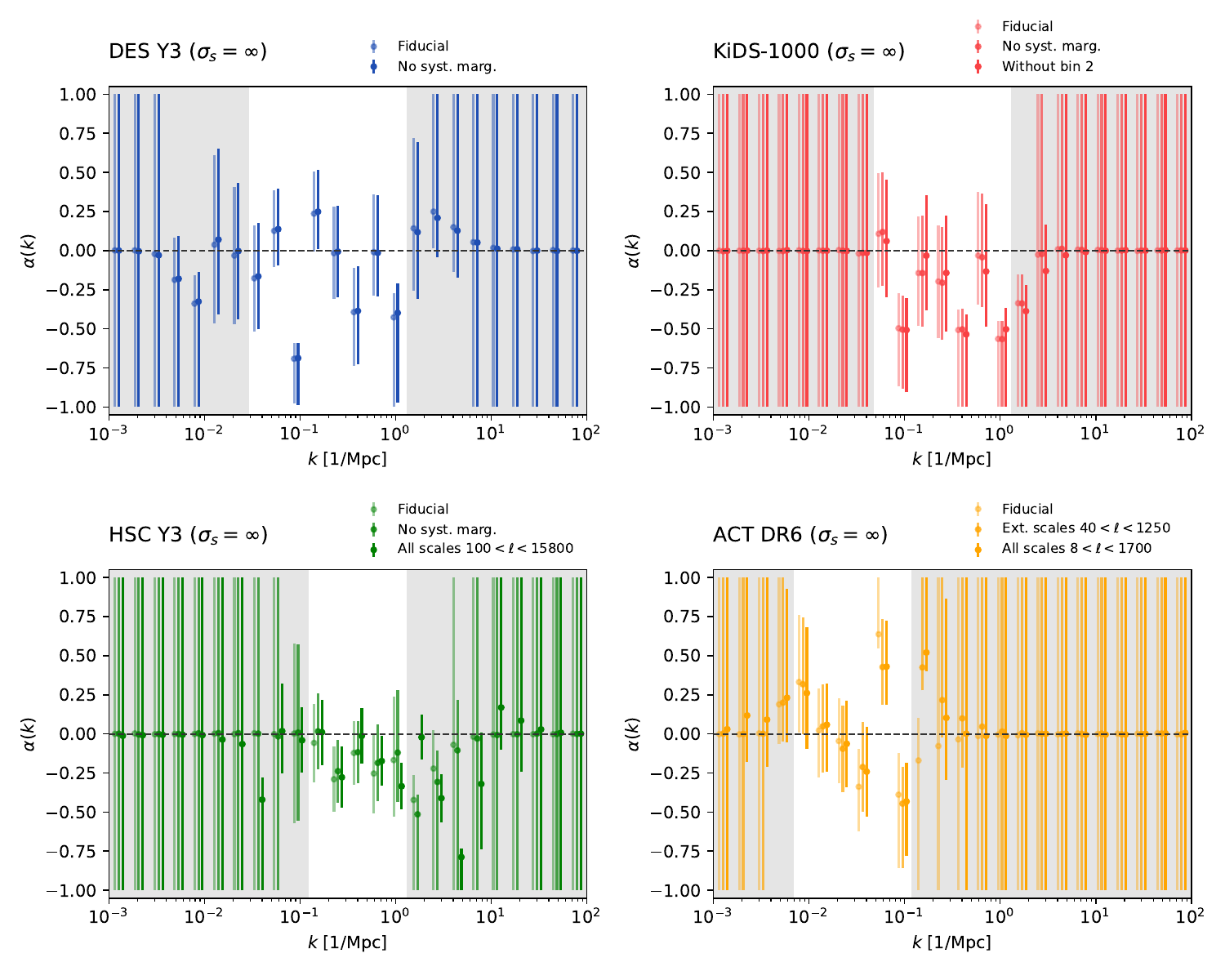}
    \caption{
    Constraints on $\alpha(k)$ obtained from the four lensing surveys considered in this work, varying several assumptions. The fiducial case is that presented in \cref{fig:all_Pk}, using analytic systematics marginalization and fiducial scale cuts. We present results obtained when ignoring this marginalization for galaxy surveys. We also explore several extended scale cuts for HSC and ACT, and the effect of removing the second redshift bin for KiDS. No smoothing prior is applied, such that data-driven variations may be appreciated.
    }
    \label{fig:all_Pk_ratio_boxes_tests}
\end{figure*}

\Cref{fig:all_Pk_ratio_boxes_tests} shows $\alpha(k)$ constraints for the four surveys varying some of the assumptions in our methodology as well as scale cuts, demonstrating the overall robustness of our results.

\section{{Impact of fiducial cosmology}}
\label{app:des_cosmo}

\begin{figure*}
    \centering
    \includegraphics[scale=0.68]{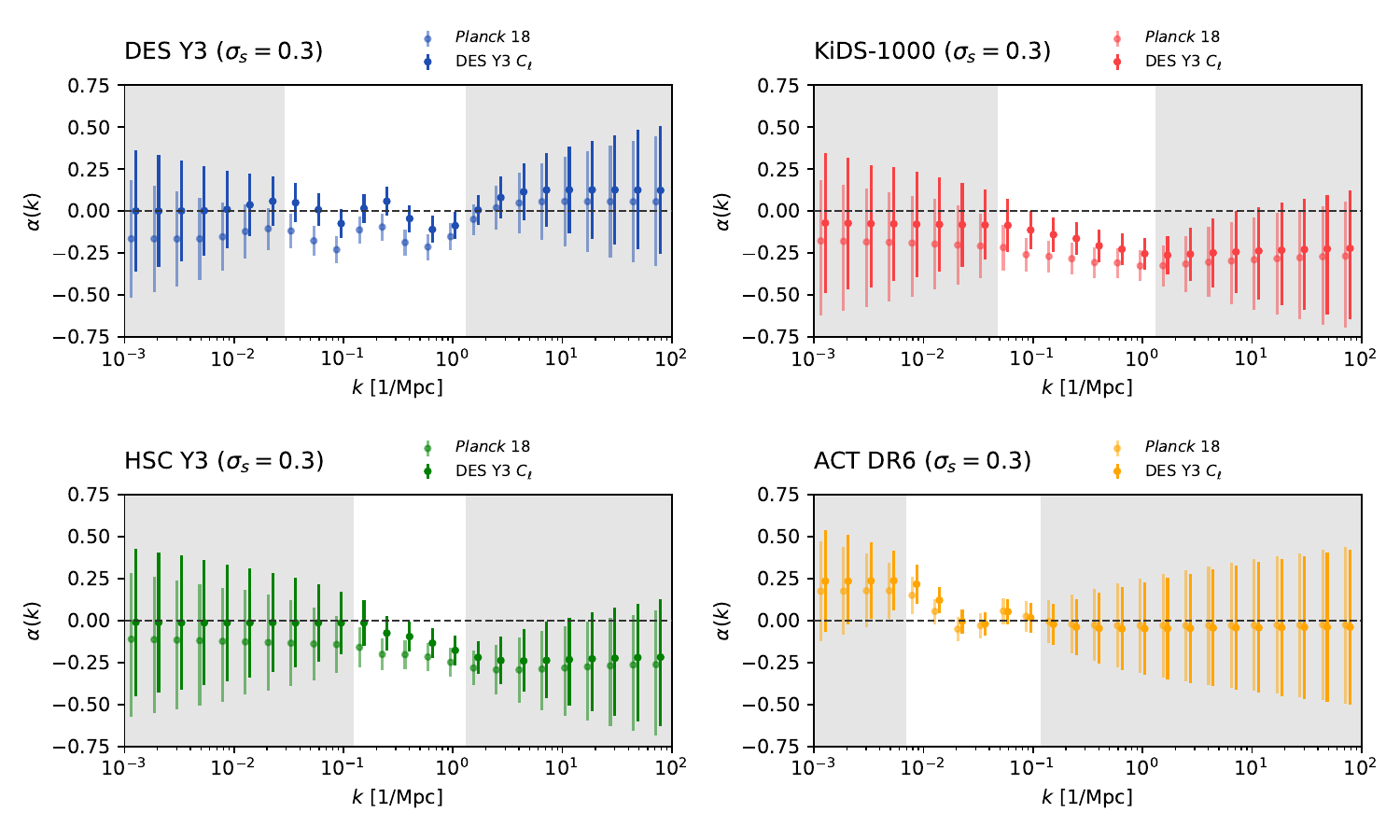}
    \caption{
    {
    Constraints on $\alpha(k)$ obtained from the four lensing surveys considered in this work, for two different fiducial cosmologies: \planck~2018~\LCDM~\citep[][$\Omega_{\mathrm{m}}=0.316$, $S_8=0.834$, as shown in \cref{fig:all_Pk}]{Planck:2018vyg} and DES~Y3 harmonic-space cosmic shear~\citep[][$\Omega_{\mathrm{m}}=0.257$, $S_8=0.803$]{2203.07128}.
    A smoothing prior of $\sigma_{\rm s}=0.3$ is applied.}
    }
    \label{fig:all_Pk_ratio_boxes_tests_cosmo}
\end{figure*}

{To examine the dependence of our results on the choice of fiducial cosmology, we repeat the analysis of \cref{sec:res_ind} using an alternative nonlinear matter power spectrum and corresponding window matrices computed at the mean cosmology inferred from the DES~Y3 harmonic-space analysis~\citep{2203.07128}, rather than from the \planck~2018~\LCDM cosmology~\citep{Planck:2018vyg}.
\Cref{fig:all_Pk_ratio_boxes_tests_cosmo} compares the resulting $\alpha(k)$ constraints for the four lensing surveys considered in this work under both cosmologies.
}

{For the galaxy-lensing surveys, adopting the DES~Y3 cosmology increases the overall amplitude of $\alpha(k)$ relative to the \planck-based results.
DES becomes consistent with $\alpha(k)=0$, as expected since the DES fiducial cosmology corresponds to the posterior mean of its own data.
For HSC and KiDS, $\alpha(k)$ remain lower than zero. This is consistent with the smaller level of tension between DES~Y3 and \planck~(see Fig.~16 of~\citealt{2203.07128}) relative to \planck and HSC, KiDS, which still requires a suppression in $\alpha(k)$, albeit smaller, at this new fiducial. 
Beyond this overall amplitude shift, we find that the scale-dependence of $\alpha(k)$ within the constrained $k$-range remains very similar between the two fiducial cosmologies, with a more pronounced discrepancy at intermediate scales ($k\sim\SI{0.1}{\iMpc}$) than at nonlinear scales ($k\sim\SI{1}{\iMpc}$), where differences between cosmologies become less significant. For ACT lensing, differences are even smaller and indistinguishable for $k\gtrsim\SI{0.03}{\iMpc}$.
}

\section{Full posteriors from ACT, DES and HSC}
\label{app:full_posterior}

\begin{figure*}
    \centering
    \includegraphics[scale=0.68]{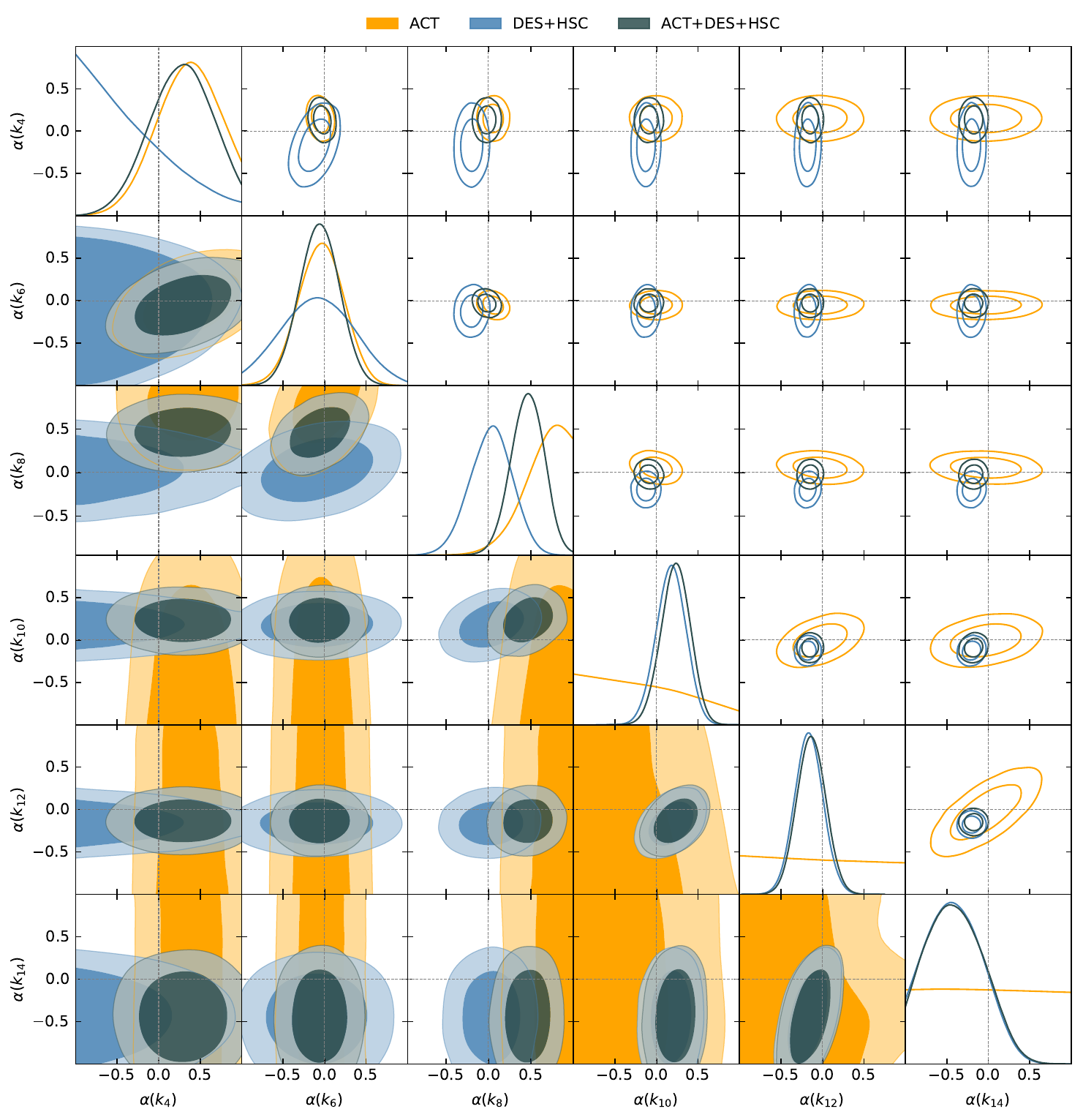}
    \caption{
    {Constraints on $\alpha(k)$ from ACT~DR6 (yellow), DES~Y3 combined with HSC~Y3 (blue), and their combination (dark gray), using the uniform prior (lower triangle and diagonal) or the standard $\sigma_{\rm s}=0.3$ smoothing prior (upper triangle). 
    For readibility, we only show every other $k$-bin in the range \SIrange{7e-3}{1.2}{\iMpc} which is constrained by ACT+DES+HSC data. 
    The effects of the prior are clearly visible in the lower triangle, while the upper one illustrates how the smoothed posteriors are Gaussian and mostly uncorrelated in the constrained range.}
    }
    \label{fig:DES_HSC_ACT_comb_Pk_corner}
\end{figure*}

\begin{figure*}
    \centering
    \includegraphics[scale=0.68]{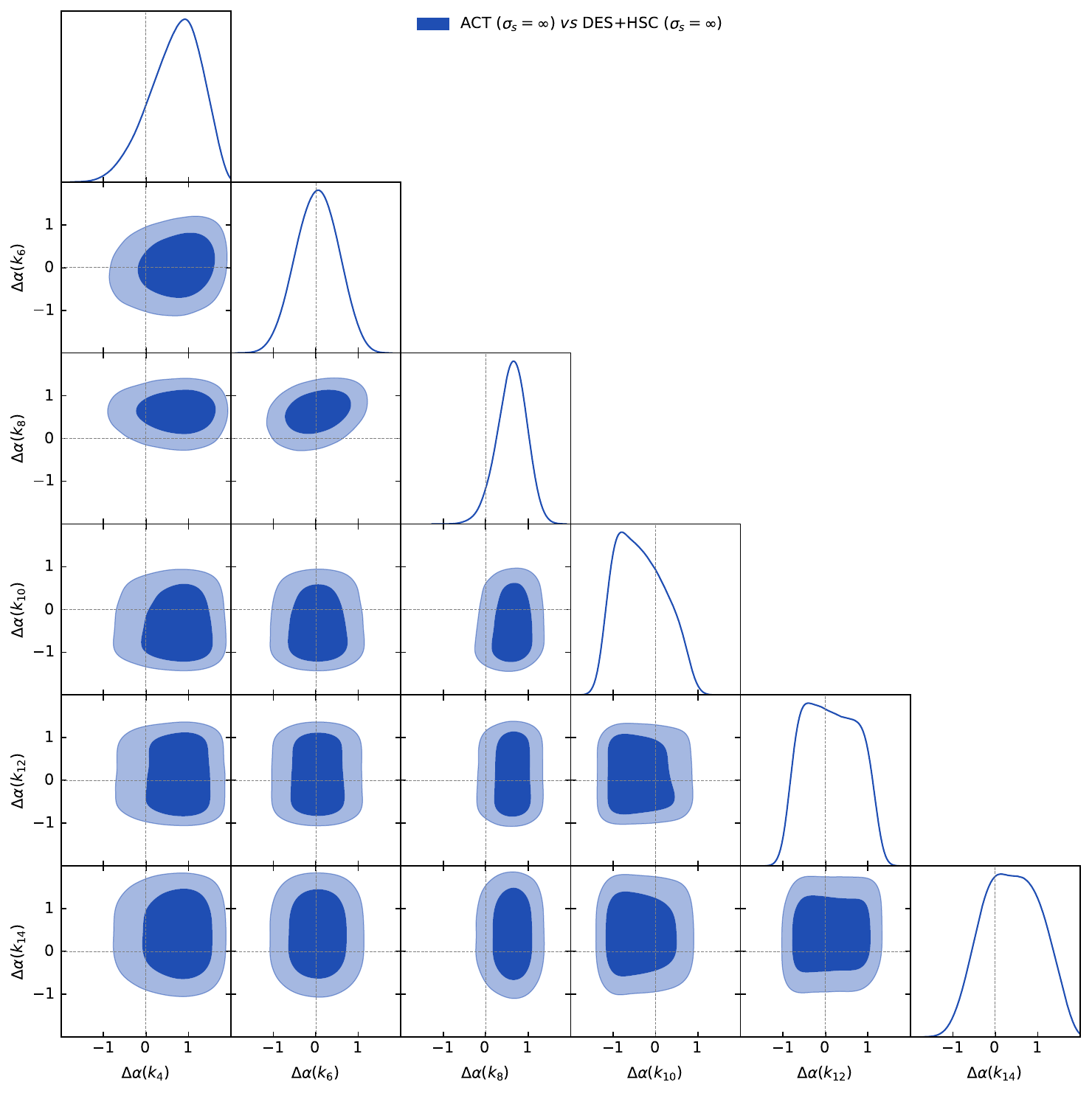}
    \caption{
    {Marginals of the parameter-difference distribution between ACT and DES+HSC (uniform prior) for the same $k$-bins as \cref{fig:DES_HSC_ACT_comb_Pk_corner}. This figure illustrates the non-Gaussianity of the distributions due to prior boundaries. For this reason, we estimate consistency using the non-Gaussian metric in \texttt{Tensiometer}, corresponding to the volume of the difference distribution within the isodensity contour crossing the zero-difference point~\citep[see][for details]{2021PhRvD.104d3504R}.}
    }
    \label{fig:DES_HSC_ACT_comb_Pk_corner_diff}
\end{figure*}

\Cref{fig:DES_HSC_ACT_comb_Pk_corner} shows the one- and two-dimensional marginal distributions for constraints obtained from ACT, DES, HSC and their combinations. 
\Cref{fig:DES_HSC_ACT_comb_Pk_corner_diff} shows the parameter-difference distribution used in estimating the level of agreement between ACT and DES+HSC data before combining them. 

\section{Comparing \LCDM constraints from $C_\ell$ and $\alpha(k)$}

\begin{figure*}
    \centering
    \includegraphics[scale=0.68]{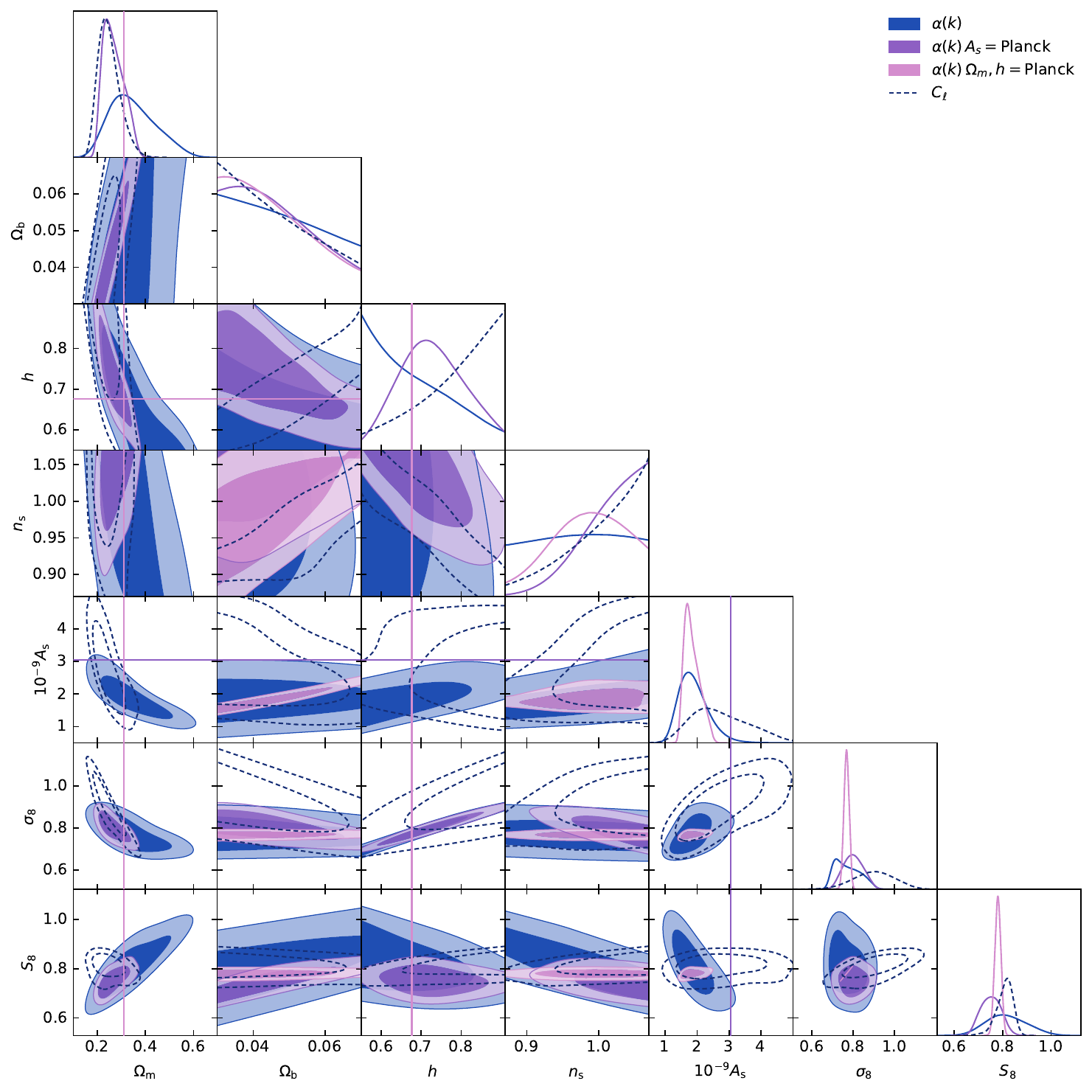}
    \caption{
    Comparison of constraints on \LCDM parameters inferred from the derived $\alpha(k)$ likelihoods and those obtained directly from the power spectra $C_\ell$ for DES Y3 data. 
    The blue contours vary all \LCDM parameters, same as the $C_\ell$ chains. 
    We also draw a comparison with fixing $A_{\rm s}$ at the \planck fiducial value in Table \ref{tab:fiducial_cosmology} to isolate shifts in the overall amplitude of $\alpha(k)$ in purple. 
    Lastly, we also separately factor out the background cosmology by fixing $\Om$ and $h$ to their fiducial values in pink. 
    For both these cases, the values at which the \LCDM parameters are fixed are marked by horizontal and vertical lines following the legend colors.
    None of these methods recover the same posterior footprint as the power-spectrum analysis. 
    }
    \label{fig:alpha_vs_Cls_full}
\end{figure*}

\Cref{fig:alpha_vs_Cls_full} compares the posteriors of subsets of \LCDM parameters when constrained with the $\alpha(k)$ likelihoods relative to those obtained from a full analysis of the shear power spectra $C_\ell$ for DES Y3 data. 
We isolate an overall amplitude shift in $\alpha(k)$ by fixing the amplitude $A_{\rm s}$ of the primordial power spectrum to \planck. 
We also separately fix the background cosmology to the fiducial \planck cosmology by fixing $\Omega_{\rm m}$ and $h$. 
Lastly, we vary all \LCDM parameters. 
None of these scenarios reproduces the same contour extents or degeneracy directions as the $C_\ell$ posteriors, demonstrating that the two approaches have fundamentally different information contents.
\\

\end{appendix}

\end{document}